\documentclass[12pt,a4paper]{article}
\usepackage[latin1]{inputenc}
\usepackage{times, mathptm}
\usepackage{graphicx}
\def\degr{^{\circ}}
\def\lsi{LS~I~+61$^{\circ}$303~}
\def\beq{\begin{equation}}
\def\eneq{\end{equation}}
\begin{document}
\ifx\href\undefined\else\hypersetup{linktocpage=true}\fi

\bigskip
\bigskip
{\centerline {\LARGE {\bf Introduction to  Astrophysics of  Microquasars}}} 
{---------------------------------------------------------------------------------------}\newline
\bigskip
\bigskip
\bigskip
\bigskip
\bigskip
\bigskip
\bigskip
\bigskip
\bigskip
\bigskip
\bigskip
{\centerline {\LARGE {\bf Einf\"uhrung  in die Astrophysik der  Mikroquasare}}} 
\bigskip
{\centerline {\LARGE {\bf Habilitationsschrift}}}\newline 

\bigskip
\bigskip
\bigskip
\bigskip
\bigskip
\bigskip

\bigskip

\bigskip
\bigskip
\bigskip
{\normalfont
zur
\newline
   Erlangung der Venia Legendi der Hohen Mathematisch-Naturwissenschaftlichen Fakul\"at
 der
{\it Rheinischen Friedrich-Wilhelms-Universit\"at Bonn}
\newline
\newline
\bigskip
\bigskip

vorgelegt von Dr. rer. nat.}  
{\Large {\bf  Maria Massi}} aus Rom
\bigskip
\newline
\bigskip
Bonn, im Dezember 2004 
\newpage
{\LARGE {FOREWORD}}

\bigskip
\bigskip
\bigskip
This review was written to fulfill the requirements for the
"Habilitation" procedure  at the University of Bonn.
It summarizes a part of the research I have conducted over the
last years in the field of microquasars 
and it should  stimulate  students' interest in this
challenging and fascinating topic.

\bigskip

The astronomical  methods used in
the range of gamma-rays, X-rays, optical and  radio wavelengths  
are first reviewed and 
then, all of them,  directly applied always to the same astronomical
 source,
the periodic microquasar {LS~I~+61$^{\circ}$303}.

\bigskip

Reviewing the literature in that field, this presentation 
 is of course biased towards my own contributions 
 for the reason mentioned in the beginning.
Therefore, I want to  
acknowledge  here  all colleagues working like me 
for years on LS~I~+61$^{\circ}$303,   first of all 
the two discoverers of the source: P.C. Gregory and A.R. Taylor,  but also 
D. Crampton and J.B. Hutching for the optical observations,
F.A. Harrison, D.A. Leahy and M. Tavani for the high energy, K,M.V Apparao,
and R.K, Zamanov for  
multiband  studies, and of course all my collaborators:
 J.M. Paredes, M.  Rib\'o, J.  Mart\'{\i},
S. Garrington and M. Peracaula.  
\bigskip
\bigskip
\bigskip

Maria Massi, December 2004

\newpage
{\LARGE {CONTENTS}}

\bigskip
\bigskip
\bigskip
\bigskip

1. Introduction.............................................................................................................4
\bigskip

 2. The Accretion-Ejection Process..............................................................................8

~~   2.1 Accretion.............................................................................................................8

~~   2.2 Magnetohydrodynamic Jet Production................................................................9

~~   2.3 Strong Magnetic Fields: The X-Ray Pulsars......................................................11

\bigskip
 3. Optical Observations.............................................................................................12

~~   3.1 The Nature of the Compact Object.....................................................................12

~~   3.2 The Nature of the Companion Star.....................................................................13

~~   3.3 \lsi:  The  Be Star ...............................................................................14

~~   3.4 \lsi: Really a Neutron Star ?..............................................................17

\bigskip
 4. X-Ray and Radio Observations.............................................................................17

~~   4.1 High/Soft State and Multicolor Disk.................................................................19

~~   4.2 The Disk-Jet Connection...................................................................................21

~~   4.3 \lsi: Soft and Hard X-Ray States......................................................29

~~   4.4 The Periodical Radio Outbursts of \lsi..............................................34

\bigskip
 5. Theory of Accretion: The Two-Peak Accretion Model.........................................35

\bigskip
 6. Gamma-Ray Observations....................................................................................38

~~   6.1 EGRET Sources................................................................................................38

~~   6.2 the Variable Gamma-ray Source \lsi.................................................40

\bigskip
 7. Radio Interferometry: Imaging at High Resolution...............................................42 

~~   7.1 The Jet Velocity............................................................................... ..................42

~~   7.2 The Precessing Relativistic Jet of LSI+61303....................................................44

\bigskip
 8. Conclusions...........................................................................................................50 
 
\bigskip
 9. Summary................................................................................................................52

\bigskip
 10.
Zusammenfassung...............................................................................................54

\bigskip
 11.
 Appendix.............................................................................................................57

\bigskip
 12. References...........................................................................................................59

\bigskip
 13. 
 Danksagung.........................................................................................................69

\newpage
\section{INTRODUCTION} \label{introduction}

\begin{figure*}[]
\begin{center}
\includegraphics[height=\textwidth, angle=-90.0]{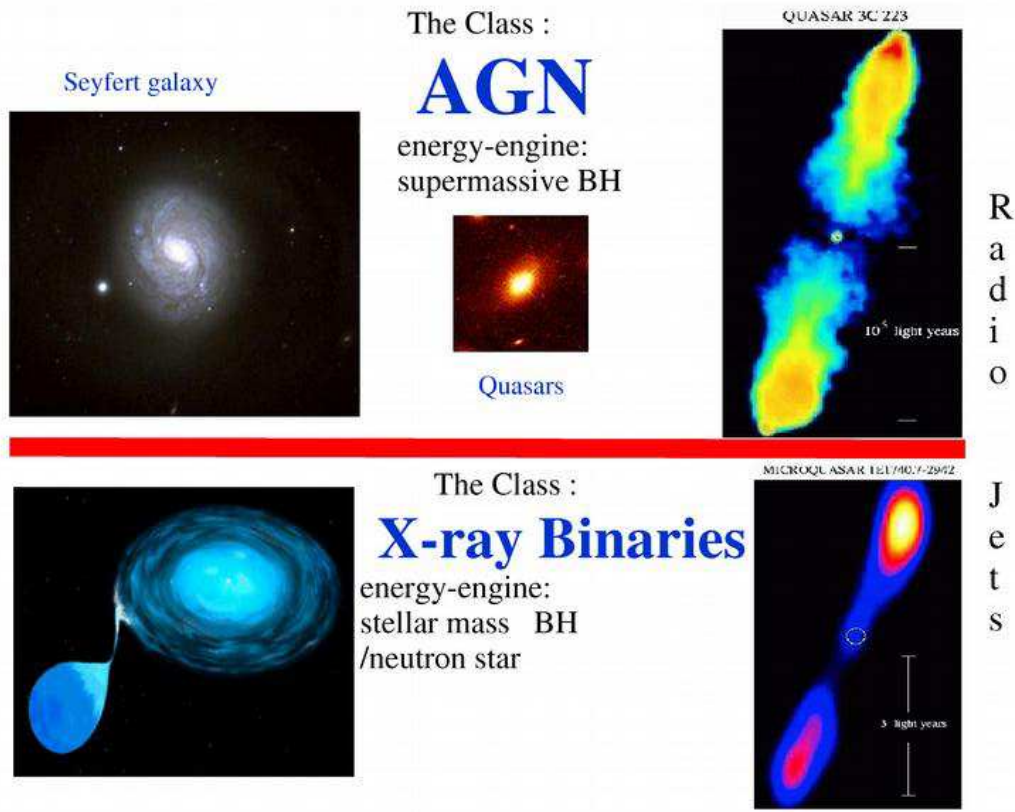}
\caption{{\bf: The AGN (Top) and X-ray binary classes (Bottom).} 
The energy engine for the AGN is a super-massive black hole
($\ge 10^6 M\odot$) accreting from its host galaxy. The
X-ray binaries are stellar systems formed by a normal star and a
degenerate object (a neutron star or a black hole  of a few solar masses)
accreting from the companion star. 
If there is   
 evidence for a radio jet 
the X-ray binary system is defined as radio-loud.}
\end{center}
\label{fig:classi}
\end{figure*}

Since the beginning of the 1980s objects like radio-galaxies, quasi-stellar radio sources
(Quasars),  Seyfert Galaxies
are simply classified as   Active Galactic Nuclei (AGN),
because the ``energy-engine'' is thought to be the same for all of them:
A super-massive black hole of millions of solar masses 
accreting from its host galaxy (Fig.~\ref{fig:classi}).
AGN with  radio-emitting lobes or jets are
called radio-loud, the others
radio-quiet
(Ulrich et al. 1997).

\begin{figure*}
\begin{center}
\includegraphics[width=0.7\textwidth]{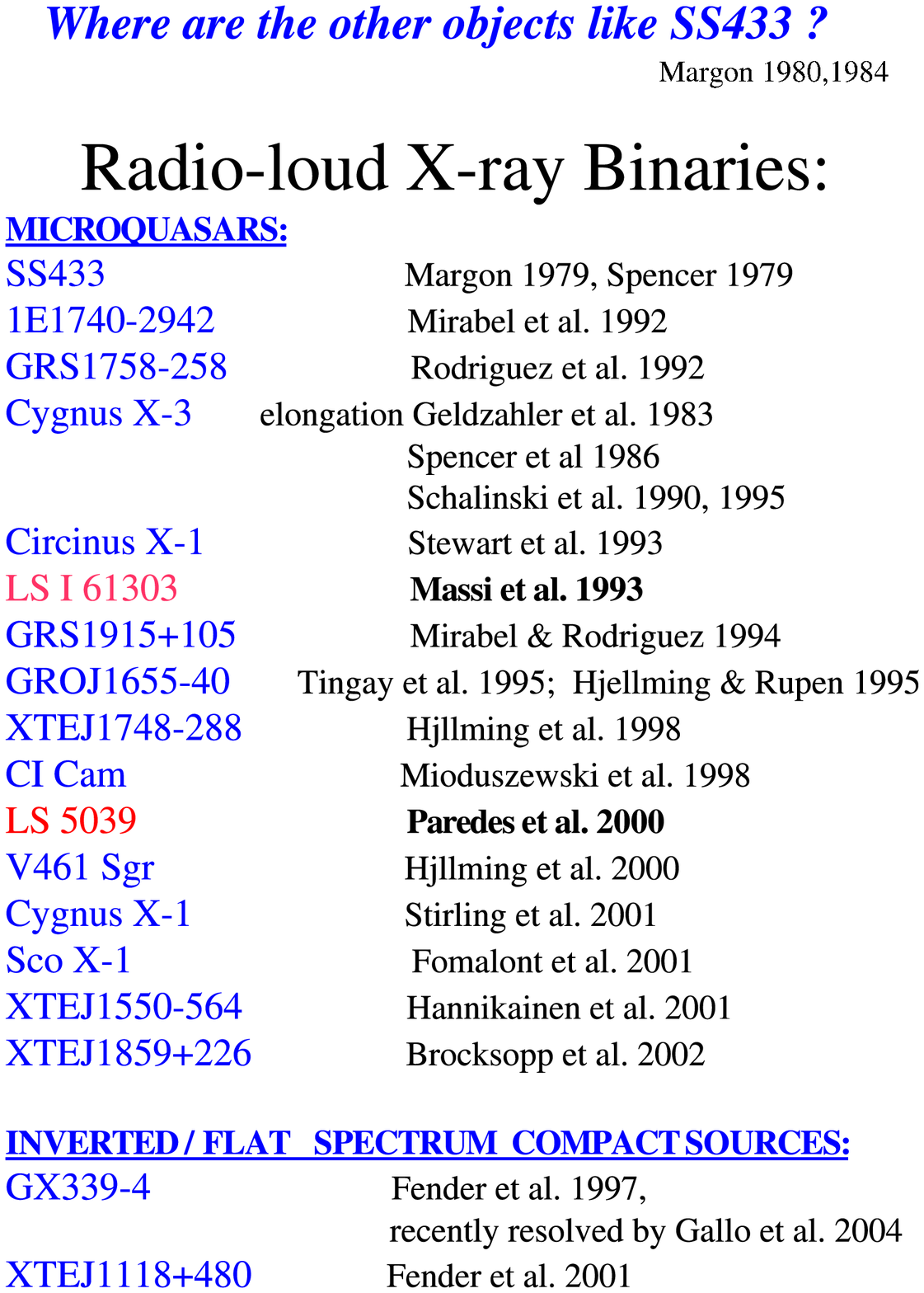}
\caption{{\bf: Radio-loud X-ray binary systems.}
The radio-loud subclass of X-ray binary systems  
includes together with the microquasars
--objects where high resolution radio interferometric techniques
like VLBI have given  {\bf direct} evidence of the
presence of collimated  jets --
also   unresolved radio sources with a flat spectrum that gives
 {\bf indirect} evidence for  continuous ejection. 
At the top: The historical sentence of Margon,
who discovered SS433, the first  galactic miniature of a quasar, unique 
for several years. Traces of elongation were evident in Cygnus X-3 
already in 1983. 1E1740-2942 was the first object with  
radio jets to be called a "microquasar".
The two sources in red \lsi and LS 5039 (both discovered by our
group) are the only
two microquasars coincident with  high-energy Gamma-ray sources.}
\label{fig:microquasars}
\end{center}
\end{figure*}

The class of  X-ray binary systems is very similar,
 the ``energy-engine''
 is  a compact object of only a few
solar masses accreting from the companion star (Fig. ~\ref{fig:classi}).
Up to now there are   known 280 X-ray binary systems 
(Liu et al. 2000, 2001),
but
only 18 of them 
 (Fig. \ref{fig:microquasars})
show evidence of a radio-jet
and therefore are radio-loud applying the same definition as for the AGN.

The radio-loud subclass of X-ray binary systems  
includes together with the microquasars
-- objects where high resolution radio interferometric techniques
like VLBI have given  {\bf direct} evidence of the 
presence of collimated  and relativistic jets (Mirabel et al. 1992) --
also   unresolved radio sources with a flat spectrum, which give 
 {\bf indirect} evidence for  continuous ejection.


\begin{figure*}[p]
\includegraphics[width=\textwidth]{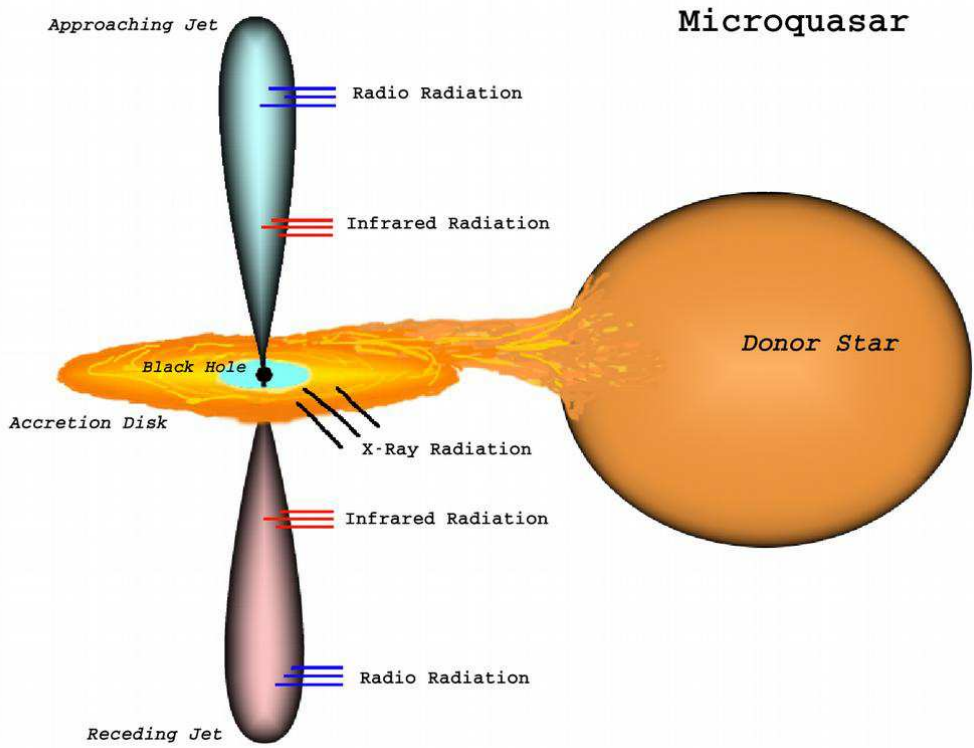}
\caption{
{\bf: The basic components of a microquasar}:    
  A spinning compact object, an accretion disk and a collimated 
relativistic jet. The compact object  here is of a few solar masses, while
in the extragalactic analog (AGN) the compact object is a black
hole of millions of solar masses accreting from its host  galaxy.
The compact object in a microquasar accretes from a normal star 
 in orbital motion around 
it.   The  study of the   periodic  velocity-shift of
 the optical spectral lines
of the companion star allows to  determine  the mass
  of the compact object and to establish whether
 it is a neutron star or a black hole.
The inner part of the disk
 emits X-rays. The inner radius is three times the Schwarzschild radius,
the outer radius a factor of 10$^3$ larger (the figure is not in scale). 
Due to magneto-rotational instabilities a part of the disk is propelled
into a relativistic jet, studied at high resolution with radio 
interferometric techiques.
In some microquasars, like SS433 and \lsi, the jets
are  precessing. If the precession brings the jet pointing towards the Earth 
the large variable Doppler boosting  mimics the variability of  Blazars 
and in this case the Microquasar is  called Microblazar (Massi 2003).
}
\label{fig:microquasar}
\end{figure*}

I here review how 
the three important basic components of a microquasar (Fig. ~\ref{fig:microquasar})   
 - a  compact object, an accretion disk and a collimated 
relativistic jet - 
have been   observed in  
 gamma-rays, X-rays, optical and radio emission. 
After a basic introduction of the
accretion-ejection processes presented in section~\ref{accretion}, 
the following sections describe the
astronomical methods: 
First, presenting  their theory and, afterwards their application
on the source LS~I~+61$^{\circ}$303.
In detail: Optical observations, reported in Sect.~\ref{optic}, 
reveal  the nature of the compact object: Neutron star or black hole.
X-ray observations, discussed in Sect. 4, deliver 
 information on the accretion disk, while 
radio observations  allow to study the jet;
the section  describes 
how the simultaneous use of  X-ray and radio tools
allow to  study  "the disk-jet" connection.
The application to  LS~I~+61$^{\circ}$303 
shows the limits of an  approximation of the  accretion 
 theory which assumes a  constant velocity of the accretor along the orbit. 
Section~5  
shows that the observational results can be  explained if  one takes into account 
the geometry of the orbit.
The better understanding of the physical processes motivates
for new observations at higher energy. 
Gamma-ray observations are discussed in Sect.~6.
The characteristics of the jet (morphology, velocity, etc) 
are derived from high resolution radio-astronomical observations and 
their typical procedures are described in Sect.~7.
Finally, the conclusion  of this  multiband approach applied on \lsi 
are presented in Sect.~8
\newpage
\section{THE ACCRETION-EJECTION PROCESS} \label{accretion}

\subsection{Accretion}

X-ray binaries are  stellar systems formed by two stars
of a very different nature:  A normal star (acting as a mass donor)
 and a compact object (the accretor) that can
either be a neutron star or a black hole 
(White et al. 1996).

Several  mechanisms have been proposed to explain the presence of a compact
object in a binary system, and they principally depend on  the mass of the
companion. If the companion is a low mass star (Low Mass X-ray Binary, LMXB)
the theory assumes, first, the   
 formation of the neutron star/BH,  which  later coupled with its
companion in a (tidal) capturing process.
As a matter of fact several  LMXB  are close to the 
core of globular clusters
or near the center of  the Galactic bulge  
(Verbunt \& van den Heuvel 1996).
The High Mass X-ray Binary (HMXB)
systems, where the companion has a mass above 5$ M\odot$,
have a galactic disc distribution characteristic that of young  stars
(population I). 
It is assumed that
 large scale mass transfer has  occurred in the system
before the supernova explosion: 
When the progenitor star of the
compact object had evolved
to a  Red Giant and had filled its  Roche lobe, 
the smaller companion accreted from it to a level that it
survived after the  explosion.
In most HMXRB systems the massive companion of the compact object is
a   rapid rotating Be star, whose formation is explained by the large amount
of angular momentum received together with matter from
the initially  more massive and therefore faster evolving companion   
(Verbunt \& van den Heuvel 1996).

Since the binary pair is in orbital motion around the common center
of gravity, the matter leaving  the companion star  
has some angular momentum ($J$), which prevents  it  from
directly falling into the accretor. 
The stream of matter orbits the compact object with a radius 
determined  by  $J$  and  the  mass of the compact object ($M_X$).
The angular momentum is  redistributed by the viscosity:
Some of the material takes angular momentum  and 
spreads outwards, whereas other material spirals inwards. In
this  way a disk is created from the initial ring of matter
(King 1996; Longair 1994  p. 135).
Gradually  the matter drifts inwards until it reaches the last stable orbit,
called ``the inner radius'' of the accretion disk ($R_{in}$), 
which for a non rotating black hole is 
approximately three times the Schwarzschild radius
($r_{\rm s}$):
\beq
r_{\rm s}={2G M_{\rm X}\over c^2}
\label{eq:rs}
\eneq
\beq
 R_{\rm in}= 3~r_{\rm s}\simeq 9~({M_{\rm X}\over M\odot})~{\rm km}. 
\label{eq:rin}
\eneq
The viscosity has two effects: Besides the transport of angular momentum 
it also acts like a frictional force resulting in the dissipation of heat.
The amount of friction depends
on how  fast the gas orbits around the compact object;
the temperature ($T_{\rm in}$) reaches its maximum at the inner disk  where  
it  rises up to (Longair 1994  p. 141)
\beq
T_{\rm in} \simeq 2 ~ 10^7 ({M_{\rm X} \over M_{\odot}})^{-1/4}   {\rm K}.
\label{eq:Tin}
\eneq
On the basis of this equation we see that 
for a microquasar of about 1 solar mass the 
matter around the last stable orbit is  heated up to tens of million degrees
therefore emitting predominantly in the  X-ray band. 
This led to the name `X-ray binaries' for this class of  objects.
The Earth´s atmosphere is opaque at these wavelengths;
therefore it is understandable that there was an impasse
(until  recent developments in X-ray astronomy) in  discovering
such stellar sources and
 (besides SS433 discovered by chance) 
 their subclass with  relativistic jets. 
On the contrary, the temperature of the last
stable orbit  around a super massive black hole of an AGN 
with a  mass of $10^9~M\odot$
is  $ T_{in}=  10^5  K$. Therefore the emission is
in the ultraviolet band
causing    the "blue bump" associated with the  "visible" Quasars.
This last fact is  an example that  
the same laws are applied to both, AGN and X-ray binaries,
deriving parameters only scaled  with the mass 
(see  Heinz \& Sunyaev 2003; Merloni et al. 2003; Falcke et al. 2003).

The accretion luminosity can be written as: 
\beq
L=\eta~ \dot{m}  c^2,  
\label{eq:L}
\eneq
where $\eta$, the efficency   
of  energy  conversion, expresses here  how compact an object 
(with radius $R$) is:
$\eta= 1/2(r_s/R)$
(Longair 1994 p. 134).
Whereas for a white dwarf  $\eta$ is only 0.0001, 
for neutron stars $\eta$ is  0.1. As a comparison, 
the  release of nuclear binding
energy occuring  in the conversion of four protons  into helium 
has  an $\eta={4 m_p - m_{He} \over 4 m_p} =7x 10^{-3}$.
Thus, accretion in neutron stars already  is an order  of
magnitude more efficient as an energy source as compared with 
nuclear energy  generation (Longair 1994 p. 134).

However, there is a limit of  energy that is possible to extract by
accretion:
If the force generated in the accretion 
disk by radiation pressure  exceeds the gravitational force of the compact
object a further accretion of gas  ceases.
The expression  for that  luminosity limit,  the Eddington luminosity,
is (Longair 1994 p. 137; Frank et al. 2002) 
\beq
L_{\rm E}=  1.3~ 10^{38}  ({M_X\over M_{\odot}}) \rm{erg~s^{-1}}
\eneq

\subsection{Magnetohydrodynamic Jet Production} \label{jetmagneto}

\begin{figure*}[htb]
\centering
\includegraphics[width=\textwidth]{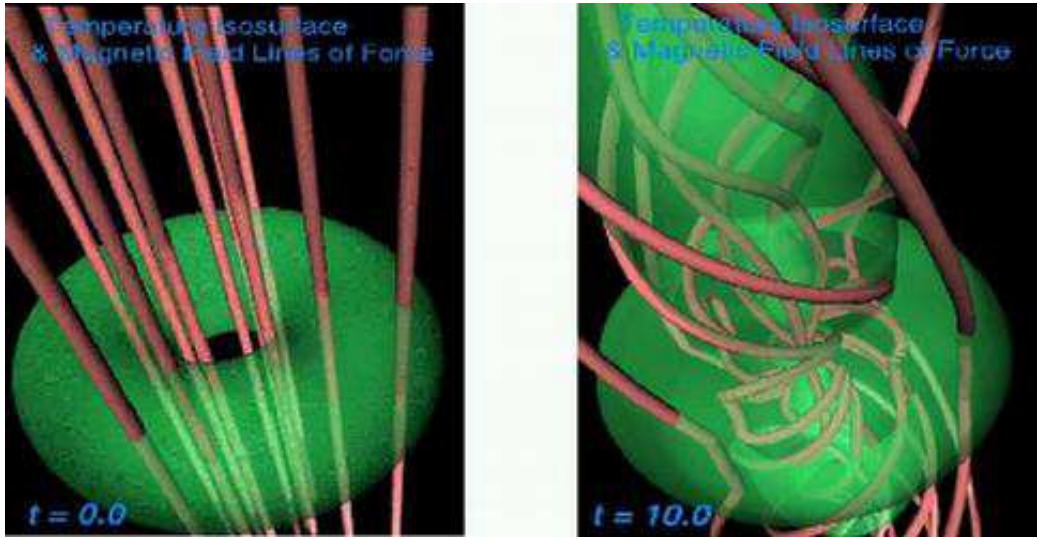}
\caption{
  {\bf: A differential rotating disk drags the field lines of a 
vertical magnetic field}.Meier, Koide \& Uchida  2001.} 
$\small  www.batse.msfc.nasa.gov/colloquia/abstracts\_ spring03/ 
presentations/meier.pdf$  
\label{fig:mei0}
\end{figure*}

In the case of a small  vertical magnetic field threading the disk
the  plasma pressure dominates the magnetic field pressure and  
the differentially rotating  disk bends
the magnetic field lines, which are passively wound up  (Fig. \ref{fig:mei0})
(Meier et al. 2001).

Due to the compression of the magnetic field lines
the magnetic pressure may  become larger than the gas pressure at
the surface of the accretion disk, where the density is lower.
At this point the gas starts to follow the twisted magnetic field lines,
creating  two spinning flows.
This extracts angular momentum (magnetic braking)
 from the surface of the disk and enhances the radial accretion.
 The avalanching material further pulls
the deformed magnetic field with it
and afterwards magnetic reconnection may happen (Matsumoto et al 1996).
 The flux tubes
open up and reconnect as  is known from stellar flares (Massi et al. 2002).

The thickness of the disc is a fundamental parameter 
in this  magneto-rotational process,
or better the extent of the poloidal magnetic field frozen in the disc
(Meier 2001; Meier et al. 2001; Maccarone 2004).
No radio jet is associated with  X-ray binaries in
High/Soft states (Sect. 4.1), where the X-ray spectrum is dominated by a
 geometrically thin (optically thick)
 accretion disc (Shakura \& Sunyaev 1973).
In the contrary numerical results show a jet being launched
from the   inner geometrically
thick portion of the accretion disc that is present (ADAF/Corona)
when the   X-ray binaries are in their Low/Hard state (Sect. 4.2)
(Meyer et al. 2000; Meier 2001).

In conclusion, 
a better understanding of  the transition from radio-quiet to radio-loud 
therefore seems to  be possible through  a better understanding 
of the  X-ray states and  their switch mechanism.

\subsection{Strong Magnetic Fieds: the X-Ray Pulsars}

\begin{figure*}[htb]
\centering
\includegraphics[scale=0.6, angle=0]{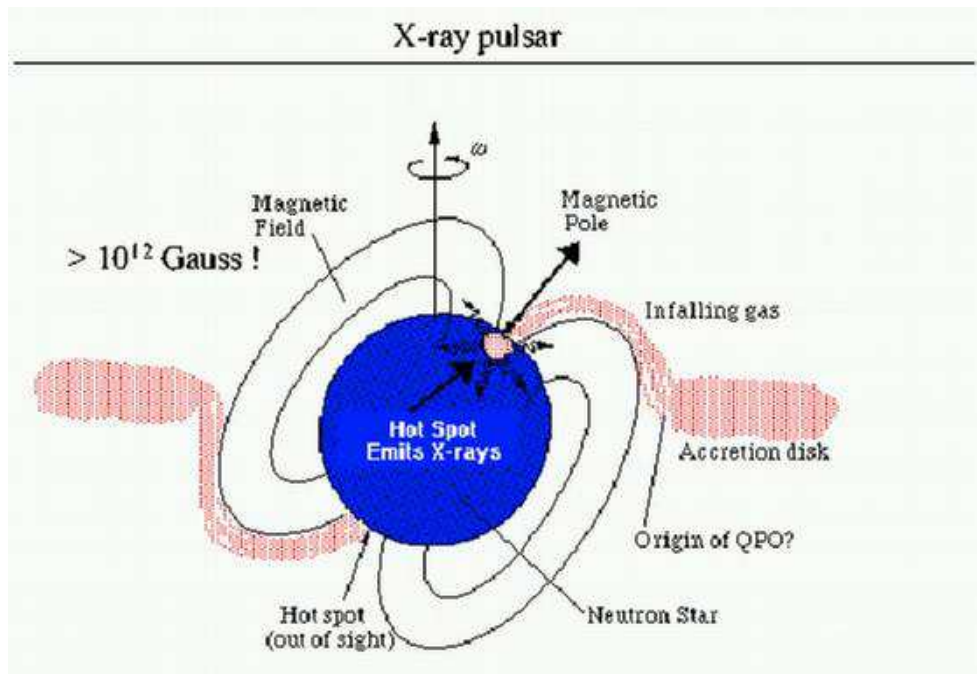}
\caption{
{\bf: Sketch of an X-ray pulsar.}}
$\small {http:// lheawww.gsfc.nasa.gov/ users/ white/xrb/ xray\_pulsar.gif}$
\label{fig:pulsar}
\end{figure*}

A radio-loud X-ray binary system may contain either 
a black hole or a neutron star with a low ($B< 10^{10}$ Gauss) 
 magnetic field.
Accreting neutron stars with  a
low magnetic field can give rise to  jet production 
because of the following reason:  
As described in the previous section only for  a 
low  magnetic field can the plasma pressure  dominate and bend the field.
On the contrary, jet formation is prevented in presence of a strong magnetic
field.
In the case of  B$>10^{12}$ Gauss,   the plasma 
is forced to  
move along the magnetic field lines,
converges  onto the magnetic poles of the neutron star and there releases
its energy creating 
two X-ray emitting caps that, in case of a misalignment of  
the rotation and the  magnetic axis, produce  X-ray pulses (Fig. \ref{fig:pulsar}).
X-ray pulsars are not associated with microquasars.
The lack of detected radio emission from  X-ray
pulsar systems is discussed in Fender et al. (1996).

\newpage
\section{OPTICAL OBSERVATIONS} \label{optic}

\subsection{ The  Nature of the Compact Object} \label{optic1}

The most reliable method to determine the nature of
the compact object is the study of the Doppler shift 
of  absorption lines in the  spectrum of its companion.
The study of the changing radial velocity during the orbital motion 
is a technique that has been  applied for  more than one hundred
 years to measure the masses of stars in  binary-systems.
The same method is applied for  systems like X-ray binaries, where one 
component is "invisible". In this case  the variations of the radial velocity of 
the normal companion during its orbit are studied.

\begin{figure*}[htb]
\centering
\includegraphics[width=0.8\textwidth]{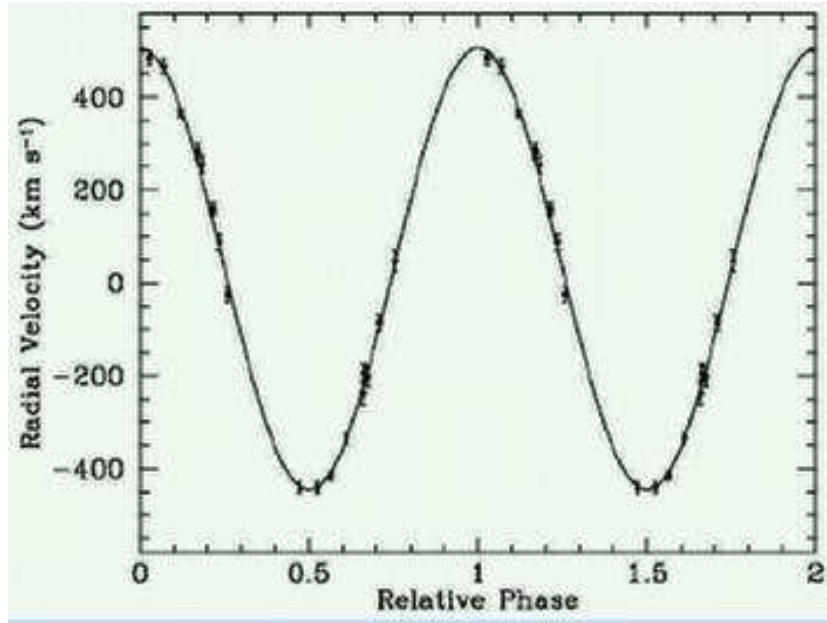}
\caption{{\bf: Amplitude of the radial velocity variations
versus orbital phase}(Filippenko et al. 1999; GRS 1009-45).  Using the Doppler shift 
of spectral lines from the companion  star orbiting around the
compact object, one determines the mass function $f$,
lower limit to the mass $M_X$ of the compact object.
Notice that for a better display, the orbital phase interval 0-1 is 
repeated twice.}
\label{fig:radial}
\end{figure*}

The   amplitude  ( $K_{\rm c}$) of the  radial velocity variations 
(Fig. \ref{fig:radial}) of the mass donor 
 and the period (P$_{\rm orb}$) of the system 
applying Newton´s/Kepler's third law
define a quantity called the ``mass function'' (Charles \& Wagner 1996),
 which is equal to:

$$f ={P_{\rm orb} K_{\rm c}^3\over 2\pi G}={M_{\rm X}^3 sin^3 i\over (M_{\rm X} +M)^2} $$
where 
$M_{\rm X}$ and $M$ are  the masses of the compact object and of the 
companion, respectively, 
$i$ is the angle between  the axis of the orbit and the line of sight and
$G$ is the gravitational constant.

The mass function alone already provides a lower limit for 
$M_{\rm X}$ corresponding to a zero-mass companion ($M$=0)
viewed at the maximum inclination angle (i=90$\degr$).
In the cases where  the inclination $i$
 and the mass of the companion $M$ are known one
can solve for the mass $M_X$ of the invisible object.

Rhoades \& Ruffini (1974),
taking the most extreme equation of state that produces
the maximum critical mass of  a neutron star, established the very upper
limit of 3.2 $M\odot$ for a neutron star.
This absolute maximum mass  provides a decisive constraint 
to observationally distinguish between  neutron stars and black holes.

\begin{figure*}[htb]
\centering
\includegraphics[width=\textwidth]{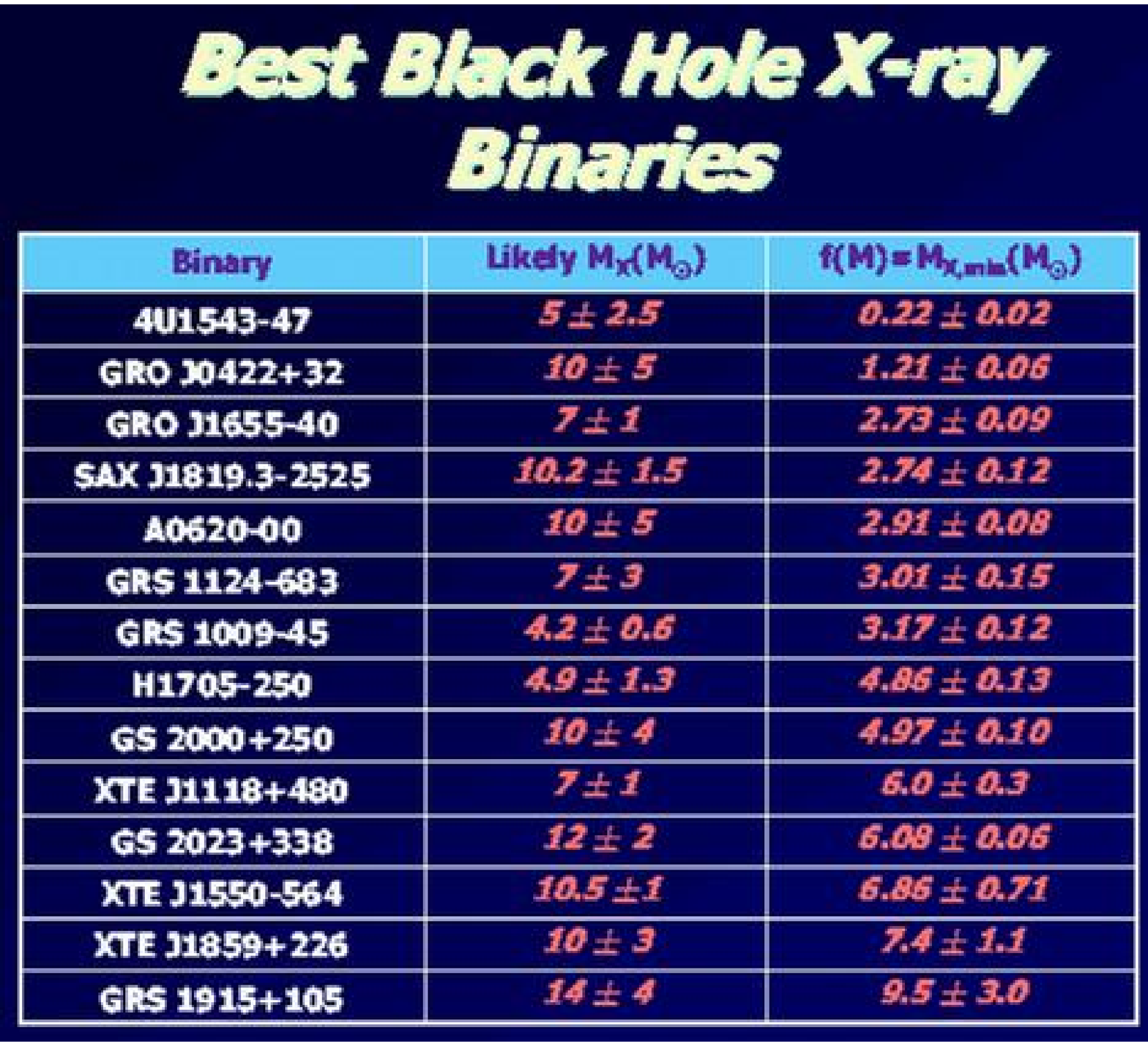}
\caption{
{\bf: Black hole candidates.} Compact objects with a mass
 ($M_X$) greater than 3$~M\odot$, upper limit for a stable neutron star.
 Ramesh Narayan.}
${\small http://cgpg.gravity.psu.edu/events/conferences/Gravitation\_Decennial/}$
${\small Proceedings/Plenaries/Sunday/Narayan/narayan\_plenary.pdf}$
\label{fig:candidate}
\end{figure*}

In Fig. \ref{fig:candidate} 
a list of some X-ray binaries is given for which both mass function
and $M_{\rm X}$ is available. 
All sources below GRS 1009-45,  with $f(M)= 3.17 M\odot$, can be  
defined  black hole candidates  on the basis of the mass function alone.
On the contrary, for  cases where  $f(M) < 3$ 
the determination of inclination and mass of the companion is mandatory
to determine the type of object.

\begin{figure*}[htb]
\centering
\includegraphics[scale=1.0, angle=0]{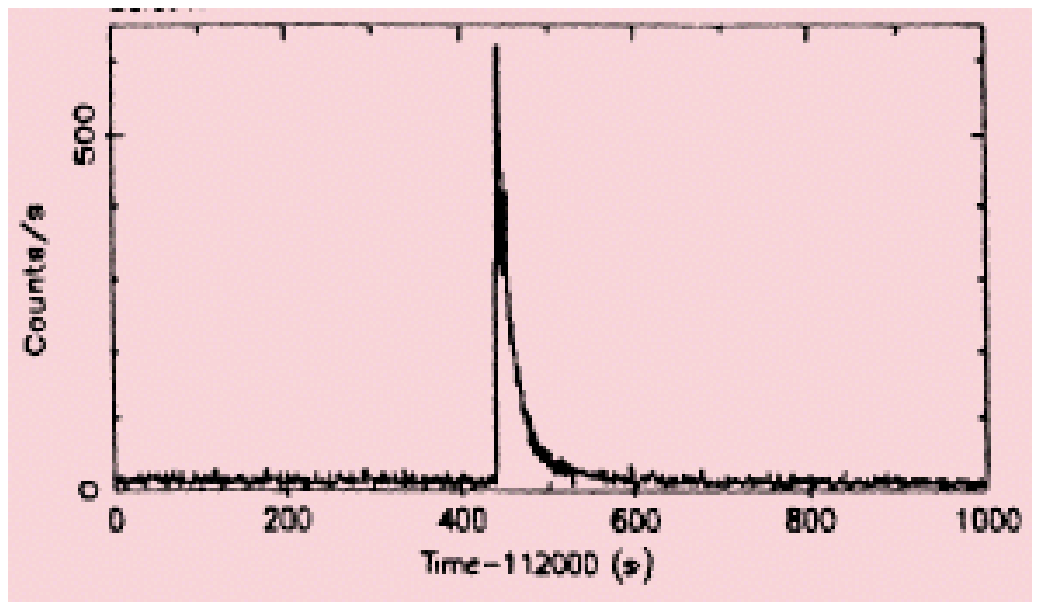}
\caption{
{\bf:Type I X-ray burst}. This type of burst comes from a  thermonuclear 
flash on the surface
 of an accreting  neutron star (Tennant et al 1986; Circinus X-1). 
No type I burst has never been observed for accreting black holes.}.
\label{fig:burst}
\end{figure*}

It is worth  mentioning here
that the accumulation of accreted material  on the surface of a neutron star
triggers thermonuclear bursts (see typical profile in Fig. \ref{fig:burst}).
These are  called bursts of Type I.
No Type I burst has ever been observed  from  a compact object where optical
observations resulted in a mass above
3 $M\odot$. That  fact  might confirm that in  black holes
there is no surface where material can accumulate (Narayan  \& Heyl 2002).
In conclusion: Observations of
Type I bursts give a direct  evidence for the existence of a neutron star.

\subsection{ The  Nature of the Companion Star}

The classification of the  X-ray binaries    into
 Low Mass X-ray Binary  and High Mass X-ray Binary
 systems  
leaves unspecified  the nature of the accreting
object and is based on the mass of the companion star
(van Paradijs \& McClintock 1996).   

A LMXB  contains a late type (K,M)
low mass donor star. The  mass transfer
takes place via Roche lobe overflow (Frank et al. 2002):
 Material streams through the inner Lagrangian point
and  will orbit the compact object at the  radius 
determined by its specific angular momentum.

In  HMXB systems, the companion is an OB star.
 OB  stars have a substantial stellar wind  
(mass loss rates $10^{-10}-10^{-5} M\odot{\rm yr}^{-1}$)
 with a velocity of
$v_{wind}\sim v_{escape} =\sqrt{{2 G M \over R}}\simeq 10^3$ km/s.
However, matter   leaves
the star in all directions, not only towards the accretor as in 
the case of Roche lobe
overflow. This accretion  therefore is less efficient 
(King 1996).
The expression for the accretion rate is (Bondi  1952):
\beq
 \dot{M} = {4 \pi  \rho_{\rm wind}(G M_X)^2\over v_{\rm rel}^3}
\label{eq:mdot} 
\eneq
where $\rho_{\rm wind}$ is the density, 
 and  $v_{rel}$ depends on the velocity along the orbit $v_{orb}$
and on the wind velocity ($v_{wind}$). 
The accretion
therefore becomes  more  efficient for denser and slower winds
present in  Be stars. In these rapidly spinning stars
together with a high velocity  (1000 km s$^{-1}$)
low density wind at high latitudes there  also exists  
a dense and slow (100 km~ s$^{-1}$) disk-like wind around the equator
having  a power law density distribution (Waters et al. 1988).

\subsection{LS~I~+61$^{\circ}$303: The Be-Star }
\begin{figure*}[p]
\centering
\includegraphics[width=10cm, height=10cm,angle=0]{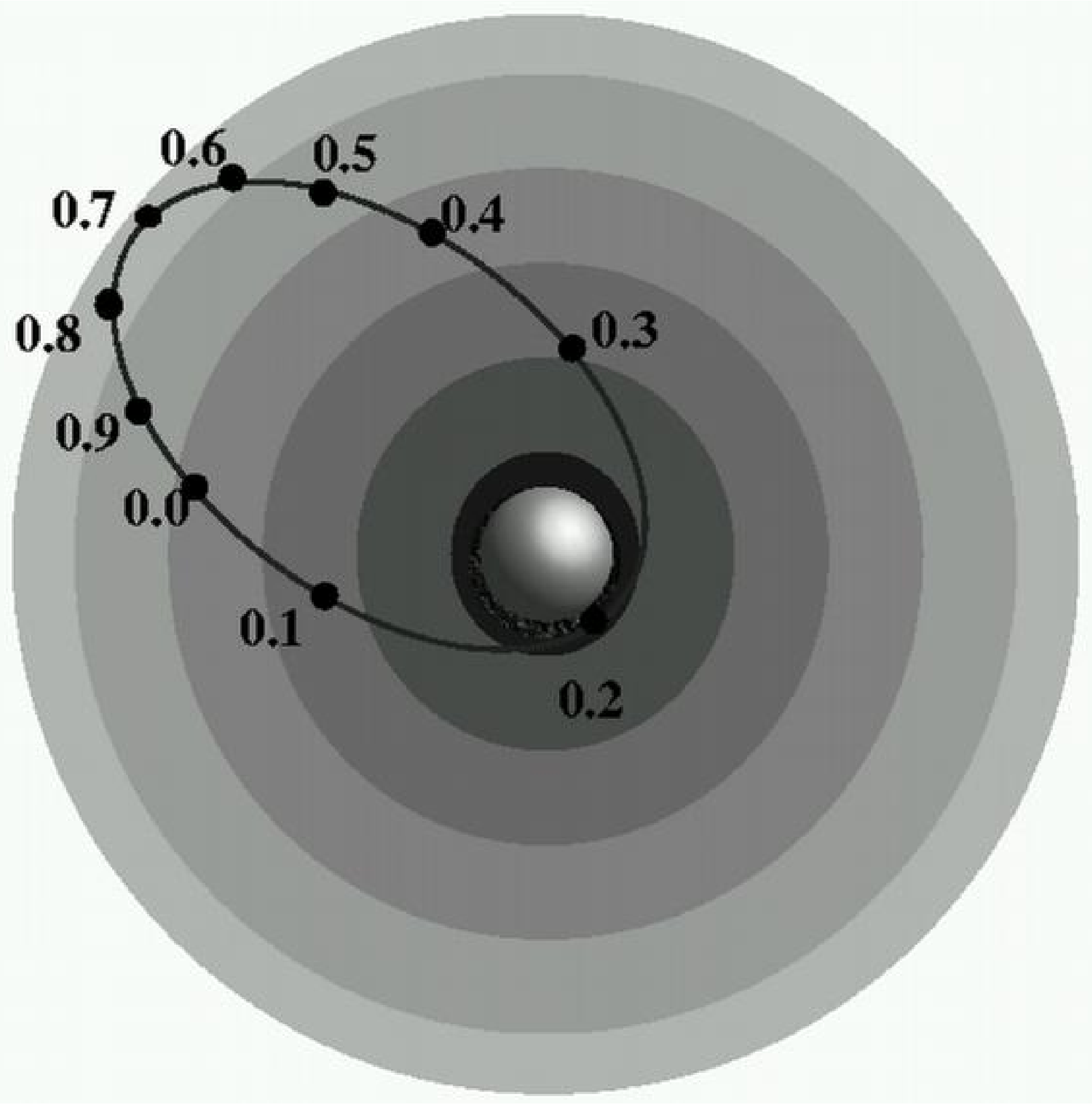}
\caption{{\bf: Sketch of the binary system \lsi.}
Orbital phases of the
compact object travelling through the wind of the Be companion star.
The accretion rate in an eccentric orbit
has two peaks: One peak
corresponds to the periastron passage ($\phi$=0.2) because of the highest density;
the second 
peak occurs in the phase interval 0.4-0.8 (i.e. around apastron)
where the drop in the  relative velocity of the compact object
compensates the decrease in density.
For supercritical accretion,  theory predicts
 matter be  ejected outwards in two
jets perpendicular to the accretion disk plane. However,
near periastron inverse
Compton losses are severe due to the very near  Be star 
(gamma-ray emission is observed   but no radio emission). 
During the second accretion peak, the
compact object  is much farther away from the Be star and both, inverse Compton losses
and wind opacity, are lower: The electrons can propagate out of the orbital
plane and   radio outbursts
are observed.
The radio emission has been resolved with VLBI observations and
in agreement with the theoretical predictions the image 
shows bipolar jets emerging from a central core (Massi 2004b).  
}
\label{fig:sketch}
\end{figure*}

The ultraviolet spectroscopy of LS~I~+61$^{\circ}$303  by Hutchings \&
Crampton (1981)  
indicates that the primary star 
is a main sequence B0-B0.5 star (L$\sim 10^{38}$
erg sec$^{-1}$, T$_{eff}\simeq 2.6~10^4$K).
Its distance is   2.0$\pm $0.2 Kpc (Frail and Hjellming 1991).
The optical spectrum is that of a rapidly rotating star 
with V$\sin$~i = 360 $\pm$ 25 km s$^{-1}$. 
The  critical rotational velocity for a normal B0 V star is
$\sim$600 km s$^{-1}$ and Be stars may rotate at a velocity that does
not generally exceed 0.9  of this, i.e.
540 km s$^{-1}$ (Hutchings et al. 1979).
Therefore the  lower limit for the  inclination of the orbit compatible
with these data is  38$^{\circ}$.
However,  Hutchings \& Crampton  
observed  shell absorptions in the strong Balmer and He I lines.
For a disk sufficiently flat this corresponds to a large inclination 
angle (i $\simeq 90 \degr$) (Kogure 1969).
The result is  a range of 
38$\degr$--90$\degr$ for the inclination of the orbit for \lsi.

LS~I~+61$^{\circ}$303 is the only X-ray binary system  showing 
variations   
compatible with the orbital period at   X-rays
(Paredes et al. 1997; Leahy 2001), at Gamma-rays (Massi 2004; Massi et al. 2004b), 
at optical wavelengths in both continuum (Maraschi \& Treves  1981;
Paredes \& Figueras 1986; Mendelson \& Mazeh 1989)  
and lines,
 (Zamanov \& Mart\'{\i}  2000; Liu et al. 2000; Apparao 2000). 
The most  accurate value for the orbital period is however 
 from radio astronomical measurements resulting in 
26.4960 $\pm$ 0.0028 days (Gregory \& Taylor 1978;
Taylor \& Gregory  1982; Gregory  2002).

The range of the mass for a  B-star is   5--18 M$\odot$.
Fits performed on near infrared data  by Mart\'{\i} and Paredes 
(1995) 
result in an   eccentricity of e$\sim$ 0.7-0.8 
and mass in the range  10--18 M$\odot$.

Finally, because of  the high eccentricity,
an  important parameter of the system is the phase at the 
periastron passage, already
determined by  Hutchings \& Crampton (1981) and very recently confirmed by
Casares et al. (2004) to be $\phi$=0.2. 
The zero phase by convention refers
to the date t$_0$=JD\,2443366.775, 
the date of the first radio detection of the system 
(Gregory \& Taylor 1978).

Figure \ref{fig:sketch} shows a sketch of the system with 
the compact object travelling (and accreting) 
through the dense, variable and structured  wind of the Be-star
along the quite eccentric orbit.

\subsection{  LS~I~+61$^{\circ}$303: Really a Neutron Star ?  }

The composite and variable nature of the spectral features of LS~I~+61$^{\circ}$303,
its long period and its high eccentricity  make it difficult to derive
meaningful radial velocities for this source.
The observations of $K_c$ 
made by  Hutchings \& Crampton (1981)  imply  
a mass function in the range 0.0028$<f<$0.043 (see also Punsly  1999). 
Following the discussion in section \ref{optic1}  we see that the low value
of the mass function is not enough to  establish 
the real nature of the compact object.
Knowledge of the  values for inclination and mass of the companion star 
also is necessary for that.
On the other hand, we have seen in the previous section that 
the inclination has a quite  large range
(38$^{\circ}$-90$^{\circ}$) and  
the  possible range for the mass (M) of the companion
 can be  10--18 M$\odot$. 

Hutchings \& Crampton (1981) have assumed $f$=0.02,
an inclination of about 70$^{\circ}$ ($\sin^3~i$=0.8)  and M=10 M$\odot$ and 
 derived M$_X$=1.2 M$\odot$.
For this reason in the literature it has generally been  assumed 
that  the  compact object in 
LS~I~+61$^{\circ}303$ is  a neutron star. 
Only Punsly (1999)  discussed the possibility 
that the compact object in LS~I~+61$^{\circ}303$ could be a black hole
and he presented a model for the high-energy emission based on it.

As discussed above and also in Massi (2004) 
the uncertainties in the parameters derived by optical measurements
are rather large;  changing the inclination   to i=38$^{\circ}$ 
 we already get   M$_X$=2.5 M$\odot$. 
If we assume a value  M=18 M$\odot$  we obtain M$_X$=3.4 M$\odot$.
Recently Casares et al (2004) have determined  an  upper limit of 
$f$=0.027 which would correpond to M$_X$=3.8 M$\odot$ (Massi et al. 2004b).
In conclusion: Accounting for the uncertainties in inclination,  
 the mass of the companion and the mass function
it cannot be ruled out that the compact object in \lsi is a black hole.

\section{ X-RAY AND RADIO OBSERVATIONS } \label{x}

An X-ray binary system is called 
{\bf transient}, if  
 at least one outburst occurs 
with a flux variation  of more than 2-3 orders of magnitude greater than the normal flux
(McClintock \& Reimmllard 2004).
  This outburst, which may last for days to months,
 is directly related to a variation of the accretion
disk and therefore is distinct from the outburst of Type I
discussed in section \ref{optic1} which lasts a few seconds.

\begin{figure*}[]
\centering
\includegraphics[width=\textwidth]{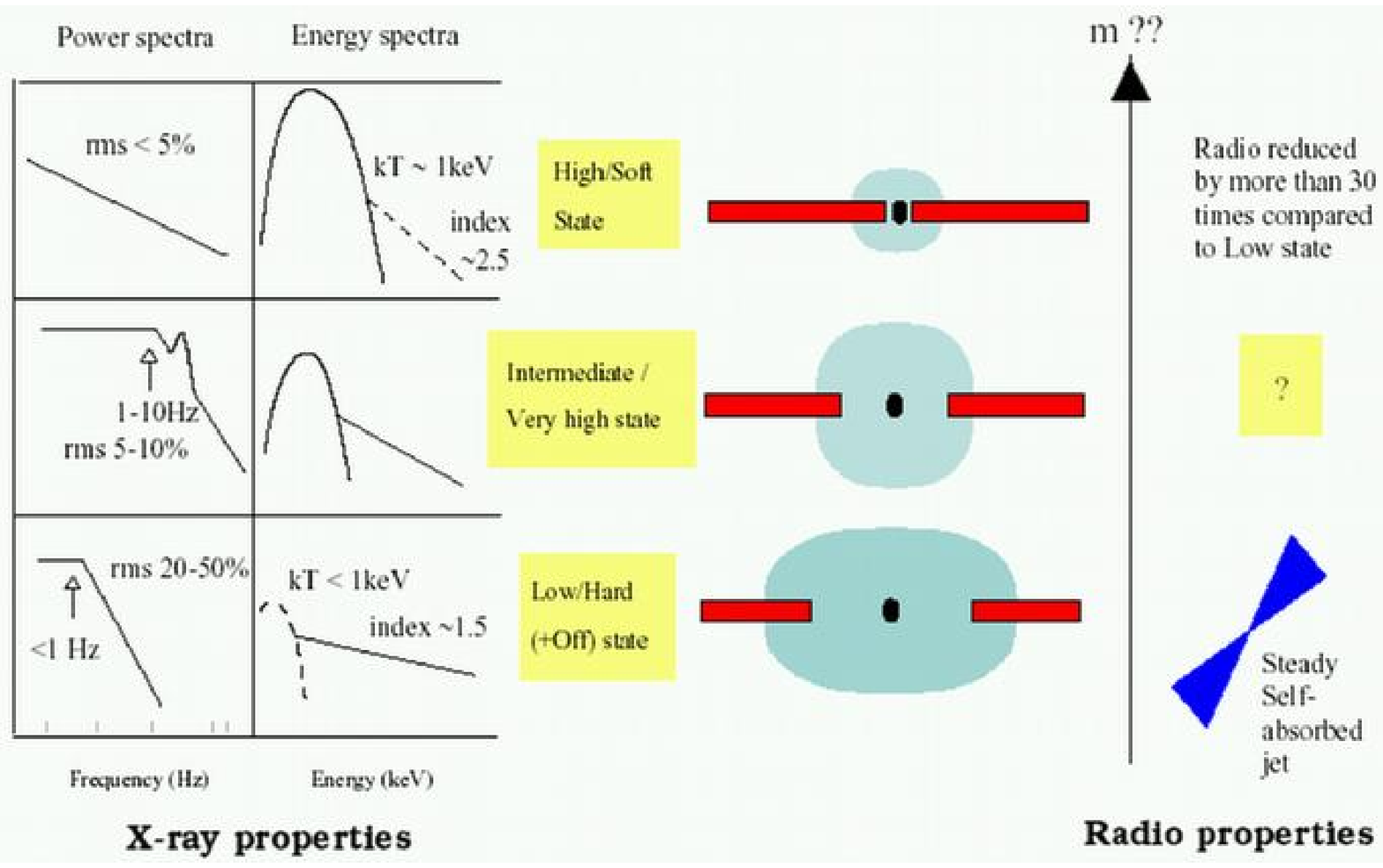}
\caption{ 
{\bf The X-ray states of a black hole .} 
Left: The states are  defined by the spectral and timing
properties. Right: Morphology of the accretion disk/Corona 
and  radio properties during the X-ray states (Fender 2002).}
\label{fig:stati}
\end{figure*}

Generally  
Microquasars have been  discovered through  high resolution radio 
observations  
immediately performed during new transients.
However,  LS 5039  was discovered (Paredes et al. 2000) 
 on the basis of a  cross-identification
in catalogs (in optical, radio,  X- and Gamma-rays)  without
any X-ray outburst calling attention to it.
Nowadays one observes changes in the X-ray 
 "states" (Fig. \ref{fig:stati}).
These states are 
defined by spectral (see below)
and timing characteristics (van der Klis 2004):
As soon as an X-ray binary is  discovered in a "Low/Hard state", it 
immediately is observed  at radio wavelengths with  
high-resolution techniques.

\subsection{High/Soft State and Multicolor Disk } \label{soft}

 X-ray binaries with a neutron star as compact object
may have  spectra  that are completely different depending on
whether the magnetic field of the neutron star is strong or
weak.

The form of the spectrum of a binary X-ray pulsar,
with a surface  magnetic field  $> 10^{12}$G
is a flat  hard power-law function
with  a sharp cut-off above a few tens of keV (Tanaka 1997;
White et al. 1996).

The  spectrum of an X-ray binary with  a weakly magnetized
 neutron star is typically formed  
by the properties of the accretion disk   and  the  neutron star  envelope.
The neutron star envelope contributes to the harder part of  the 
spectrum and has  a temperature of $\sim$ 2.5 keV.
Mitsuda and collaborators (1984), 
assuming  an optically thick disk,
where  the energy generated by viscosity 
is  locally dissipated in blackbody radiation,
have represented the disk spectrum as a superposition of spectra
with  temperatures varying from a low 
value T$_{\rm out}$ at the outer
edge to a maximum T$_{\rm in}$ at the inner edge (i.e. at the
the inner radius  $R_{in}$) of the disk.
This is the reason why the  disk is generally  called a 
 multi-temperature or multi-color  disk.
By means of this model T$_{in}$  and $R_{in}$ can
be determined through the softer part of 
the observed spectrum ($f(E)$), which is  represented  by:
\beq
f(E)={8 \pi R_{in}^2 \cos~i~\over 3 D^2} \int_{T_{\rm out}}^{T_{\rm in}}
 ({T\over T_{\rm in}})^{-11/3} B(E,T) {dT\over T_{\rm in}}
\eneq
here $i$ is the inclination angle of the disk,
$D$ is the  distance and  $B(E,T)$ is the Planck
function
(Shakura \& Sunyaev 1973; Mitsuda et al. 1984; 
van Paradijs \& McClintock 1996; Tanaka 1997).

\begin{figure*}[htb]
\centering
\includegraphics[width=8.8cm]{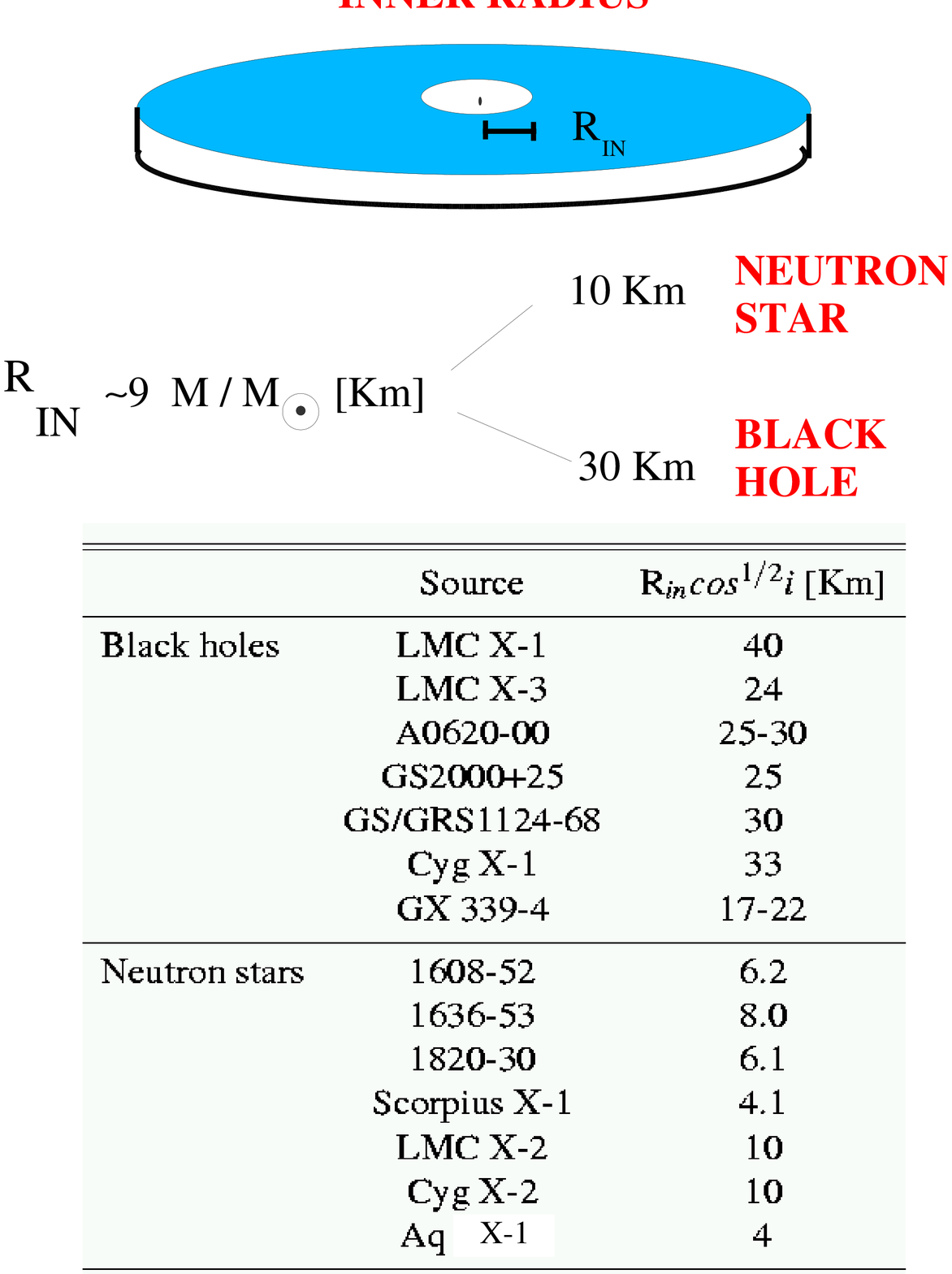}
\caption[]{{\bf: Best fit  of  R$_{in}cos^{1/2}i $ for black holes and neutron stars}.
 From Tanaka (1997). Expected values are $\sim$ 10 km  and $\sim$ 30 km 
 for a neutron star and
 a non rotating black hole of 3 solar masses respectively.}
\label{fig:tana0}
\end{figure*}

X-ray binaries  known to contain a black hole,
 proved by measurements of the mass function resulting in a
mass $\ge$ 3 M$_{\odot}$,
 have
 spectra with
 a soft component accompanied by 
a hard power-law tail 
(Tanaka 1997).
The soft component is   described  by the  multicolor blackbody
spectrum  given  above
and therefore it is associated with the accretion disk around the black hole.
This X-ray state is defined: High/Soft.
In this respect it is quite interesting to compare the different
values for T$_{in}$  and $R_{in}$ derived in the two cases of  neutron
stars and black holes (Tanaka 1997).
In  
Fig. \ref{fig:tana0}
the values of R$_{in}cos^{1/2}i $ obtained from the fits for
accretion discs around black holes and neutron stars are collected:
The projected  inner  radius R$_{in}$ of accretion disks
around    neutron stars  always results in values of $\le$10km, while 
  the
 values for black hole binaries all are  larger by a factor of 3-4
than those for neutron stars. This shows that these compact objects
are indeed more massive than 3$M\odot$ as expected following  the relationship
$R_{in}\propto M_x$ (Eq. \ref{eq:rin}).

The temperature T$_{in}$ for
disks around  black holes is always found to be less than
$\sim$1 keV, significantly lower  than that for disks around
 neutron stars with similar luminosities.
Also this difference is understood  in terms of the 
difference in the mass  M$_X$ of the compact object:
$T \propto  (1/M_x)^{1/4}$ keV (Eq. \ref{eq:Tin}).

\begin{figure*}[]
\centering
\includegraphics[width=\textwidth]{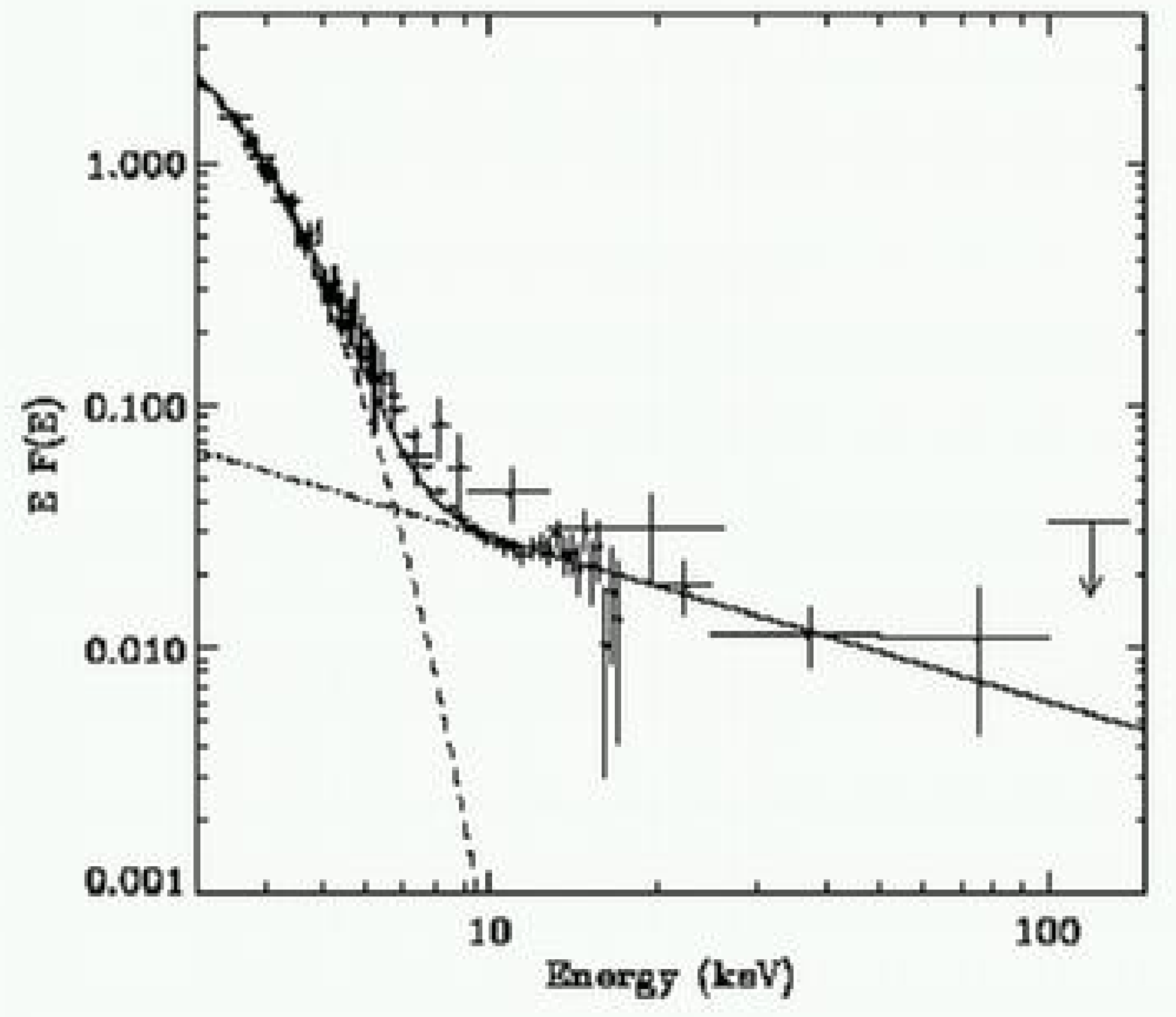}
\caption{{\bf Spectrum of XTE J1720-318 in a
Soft state.} Cadolle  Bel et al. 2004.
Dashed: Multicolor Disk with R$_{in}$=84 km.
Dotted dashed: Power law  photon index 2.6
Thick: Total model.}
\label{fig:highsoft}
\includegraphics[width=\textwidth]{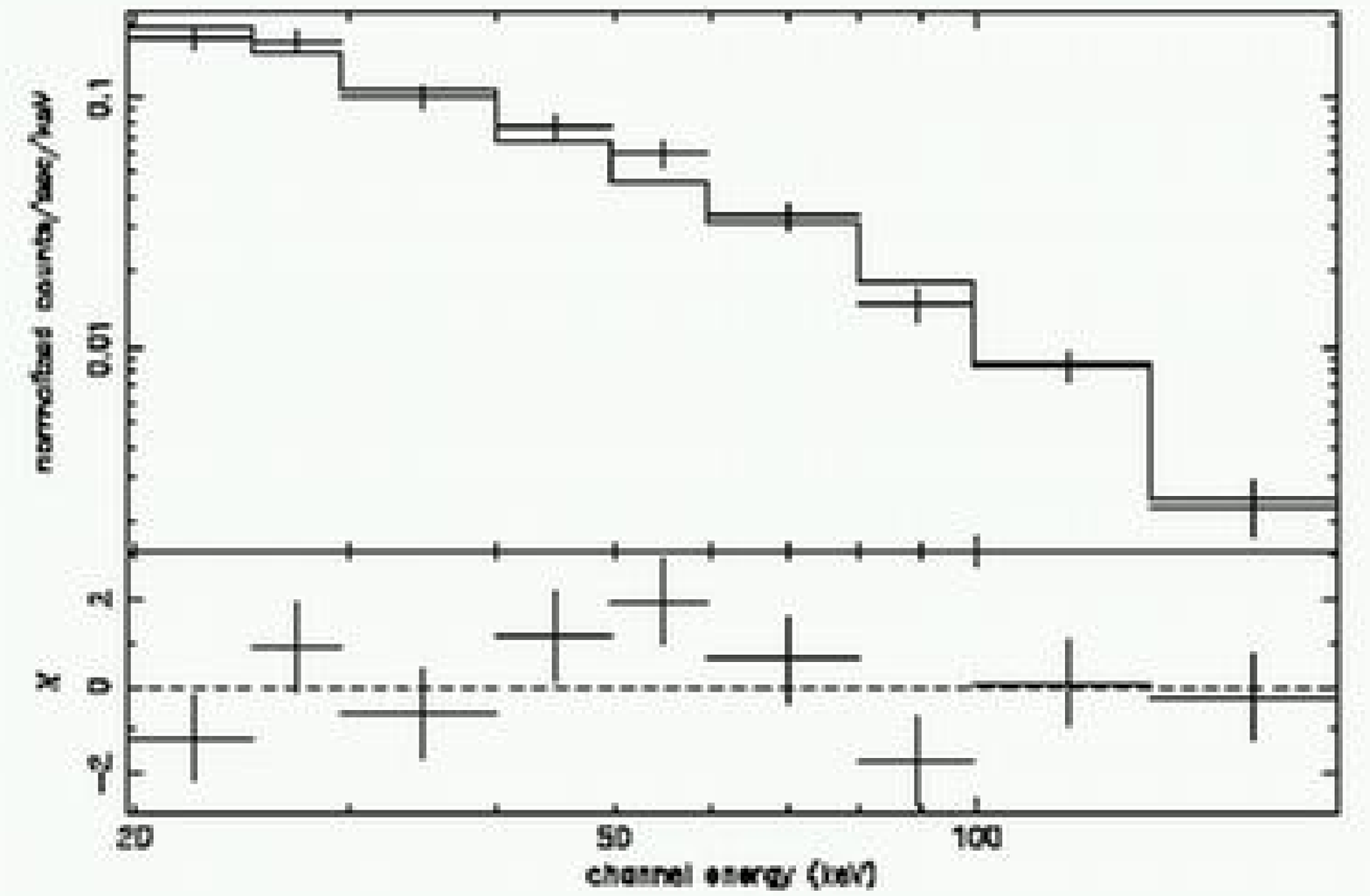}
\caption{{\bf Spectrum of XTE J1720-318
during a Low/Hard state.} Cadolle  Bel et al. 2004.
Top: Power law  photon index 1.8. Bottom: Residuals}
\label{fig:lowhard}
\end{figure*}

No blackbody component is present in the X-ray spectra of black hole
X-ray binaries, which is consistent with the absence of a solid surface in
a black hole. The  second spectral  component
in black holes  in the  High/Soft state 
is  a weak power-law with spectral index $\Gamma$, defined by the photon flux
$\propto$E$^{-\Gamma}$ (Fig. \ref{fig:highsoft}). A  photon index 
of 2.0-2.5 has
 been determined by Tanaka (1997)
for a sample of 5 black holes.
Esin and collaborators  quote a range from 2.2 to 2.7.
The recent review by McClintock and  Remillard (2004)
gives  a photon index ranging from 2.1 to 4.8 for 10 black holes.

The power law component in the X-ray spectra of accreting
black holes in their High/Soft state
 is generally interpreted as  the result of inverse Compton 
 up-scattering
 of low-energy disc photons  
by electrons with a power-law or at least a hybrid distribution
(consisting of both thermal and non-thermal electrons)
that can be located in coronal regions (possibly flaring)
above the disc (Coppi 2000; Zdziarski et al. 2001).

Evidence for the existence of an  accretion disc corona comes 
 from systems 
seen almost edge on: The strong central  X-ray source (i.e. the inner disc)
remains hidden behind the disc rim, but X-rays are
still seen.
The source of emission must be  quite extended because the eclipse 
by the companion star is only partial (White et al. 1996). 
The origin of  accretion disc coronae is described
 by buoyancy of  magnetic fields amplified in the disk 
(see Miller \& Stone 2000 and references therein).

\subsection{The Disk-Jet Connection } \label{hard}

Both  X-ray binary systems containing  weakly-magnetized neutron stars 
and those containing black holes  change  their
spectral shapes (Tanaka  1997). The most drastic change is a transition 
from  a  spectrum with the  accretion  disk component
 as that discussed in the previous section
 (Fig. \ref{fig:highsoft}),
to  a spectrum without it and showing 
 a single power-law component (Fig. \ref{fig:lowhard}).

In the sample discussed by Tanaka (1997)
the photon index  ($\Gamma$) in this state
varies in the range from  1.4 to 1.7 for  systems containing
 black holes 
 and is equal to 1.8 for systems with  neutron stars.
 Esin and collaborators (1998) quote the  range from
1.4 to 1.9.
McClintock and Remillard in their recent review (2004) give 
$\Gamma$=1.5--1.9 (excluding GRS 1915+105) for 9 black holes.
In conclusion, the power law in this state
is definitely less steep than in the High/Soft state.
This spectral state, 
 present also in systems with  neutron stars, 
is defined in the literature as Low/Hard
only for   systems with black holes.

It has been established that when an X-ray binary system is radio-loud 
and in particular with a flat or inverted radio spectrum
 (i.e. spectral index $\alpha \ge 0$
 with flux density $S \propto \nu^{\alpha}$) then
it is always in its
Low/Hard state (Fender 2004 and references therein).
Emissions in the radio band and at hard X-rays 
are related by:
 $L_{radio}\propto L_X^{0.7}$ 
( Corbel et al 2003; Gallo et al. 2003). 

\begin{figure*}[]
\begin{center}
\includegraphics[width=14cm, height=14cm]{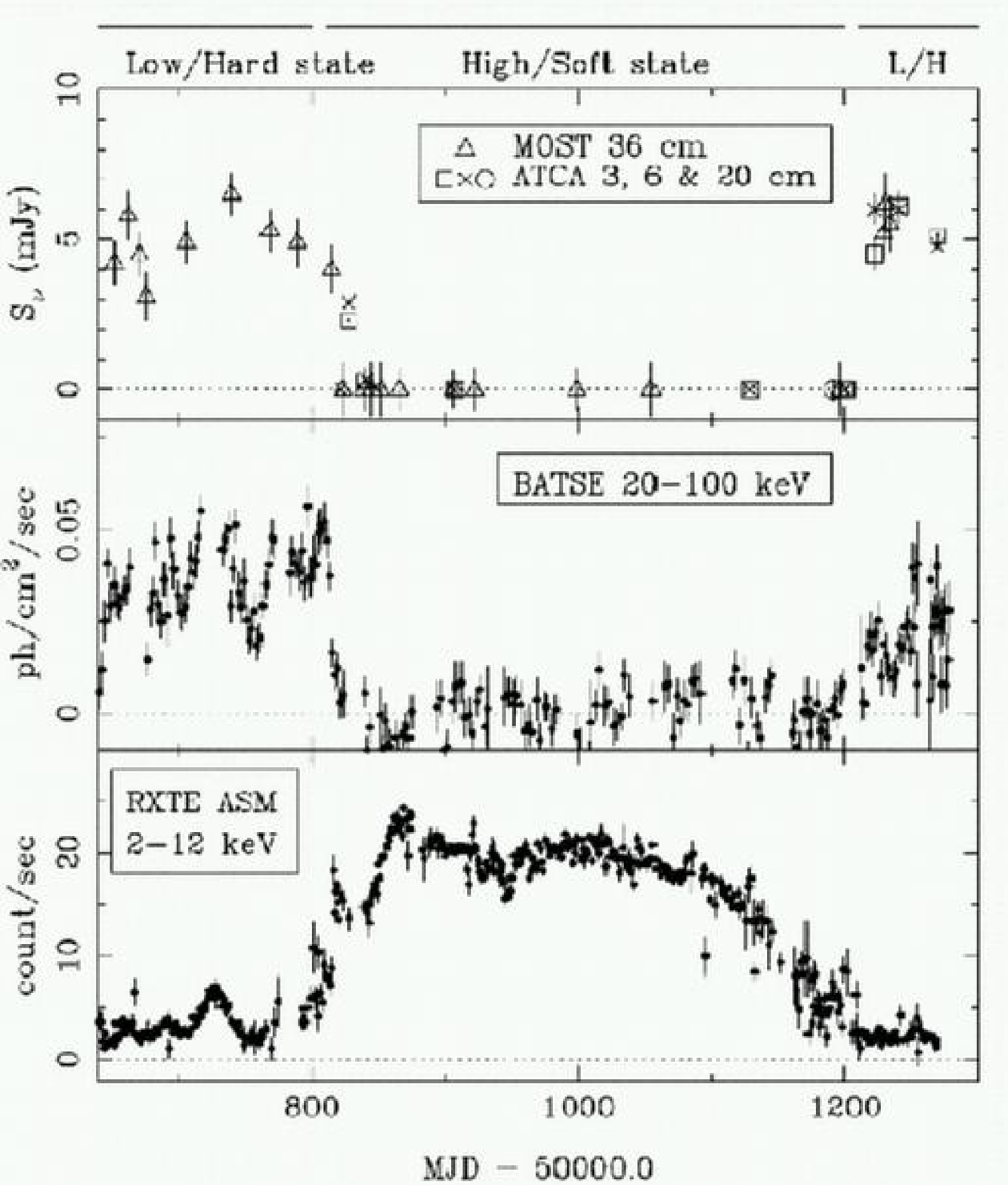}
\end{center}
\caption{
{\bf Radio (Top), hard (Middle), and soft X-Ray (Bottom)
  monitoring of GX 339-4.} Fender et al. (1999).
Radio emission and Hard/Low state are both
strongly anticorrelated with  the High/Soft state, where the
 radio emission is suppressed.
}
\label{fig:gx339}
\end{figure*}

Figure \ref{fig:gx339}  shows a multiband monitoring 
of GX 339-4.  At  the beginning of the observations
both  radio emission and  hard-X ray emission 
 (Hard/Low state) are present,
 whereas the emission in the softer X-ray band is quite
weak.
When the system switches to the High/Soft state, then  {\bf both}
radio and hard X-ray emission become quenched.

Finally,
as GX 339-4 switches again
into its Low/Hard state, radio emission is again observed
(Fender et al. 1999). 
From the plot it is clear that the radio and the hard X-ray fluxes
are strongly anticorrelated with the soft X-rays.

As stated above the radio emission during a Low/Hard state
has a  spectrum which is  flat or inverted.
\begin{figure*}[htb]
\centering
\includegraphics[scale=0.5, angle=0]{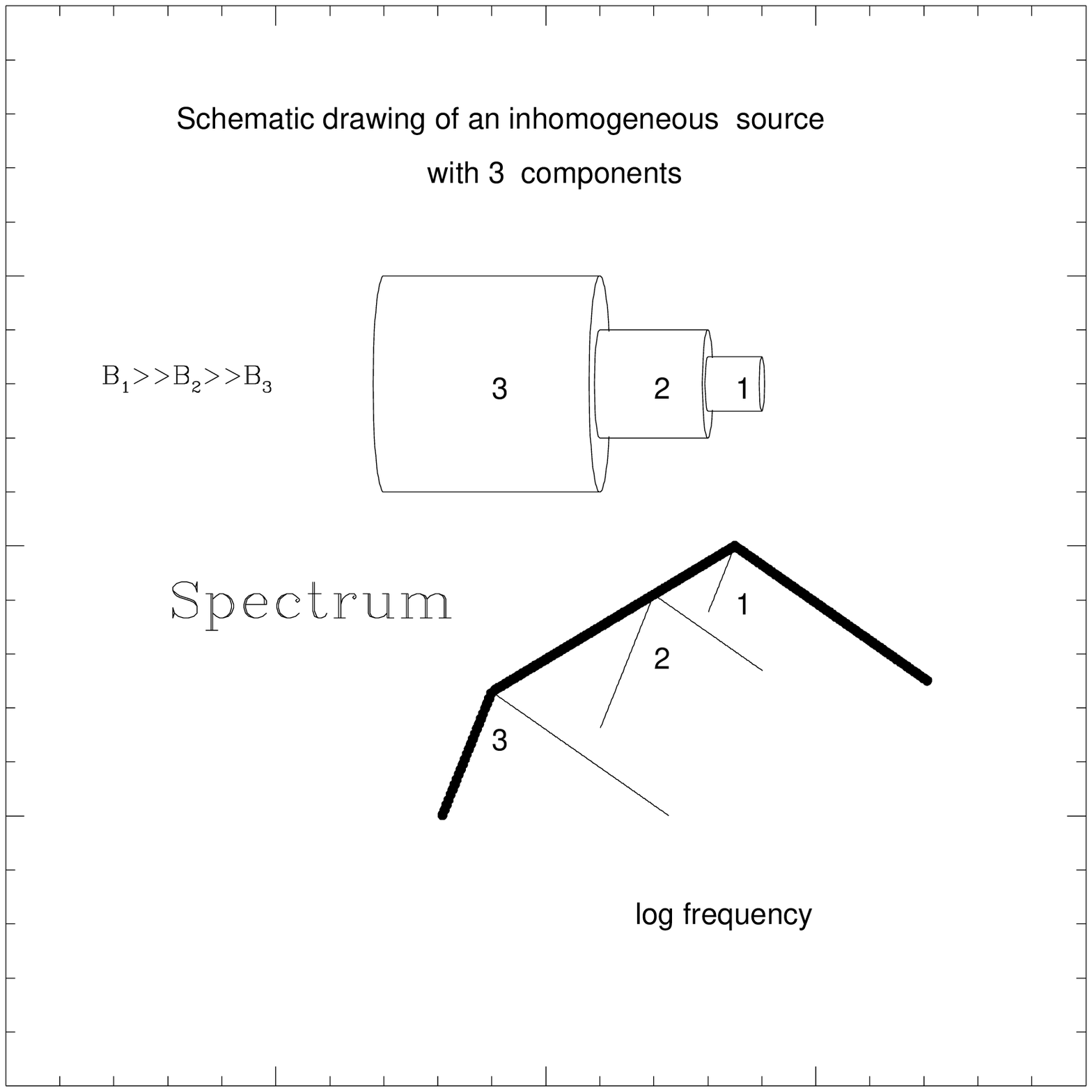}
\caption{{\bf: Inverted spectrum  
of an  expanding  continuous jet.}
A  conical jet
is here represented
by  three regions,  each with a magnetic field  B$_{\rm i}$. Each region
produces a  synchrotron spectrum 
 with an optically thin and an optically thick part 
joining at $\nu_{peak_i}\propto B_i^{0.7}$ (i.e. for electron energy
index $p$=2).
 The composite spectrum 
from the three spectra develops therefore a central part at 
intermediate slope (Massi 1999).} 
\label{fig:cilindri}
\end{figure*}
As shown in the sketch of Fig. \ref{fig:cilindri} one can  imagine 
a continuous  jet in adiabatic expansion (conical jet) 
as formed by contiguous cylinders
of increasing radii and decreasing B$_i$.
With  each  cylinder   
a canonical  synchrotron spectrum is associated with  an optically thin part
$S\propto \nu^{(1-p)/2}$ (where $p$ is the electron energy index)
 and an optically thick part $S\propto \nu^{2.5}$,
with the two  parts of the spectrum joining
around $\nu_{peak}\propto B^{(p+2)/(p+4)}$ (van der Laan 1966; Dulk 1985).
Hence, the composite spectrum in case of a prolongated emission
will have an optically thin part (that of cylinder 1)
 and a thick part (cylinder 3) at the two
opposite ends, but
 will  also develop a central part with an intermediate
or even flat slope (Torricelli et al. 1998; Massi 1999).
A flat spectrum therefore reveals   a continuous jet. That has
observationally been proven by direct imaging a radio jet in 
 Cyg X-1 during its  Low/Hard state (Stirling et al. 2001).

\begin{figure*}[htb]
\centering
\includegraphics[width=\textwidth]{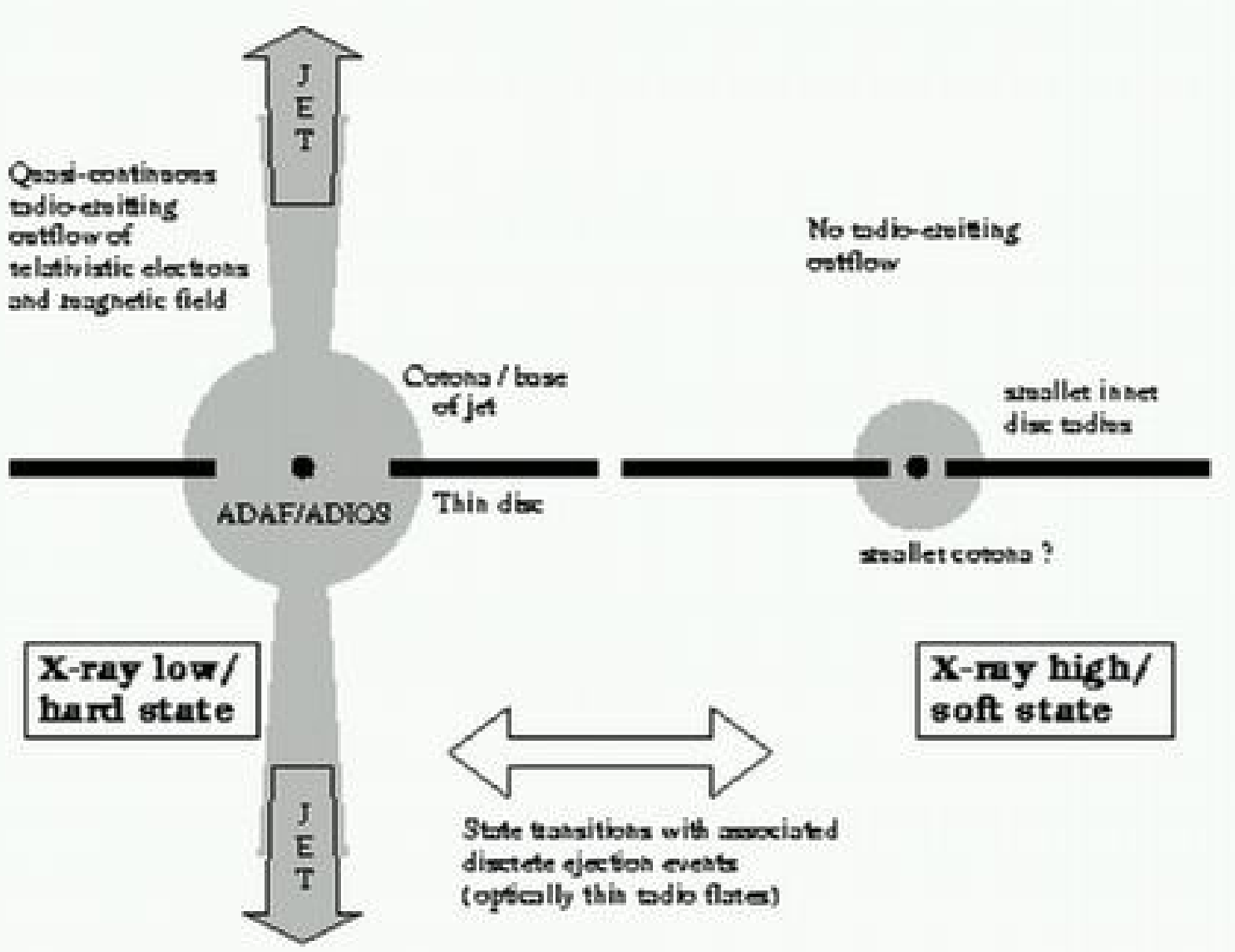}
\caption{ 
{\bf: Sketch of possible disk configurations during the High/Soft and Low/Hard states.} 
Fender et al. 1999.
In the High/Soft state the inner radius  reaches the last stable orbit $R_{in}\simeq 3 r_s$ and no radio jet is present.
In the Low/Hard state a radio jet is present and  
$R_{in}$ is larger than before (ADAF/Corona/base of jet, are all possible models for what is filling now that space). 
}
\label{fig:model-2}
\end{figure*}

With sufficient sensitivity  in the range  of a few tenths of  keV 
 in the Low/Hard state  it has been 
possible to observe a second  component due to  the 
accretion disc and therefore to measure the inner disk radius.
The $R_{in}$  resulted to be 
$\ge 50 r_s$ 
 (McClintock et al 2001b; McClintock \& Remillard 2004).
In the High/Soft state (sect. \ref{soft} and  Fig. \ref{fig:tana0}), 
the measured inner radius of the accretion disk corresponds to
$R_{in}\simeq   3 r_s$.
Either the inner disk has been removed or
during the transition High/Soft to Low/Hard
the inner part of the disk
has made a transition to a cooler state, which makes it effectively
invisible in X-rays.
The transition from a soft thermal state  to a hard
power-law  state  
therefore corresponds to  a change in the disk structure, evolving from
a state mainly characterized by the emission from the  inner part
of the disk to a state, in which this  inner-most region has been 
strongly modified 
(Tanaka 1997; Belloni et al. 1997).
Since the magneto-hydrodynamic theory of jet production (Sec. \ref{jetmagneto})  assumes
a large vertical magnetic field component, the inner-most region  is expected
now to be  geometrically thick and therefore strongly different
from the geometrically thin (optically thick) case  of the High/Soft state.
 However, how in detail is 
 the space within the inner radius filled?   What is
 the origin of the power law emission in the Low/Hard state
and why is it so well correlated to the radio 
 emission? This all  is still matter of debate
(Fig. \ref{fig:model-2})
(see the reviews by McClintock \& Remillard 2003 and  Fender 2004). 

\begin{figure*}[]
\begin{center}
\includegraphics[viewport=90 10 600 700  width=0.1\textwidth, height=1.1\textwidth, angle=0.0]{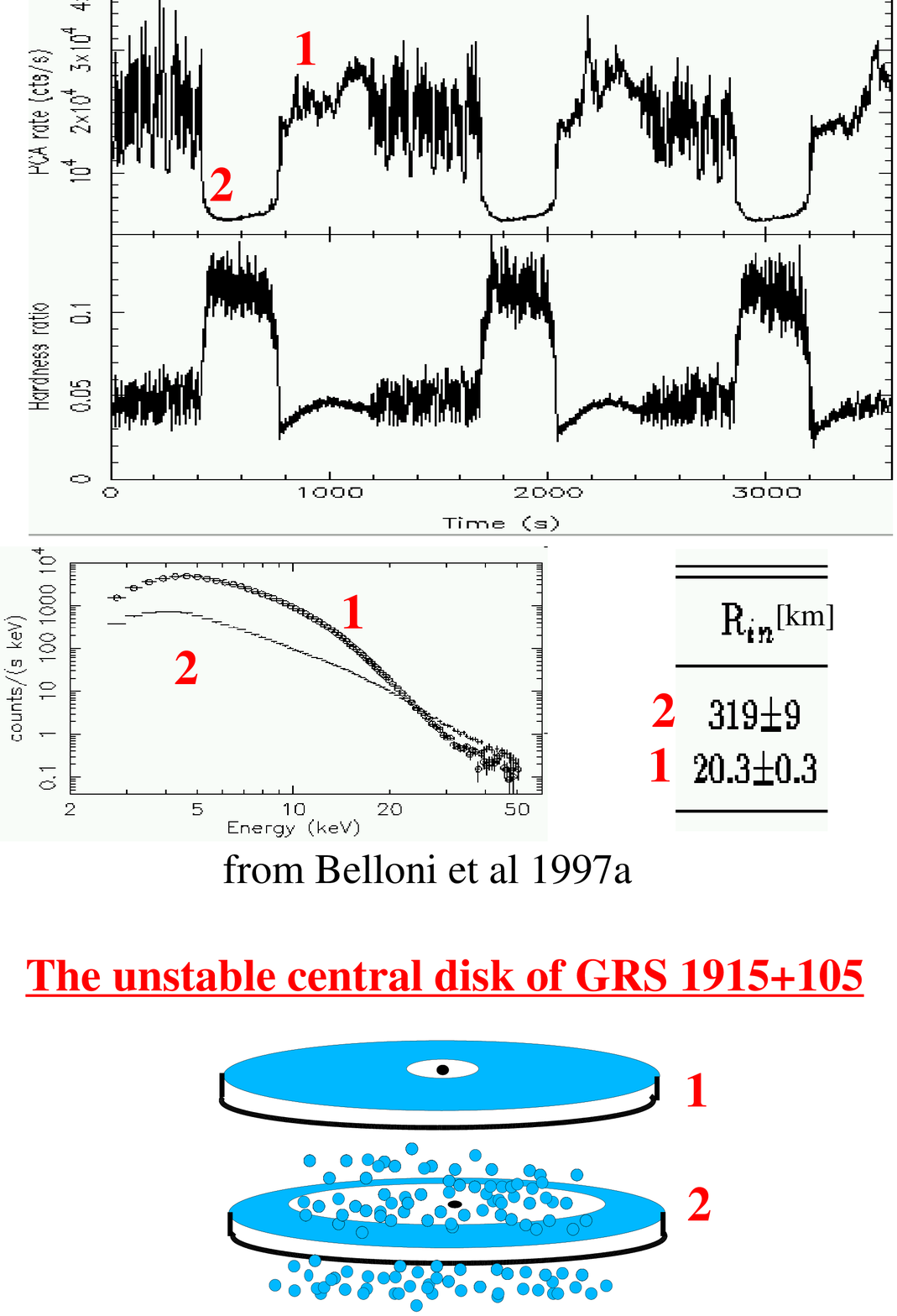}
\caption{ 
{\bf: The unstable central disk of GRS 1915+105}. Top: The 2.0-13.3 keV
PCA  light curve  and the corresponding hardness ratio (13.3-60 keV/2.0-13.3 keV) (Belloni et al 1997a).
 The signal oscillates between a harder (i.e. higher hardness ratio) state
 called here "2" and a softer state ( called "1").
Middle: PCA energy spectrum for average 1  (circles) and 2 (points) states
(Belloni et al 1997a). 
Table: Best-fit R$_{in}$ (km) by Belloni et al. (1997a). 
Bottom: sketch of the author, not in scale.
The inner part of the disk
has made a transition ($1\rightarrow 2$) to a cooler state, which makes it effectively
invisible in X-rays.
Owing to approximations in the disk blackbody model the value of the
inner radius might be unaccurate (Merloni et al 2000), however such a large
variation cannot be due to these approximations (Fender \& Belloni 2004).
}
\label{fig:belloni0}
\end{center}
\end{figure*}

\begin{figure*}[htb]
\centering
\includegraphics[width=10cm, height=10cm, angle=-90.0]{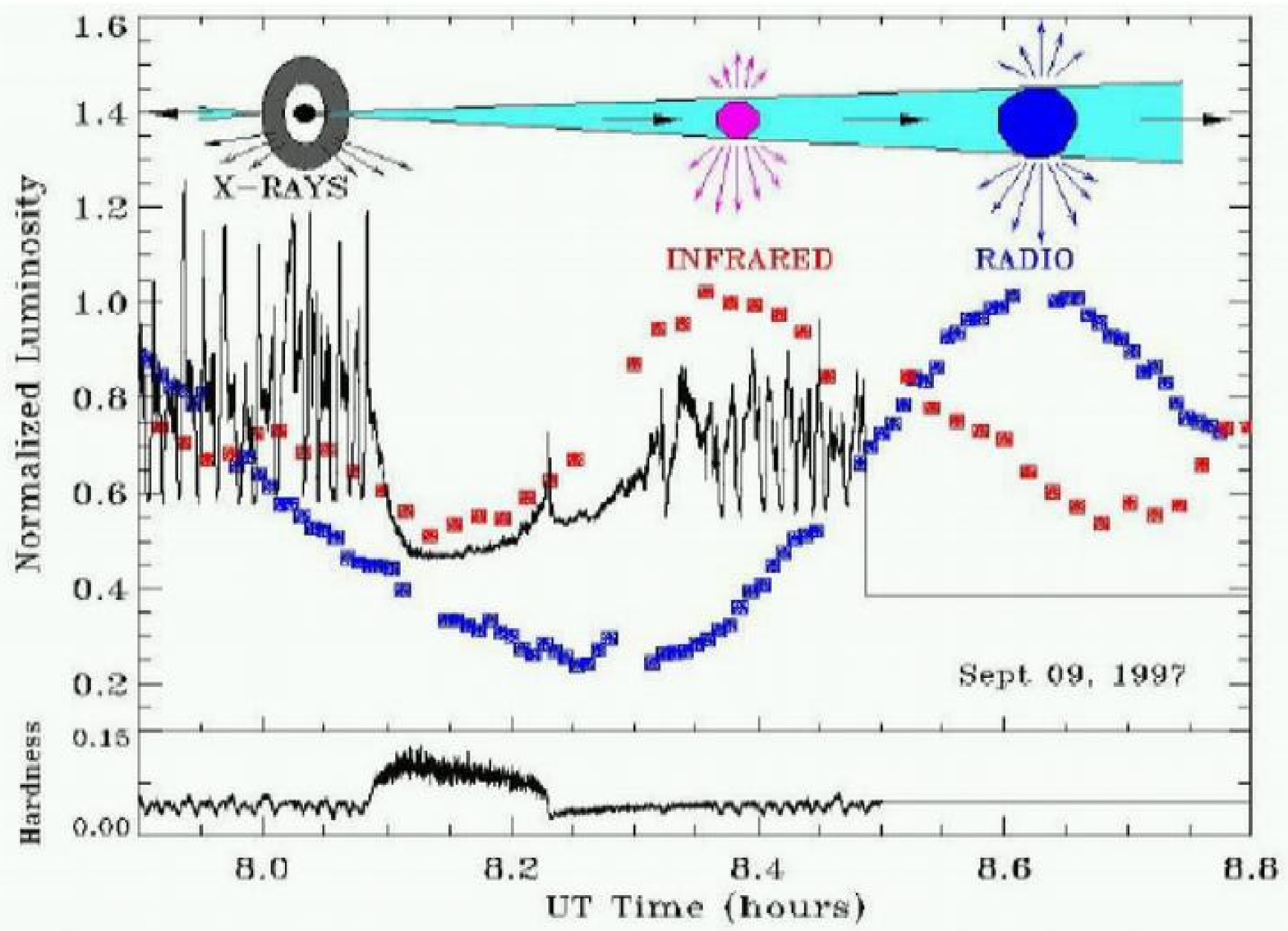}
\caption{{\bf: Accretion instabilities and jet formation in GRS 1915+105.}
Mirabel et al. 1998.
The infrared flare starts during the recovery from the X-ray dip, when an X-ray spike was observed. These observations show the connection between the rapid disappearance and follow up replenishment of the inner accretion disk seen in  X-rays, and the ejection of relativistic plasma clouds observed as synchrotron emission at infrared and radio wavelengths. The hardness ratio (13- 60 keV)/(2-13 keV) is shown at the bottom.
} 
\label{fig:mirabel}
\end{figure*}

Together with the two extreme  states High/Soft (radio quiet)
 and Low/Hard (continuous ejection) there 
are two intermediate (concerning their hardness) states:
the Very High State (VHS) and the Intermediate State (IS).
Multiple  recurrent oscillations in X-rays 
in the source GRS 1915+105 (Belloni et al. 2000) are due to 
different  VHS-like states 
 called A,B and C, where ejections- emitting in the radio band -
  occur in the
hardest (C) state (reaching a Low/Hard state with $\Gamma \sim 1.8$).
Synchronized  variations of the inner radius have been
observed to occur during this oscillations (Belloni et al. 1997,1997b).
Figure \ref{fig:belloni0} shows the spectral change corresponding to
a variation of $R_{in}$ 
 from  20 km  
(value compatible with  the last stable orbit 
around   a rotating black hole) to more than 300 km.  
During similar  episodes of cyclic variations of the
inner disk in GRS 1915+105,  Mirabel and collaborators (1998) 
could follow
the onset of a flare - first at infrared wavelenghts and then
at radio wavelengths - with a delay
consistent with synchrotron radiation from expanding magnetized clouds
of relativistic particles (Fig.\ref{fig:mirabel}).
The  straightforward interpretation was that
during the disappearance of the inner disk a
 relativistic plasma cloud was expelled.
The mass of the ejected cloud has been estimated to  $\sim10^{19}$g
(Mirabel 1998),
 while the matter  which disappeared from the
inner disk in one dip
 of similar time length 
(Belloni 1997), 
has been estimated to   $\sim10^{21}$g.
This fact could imply that only a very small fraction of the mass
 is ejected, whereas is not clear which  
fraction fell indeed onto the compact
object (Mirabel and Rodriguez 1999). 

These small oscillations  are now well established,
they have   recurrence time-scales of tens of minutes 
and have been observed  in X-ray 
(interpreted as possible  draining and refilling of
 the inner disk) and at infrared, millimeter and radio wavelengths
 (interpreted as repeated ejection events) 
(Fender et al. 1999: Eikenberry et al. 1998; Fender et al. 2002). 
Finally, Fender and coauthors (1999: 2002) have shown that
these low ($\sim$ 40 mJy in radio) amplitude 
 oscillations can happen during the decrease of flux in major flare events. 

In conclusion:
\begin{enumerate}
\item
A Soft state, characterized by disk emission
 (at temperatures $\sim$ 1 keV 
contributing mostly at 1-2.5 kev) and a 
power law component
steeper than $\Gamma\ge 2$ implies  a  radio-quiet X-ray binary system.

\item
Quasi-periodic oscillations
with  time-scales of several minutes  in X-ray and in the
infrared  or  radio band
 are a signature of episodic  disk-removal/variations
 and plasma bubble ejection.
In  case of such  isolated small ejections, one
can follow the adiabatic expansion of the  cloud and
monitor  radiation becoming optically thin at progressively
lower frequencies.

\item
 When the X-ray binary system 
is persistently emitting a radio jet
 the X-ray spectrum has a power law component with  $\Gamma\simeq 1.6$. 
 This X-ray state is called  Low/Hard.
The  superposition (Fig. \ref{fig:cilindri})
 of  spectra of different contiguous jet regions
 with different self-absorption cutoffs
result in   a composite flat spectrum
 (i.e. S$\propto \nu^{\alpha}$,
with $\alpha\sim 0$)  through and beyound  radio wavelengths.
Emissions in the radio band and at hard X-rays 
are related by:
 $L_{radio}\propto L_X^{0.7}$

\end{enumerate} 

\subsection{ {LS~I~+61$^{\circ}$303}: Soft and Hard States} 
\label{x:lsi}

There are three  X-ray observations of LS~I~+61$^{\circ}$303:
 with ROSAT  (Taylor et al. 1996),
ASCA (Leahy et al. 1997)
and RXTE  (Harrison et al. 2000).
The ROSAT observation was performed over a total orbital cycle
 in the energy range
from 0.07 to 2.48 keV.  
A single component fit was made, either a black body or 
a power-law and  average fitted results were presented: A 
temperature of  0.26 keV and a power law index of $\Gamma$=2.
\begin{figure*}[htb]
\centering
\includegraphics[width=10.0cm, height=10cm, angle=0]{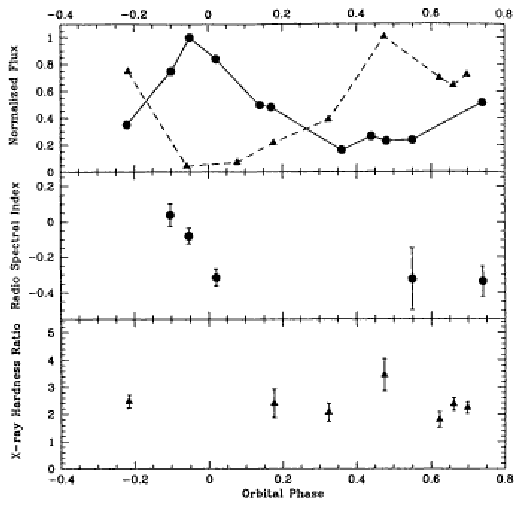}
\caption{{\bf: X-ray and radio flux density and spectral variations of \lsi}.Taylor et al. 1996. Top: The 4.9 GHz radio flux density (solid line and circles) and the
X-ray flux (dashed line and triangles). Middle: Radio spectral index.
Bottom:Hardness ratio defined as the ratio of photon counts in the energy range 1.0--2.48 keV to those in the 0.07--1.0 keV range.
}
\label{fig:taylor1}
\end{figure*}
The hardness-ratio is calculated all around the orbit and 
Taylor et al. (1996) notice
the  hardening of the X-ray emission during the onset 
of the second radio outburst (Fig. \ref{fig:taylor1}).
\begin{figure*}[htb]
\centering
\includegraphics[width=11cm, height=11cm, angle=0]{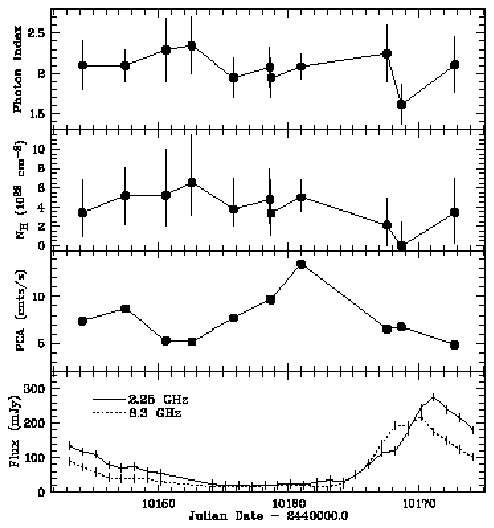}
\caption{{\bf: Variations of the X-ray spectral fit parameters (top two panels)
with Time.} Greiner \& Rau 2001. The errors correspond to 3$\sigma$. The bottom
panels show the count rate from 2.3-25 keV and the radio flux (Green Bank Interferometer).
The photon index (Top panel) drops to the value of 1.6 at JD (2440000.0+) 10169,
in coincidence with the onset of the  radio outburst.}
\label{fig:rao}
\end{figure*}

Greiner and Rau (2001) calculated  
the photon index around the orbit based on  RXTE data 
(measured range 2.3--25 keV) (Fig. \ref{fig:rao}).
It is quite remarkable that  the fit
(only a single component, i.e. a power law)
gives  $\Gamma=$2.0-2.4 all around the orbit 
(i.e.  a  High/Soft state) except for one point - simultaneous with
 the onset of a  radio outburst -, where the 
photon index is $\Gamma$= 1.6 (i.e. Hard state).    
This seems  to be the typical case of the theory presented in the
previous section for a Low/Hard state. 
The X-ray peak occured
at phase $\sim $0.48 whereas the radio outburst came
almost 6 days later (Fig. \ref{fig:rao}).
 However, in the light of developments
in the theory of the disk-jet connection (reviewed in the previous sections)
 the spectral switch to 
the hard state is relevant for the ejection; 
this switch  must precede the onset of the
radio outburst and  RXTE data
actually show this for LS~I~+61$^{\circ}$303
(Massi 2004).

If a source  remains in the Low/Hard state,
it is radio-loud and the spectrum is flat. How does the spectrum of 
 LS~I~+61$^{\circ}$303 behave taking into account that the Low/Hard state
seems  to be of a quite short duration ?
Indeed,
a flat spectrum has been measured by  Taylor et al. (1996):  in Fig.~\ref{fig:taylor1}-middle
 the radio spectral index is equal to zero  before the radio peak.
  When the ejection phase is terminated
the figure very nicely shows that 
the flat spectrum evolves into an optically thin one
(i.e. $\alpha$ changes from 0, due to the composite spectrum,
to  -0.3 related to the optically thin part of the ``last''emitted
bubble or  cylinder as in Fig. \ref{fig:cilindri}).
As shown by Paredes et al. (1991)
the  flat spectrum in  LS~I~+61$^{\circ}$303
can be reproduced by a model of an adiabatically
expanding cloud of synchrotron-emitting relativistic electrons
only if 
a continuos ejection of particles (lasting two days) is taken into account
as well as 
adiabatic expansion losses.

\begin{figure*}[htb]
\centering
\includegraphics[width=14cm, height=11cm, angle=0]{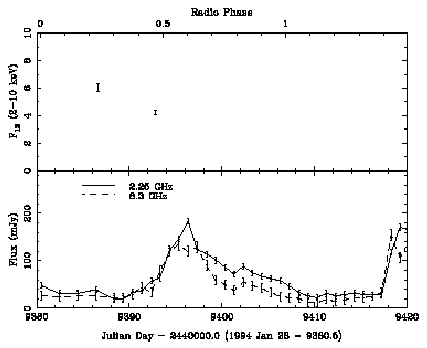}
\caption{{\bf: X-ray and radio monitoring of \lsi.} Harrison et al. 2000.
 Two-frequency radio light curves (bottom) taken simultaneously with the two ASCA pointings (flux levels shown in the top panel). Both ASCA pointings have photon indexes $\Gamma$= 1.5--1.9.
The first pointing is at orbital phase $\phi=$0.2 which is the periastron passage, where super-accretion
is predicted. The second ASCA pointing is at  the onset of a radio outburst (bottom panel). 
}
\label{fig:harrison}
\end{figure*}

Leahy et al. (1997) reported  two ASCA  observations,
where the photon index 
$\Gamma$ was  1.63-1.78 
at orbital phase $\phi$=0.2,
which is the periastron passage (where an ejection is predicted),
and $\Gamma$=1.75-1.90  
at orbital phase  0.42
coincident with the {\bf onset} of a
radio outburst (as shown in Fig. \ref{fig:harrison}).
In conclusions, both ASCA values
give a  photon index consistent with a Low/Hard state
during  predicted/observed ejections.

\begin{figure*}[]
\includegraphics[width=\textwidth]{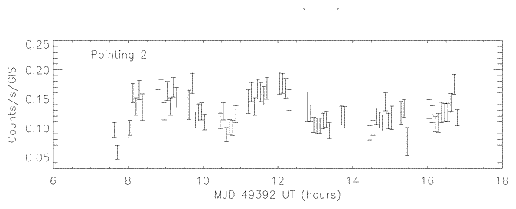}
\centering
\caption{{\bf: Short-term variability in \lsi.} Harrison et al. 2000. Count rate in the 15 keV band  for the  ASCA pointing 
simultaneous with the radio outburst (see text).}
\label{fig:qp}
\end{figure*}
\begin{figure*}[]
\centering
\includegraphics[width=\textwidth]{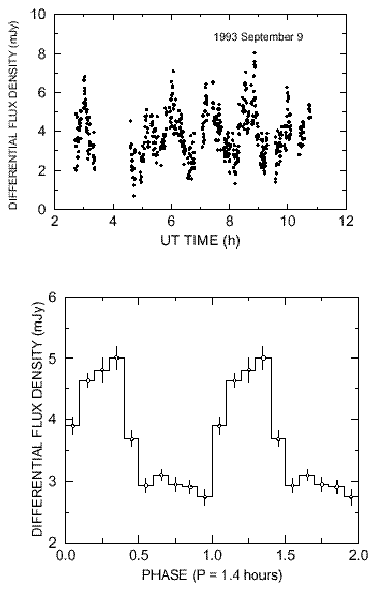}
\caption{{\bf: Radio microflares.} Peracaula et al. 1997. 
Top: Radio oscillations with mJy amplitude. Bottom: Mean radio light curve of \lsi obtained by folding with a period of 1.4 h. 
For a better display the orbital phase interval 0-1 is repeated twice.}
\label{fig:marta0}
\end{figure*}

Harrison and collaborators (2000)
have performed a 
periodicity analysis  
of the two  ASCA observations. The result is
a clear
periodicity 
 (Fig. \ref{fig:qp}) in the  ASCA pointing   
related to the onset of a radio outburst while no periodicities are found
in the other more sparcely sampled pointing.
The X-ray oscillations occured
at phase 0.42 before the radio peak, which
 (Fig. \ref{fig:harrison})  occured  at phase
 0.5-0.6.
The photon indexes given above reflect an average of all the data; 
therefore we cannot check, if the Low/Hard state is  stable
or if  there is  a continuos switching between a sort of  Very High States
 (i.e.  A,B,C) like for GRS 1915+105 (Belloni et al. 2000) 
or if the  Low/Hard state is indeed  reached only in the hardest interval.  
Such  oscillations are also present  
in  the radio band.
Peracaula and collaborators (1997) have performed
a period-analysis 
for three radio observations: two in  a decreasing phase of 
 large outbursts
and one at a high, but quiescent flux level.
While in the last data set there  was  no evidence for a 
periodicity, on the contrary, a period of P~=~84 minutes 
and   significant power 
also at   harmonics of P/2 (i.e.  $\sim 40$ minutes again) 
and 2P have been found in the two 
data sets related to the decay of radio outbursts 
(Fig.  \ref{fig:marta0}).

In conclusion:
\begin{enumerate}
\item
There is no estimate of any  inner radius  for the accretion disk in \lsi.  The multicolor disk mostly emits below 2.5 keV and this range is excluded
by the RXTE  analysis.
\item
The available X-ray observations for \lsi
reveal  transitions to Hard/Low states at the onset of  radio outbursts 
as is  expected in the context of the disk-jet connection (Massi 2004).
\item
A prolongated  ejection of particles generates a flat spectrum
in the late portion of the rise in flux.  
\item
X-ray oscillations and radio-oscillations 
are present at the early rise phase and during  
the  decay of radio outbursts (Peracaula et al. 1997: Harrison et al. 2000)..
\end{enumerate}

\subsection{The Periodical Radio Outbursts of \lsi} 
\label{x2:lsi}

The greatest  peculiarity of LS I +61$^{\circ}$303  is
its periodic radio outburst activity  with P=26.496 days
(Gregory 2002)
\begin{figure*}[]
\centering
\includegraphics[height=\textwidth, angle=-90]{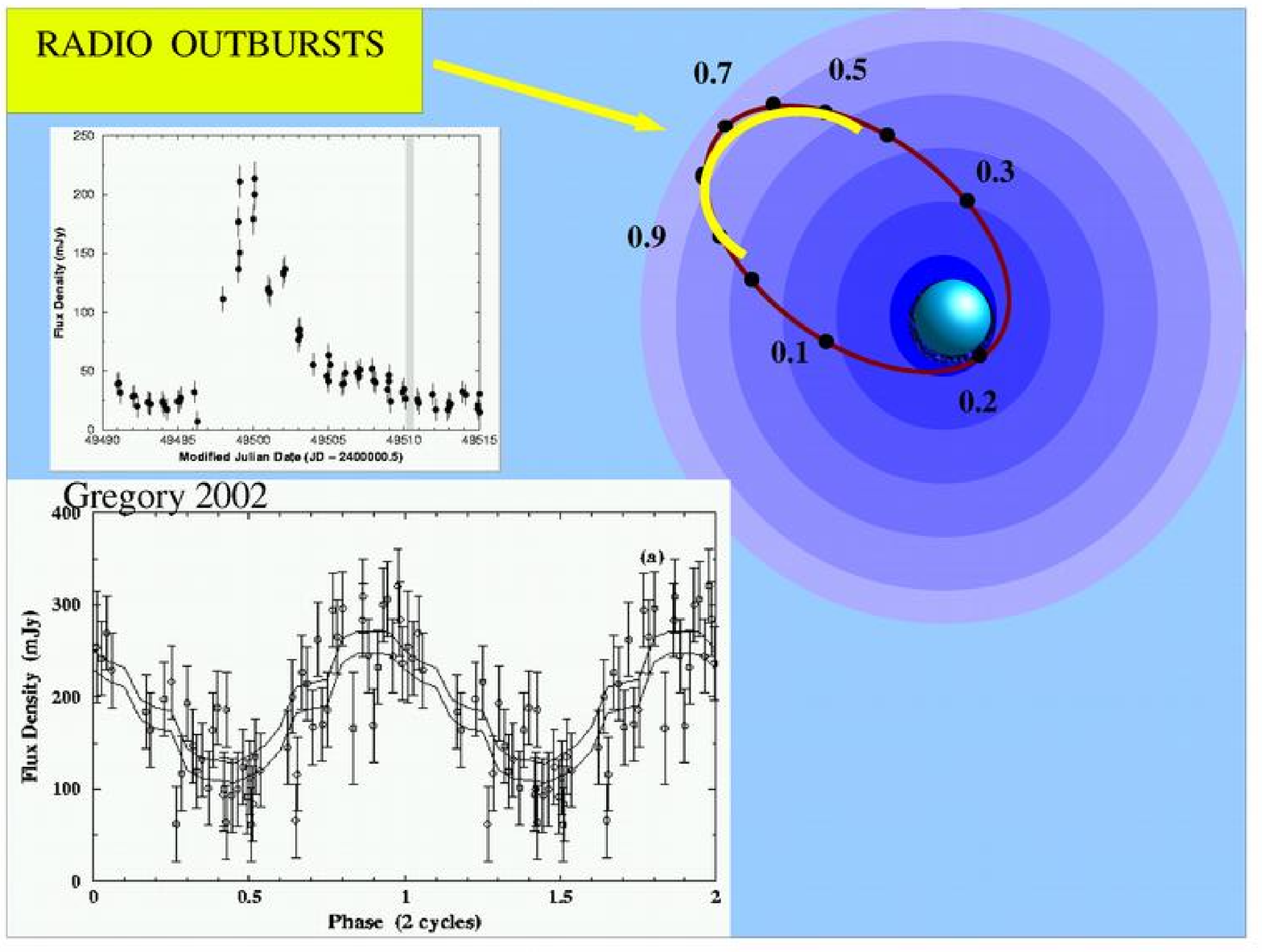}
\caption{{\bf: Radio outbursts in \lsi.} Right: Orbital phase interval (in yellow) where the periodical radio
outbursts  occur. Left-Top: Typical radio light curve during a strong outburst.Left-Bottom: The 4.6 yr periodic modulation of the  amplitude of the outbursts.  }
\label{fig:radioburst}
\end{figure*}
(see in Fig.\ref{fig:radioburst}~Left-Top  a typical radio
 light curve).
The second peculiarity of \lsi~ is that the amplitude of each outburst is not randomly varying, but itself periodic
with a periodicity of 4.6 years (Fig. \ref{fig:radioburst}~Left-Bottom)
(Gregory 1999, 2002).
\begin{figure*}[]
\centering
\includegraphics[width=7.9cm, height=7.9cm, angle=0.0]{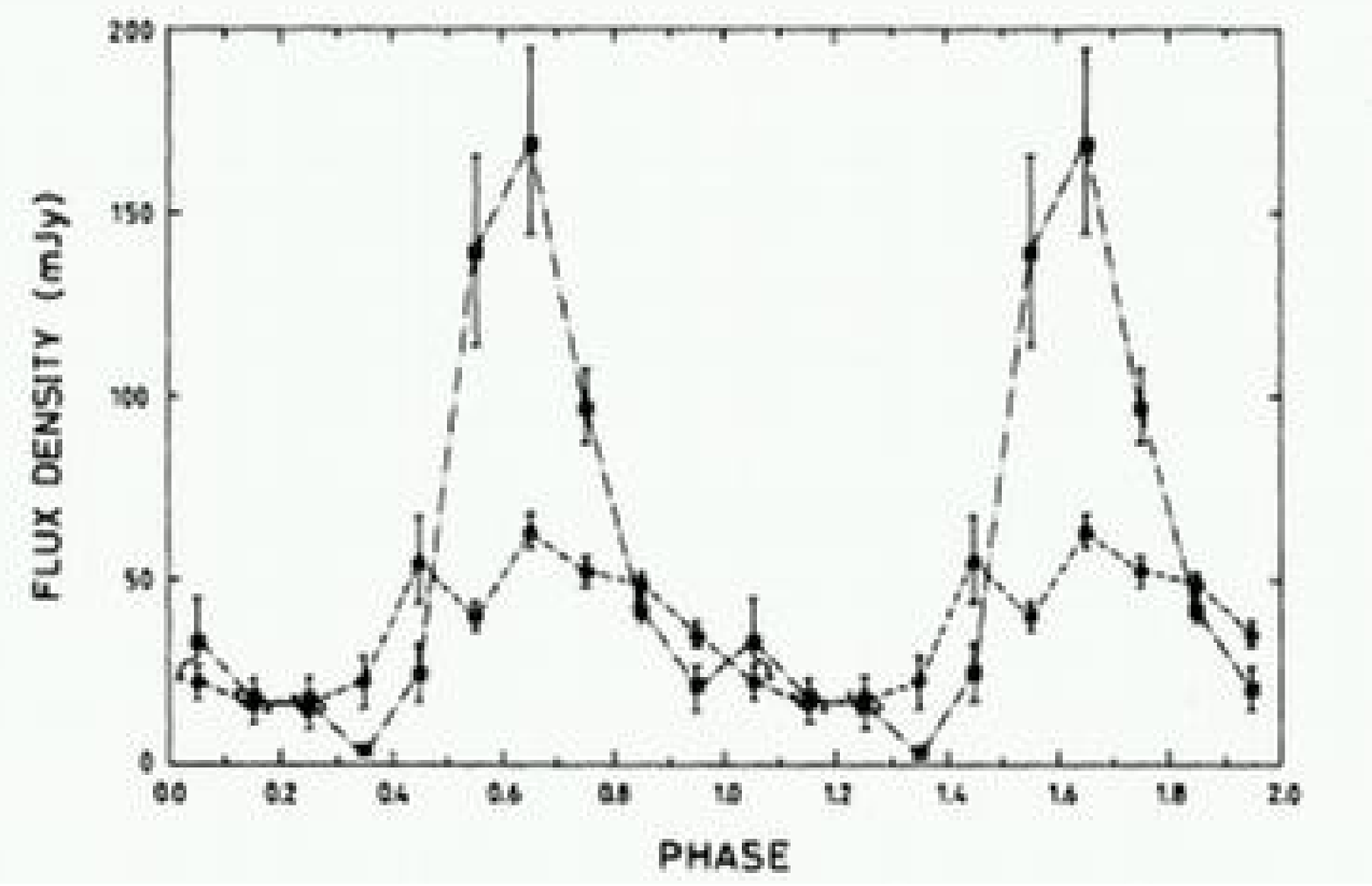}
\caption{{\bf: Average radio light curves.} Paredes et al. 1990.
 The average curves are relative to  the maxima (dash-dotted) and to
 the minima phases of the long-term  modulation.
Low outburst have a broad distribution 0.4--1.0, whereas
strong outbursts occur in the range of orbital phases 0.5--0.8.
.}
\label{fig:paredes0}
\end{figure*}
The orbital phase  
 $\phi$ at which these
outbursts occur is modulated (Gregory et al. 1999) and varies  within
 the interval
0.5--0.8 (Fig.~ \ref{fig:radioburst}~Right
 and Fig.~\ref{fig:paredes0})(Paredes et al. 1990).

The theory of accretion (Eq. \ref{eq:mdot}) predicts maximum accretion
 where the density is highest. The maximum density is obviously 
at the periastron
passage,   because the density of the wind  there is 
the largest and in addition
there  probably occurs direct accretion  from the  
the star (Roche Lobe overflow).
As shown in Fig. \ref{fig:radioburst}~Right the  $\phi$ at
 periastron passage is   0.2 and therefore practically is opposite to the
orbital region, where the radio outbursts occur.  Therefore
one of the fundamental questions concerning the periodic radio
outbursts of LS~I~+61$^{\circ}$303  has been for years:
Why are the radio outbursts shifted with respect to the periastron passage ?

\section{THEORY OF THE ACCRETION: THE TWO PEAK ACCRETION MODEL}
\label{teoria}

\begin{figure*}[]
\centering
\includegraphics[height=\textwidth, angle=-90]{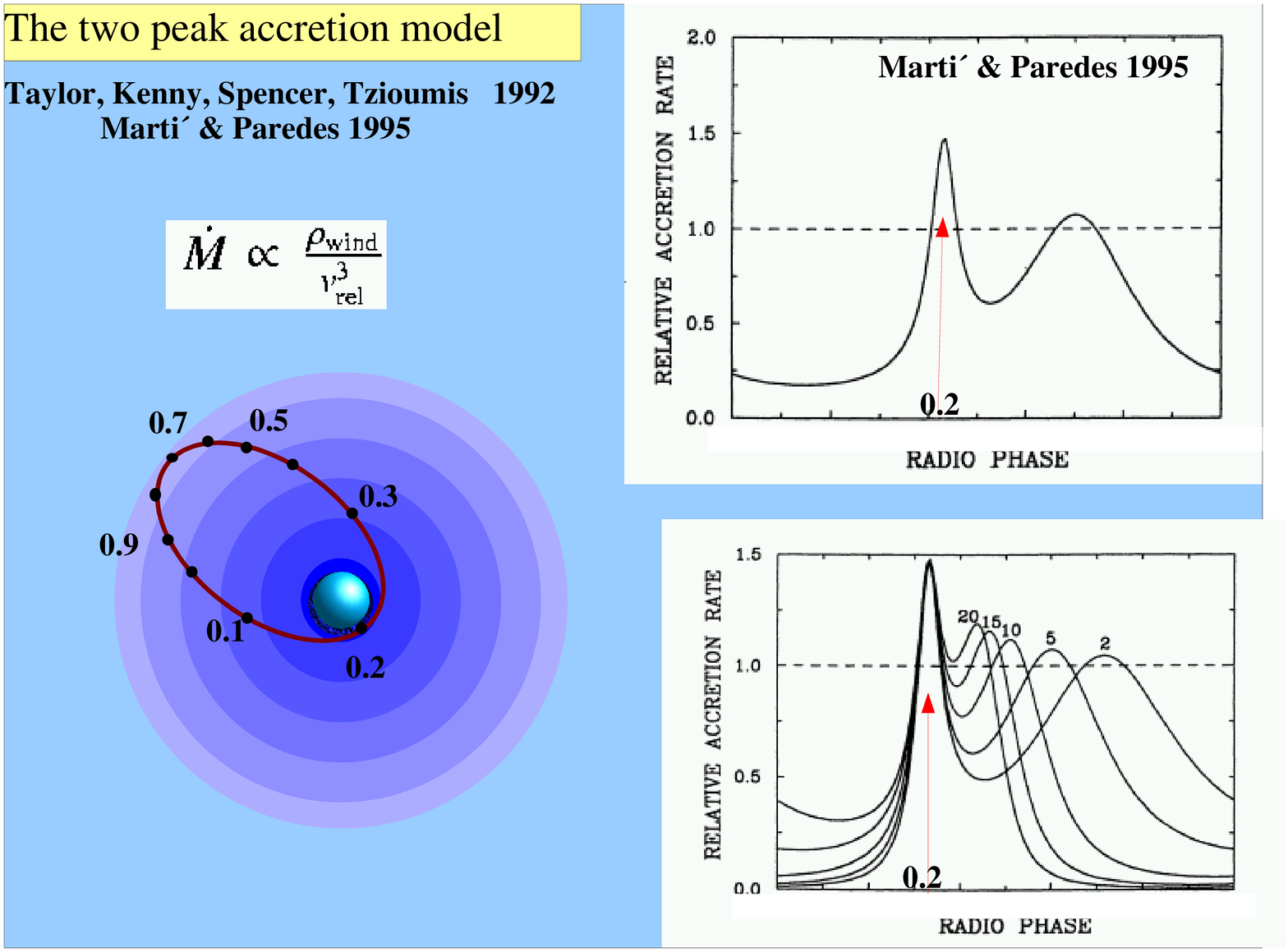}
\caption{{\bf: The  accretion model for an eccentric orbit.} Mart\'{\i} \& Paredes 1995. Top: Accretion rate
versus stellar wind. The vertical axis is in units of the Eddington accretion
limit, whose limit is indicated by the dashed line. Bottom: Accretion rate for different velocities of the stellar wind. The values are in km s$^{-1}$.
Note how the second super-critical event shifts gradually towards earlier orbital phases
for high values of the wind velocity.} 
\label{fig:marti0}
\end{figure*}

Equation  \ref{eq:mdot} gives   only one accretion peak
 for variable
density and constant velocity $v_{rel}$.
However, the orbit of \lsi is quite eccentric and therefore
with a strong variation of the velocity  along the orbit.
Taylor et al. (1992) and Mart\'{\i} \& Paredes (1995) have shown that in 
this case 
the accretion rate $\dot{M} \propto {\rho_{\rm wind}\over v_{\rm rel}^3}$
develops indeed two peaks: The
first peak corresponds to the periastron passage (highest density), while
the second peak occurs when the drop in the relative velocity $v_{\rm rel}$
compensates the decrease in density (because of the inverse cube dependence)
(Fig. \ref{fig:marti0} ~Top~{\it Right}).

This also implies that, while the first peak always occurs at periastron passage, the second peak may move to different points in the orbit, if 
 variations 
in $\rho_{\rm wind}$ or  $v_{\rm rel}$ occur.  
Figure \ref{fig:marti0} ~Right-Bottom 
shows how  for increasing values  of the  wind-velocity  
 the second peak shifts toward the first peak
(Mart\'{\i} \& Paredes 1995).
\clearpage

\begin{figure*}[htb]
\centering
\includegraphics[width=9.0cm, height=9cm, angle=0]{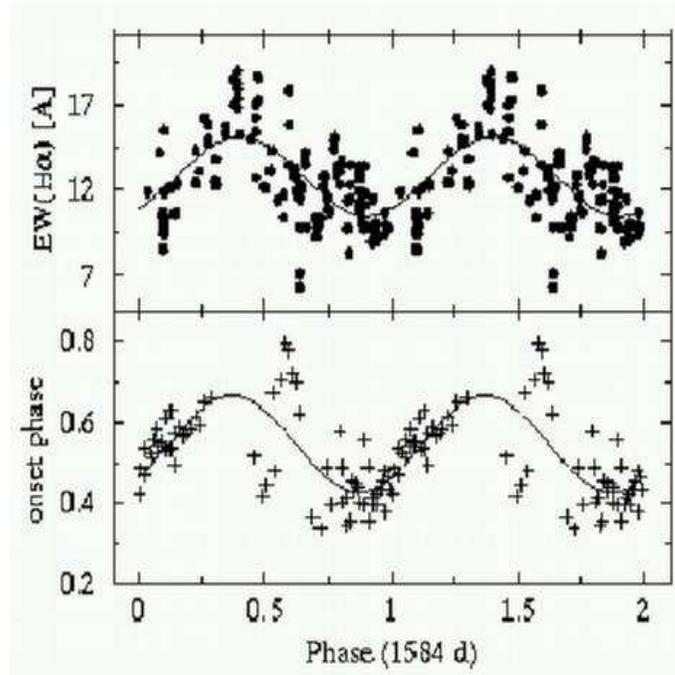}
\caption{{\bf: Correlation between  H$\alpha$ emission
line variations and the radio outburst orbital phase variation.}
Radio and H$\alpha$ parameters folded on the long (here $\sim $ 1584 d) 
period of modulation of the amplitude of the radio outbursts.
Top:
 Total equivalent width of the H$\alpha$ emission line 
Bottom: Averaged values of the onset phase.} 
\label{fig:zamanov0}
\end{figure*}
On the other hand    variations in the mass loss of the Be star have been
 well established by  H$\alpha$ emission line observations 
(Zamanov \& Mart\'{\i} 2000).
Gregory \& Neish (2002) suggest a periodic outward
moving density enhancement (i.e., shell ejection) in the Be star wind.
The variation of  H$\alpha$ emission line (Zamanov \& Mart\'{\i} 2000) 
is   periodic with a comparable  scale (1584 d)
as the radio modulation and it is in phase with the onset
of the outbursts (Fig. \ref{fig:zamanov0}). The orbital shift in the phase of
the radio outbursts  is therefore related to variations of the wind 
parameters.

Finally, Mart\'{\i} \& Paredes (1995) have shown that during both peaks the accretion
rate is above the
Eddington limit and therefore one expects that matter is ejected twice
within the 26.496 days interval.

In conclusion, 
 radio outbursts displaced from periastron passage correspond to
the second peak of the two-accretion/ejection peaks.
The remaining problem  therefore is:
Why is the first outburst at periastron passage in the radio band missing ?

\newpage
\section{GAMMA-RAY OBSERVATIONS} 
\label{gamma}

\subsection{EGRET Sources} 
\label{egret}

\begin{figure*}[htb]
\centering
\includegraphics[width=14.5cm, angle=0]{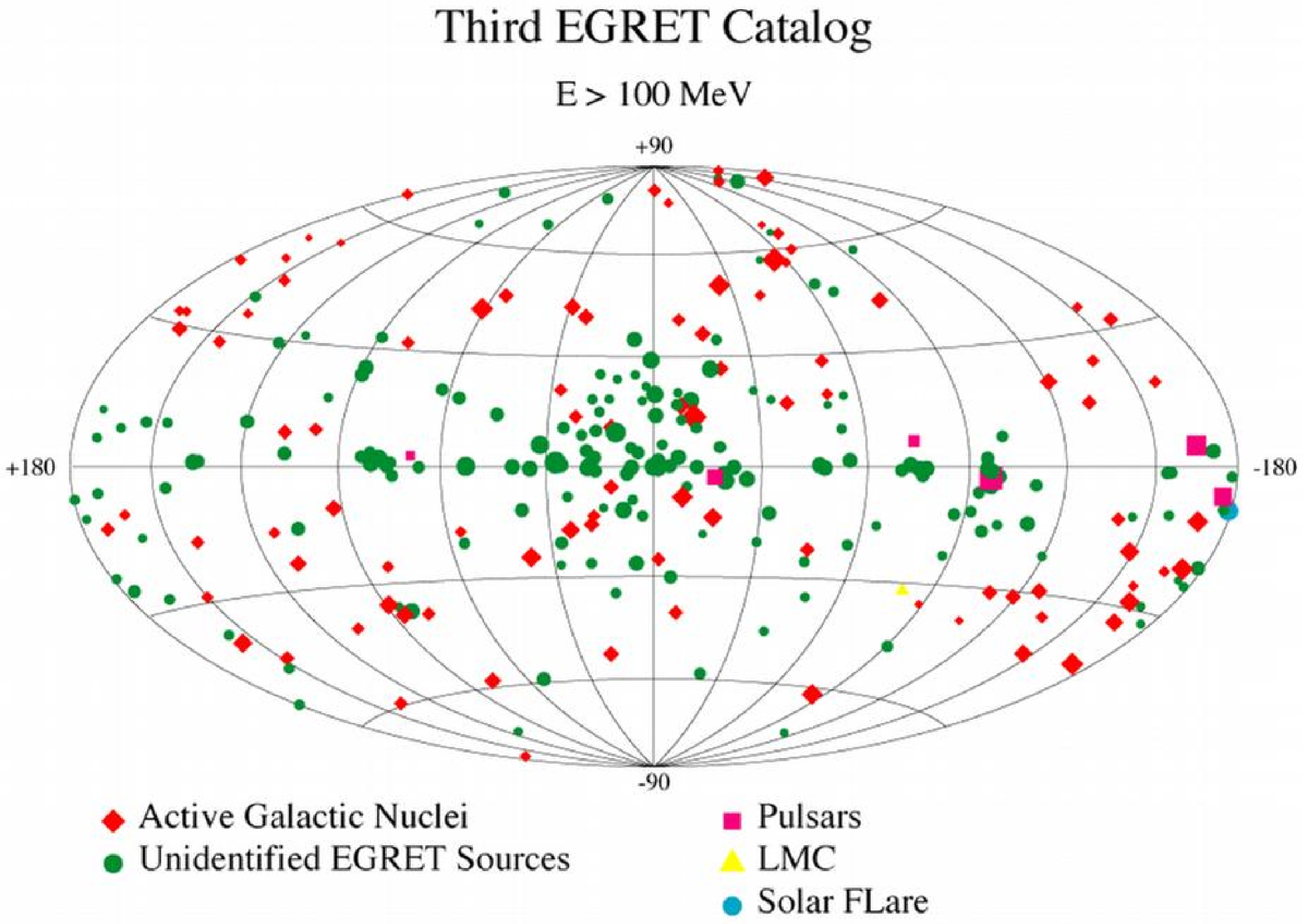}
\caption{{\bf Third EGRET catalog.} http://cossc.gsfc.nasa.gov/egret}
\label{fig:egret2}
\end{figure*}
\begin{figure*}[]
\centering
\includegraphics[width=5.0cm, angle=0]{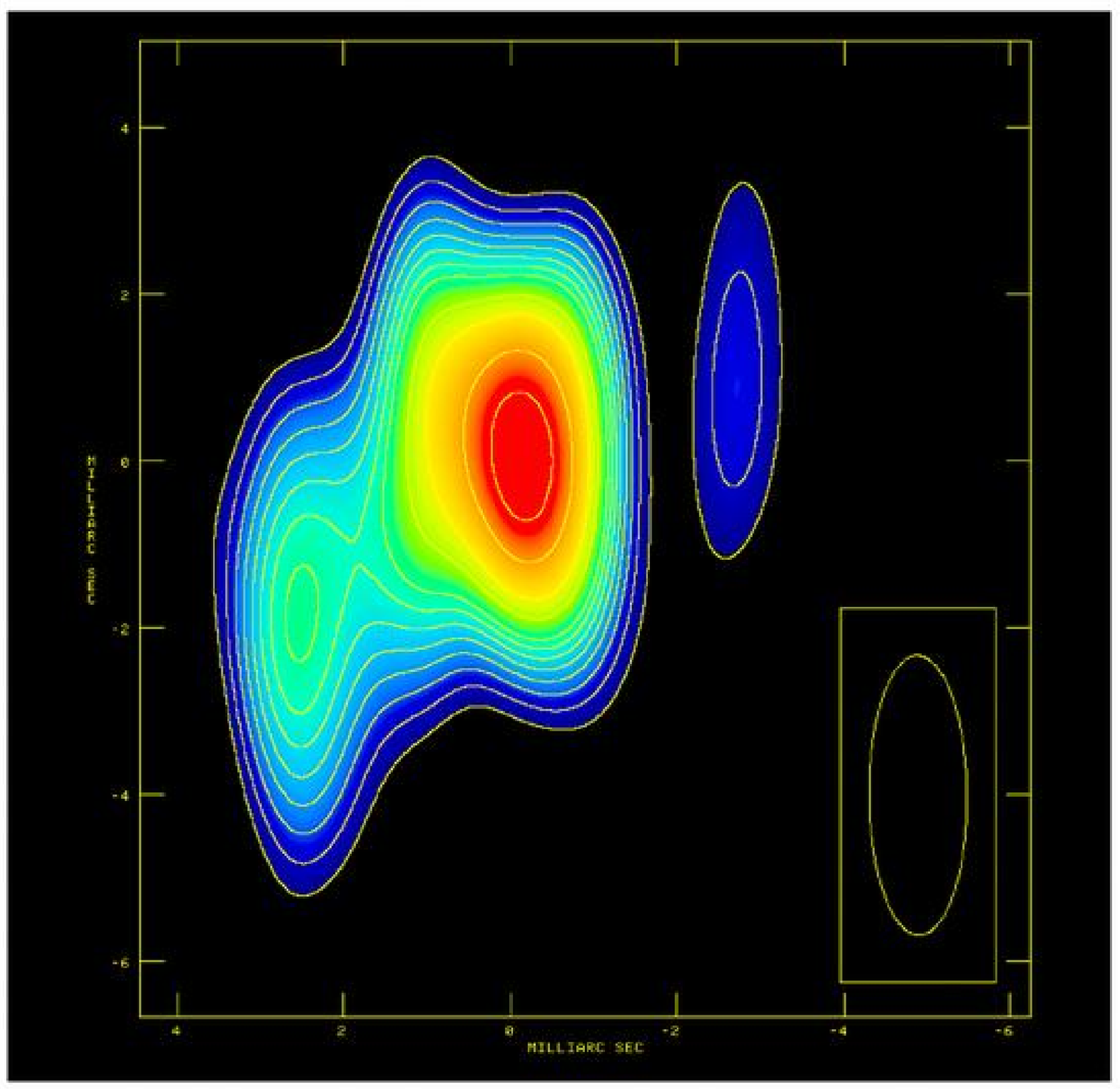}
\includegraphics[width=5.0cm, angle=0]{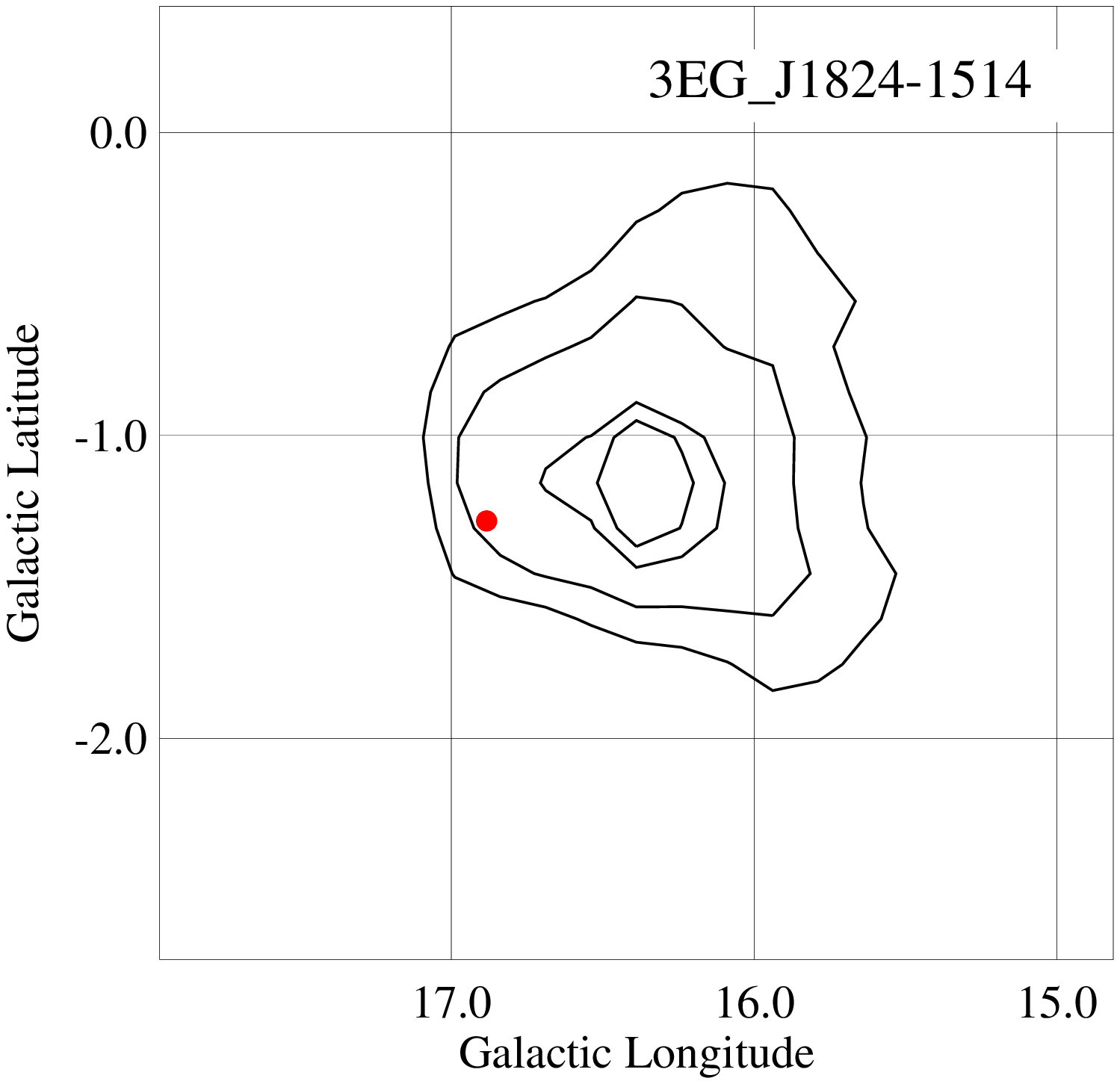}
\includegraphics[width=5.0cm, angle=0]{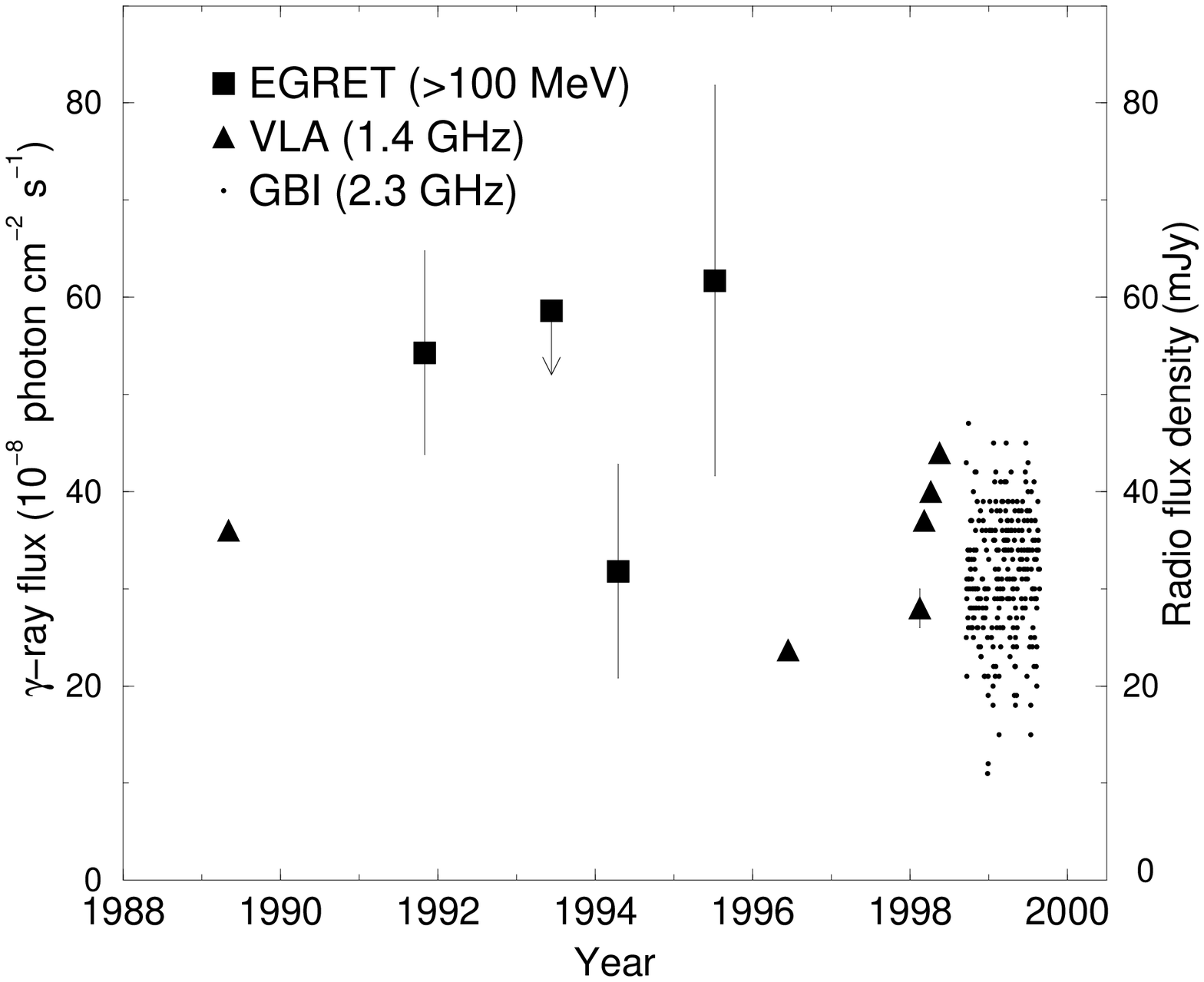}
\caption{{\bf A high-energy gamma-ray-emitting persistent microquasar:} Paredes et al 2000. 
Right:
The location EGRET-map of Hartman et al. (1999).
The contours are not intensity contours, but statistical ones
representing the 50$\%$, 68$\%$, 95$\%$ and 99$\%$ probability that a
single source lies within the given contour. The red dot inside the 95$\%$
confidence contour, whose radius is about half degree, is the position of
  LS~5039. Left: VLBA-map of LS 5039. The 
presence of radio jets in this high mass X-ray binary is the main
evidence of an accretion process resulting in the ejection of relativistic
particles.  The overall length of the source is about 18~AU. 
Bottom: EGRET  fluxes and radio observations showing a roughly persistent level of 
emission over a decade in both bands.} 
\label{fig:paredescolor}
\end{figure*}

The Third EGRET Catalog contains about 170 not yet identified
 high energy Gamma-ray sources (E $>$ 100 MeV) ((Fig. \ref{fig:egret2}). 
 The discovery of the coincidence 
of the  microquasar LS5039 (Fig. \ref{fig:paredescolor})
with an unidentified EGRET source by Paredes and collaborators (2000)
 has opened
the possibility that other EGRET sources could be microquasars as well. 
Gamma-rays can be produced by external Compton scattering of 
stellar UV photons of the massive companion by the relativistic electrons
of the jet. 
LS5039 is a persistent radio emitting source and the Gamma ray flux,
with all uncertainties reflected by the poor sampling, still reflects
this persistence (Fig. \ref{fig:paredescolor})(Paredes et al 2000). 
Therefore, for a periodic source like \lsi periodic Gamma-ray emission would
be expected. 

\subsection{The Variable Gamma-Ray Source \lsi}  \label{gamma:lsi}

\begin{figure*}[]
\centering
\includegraphics[width=9.0cm, angle=-90]{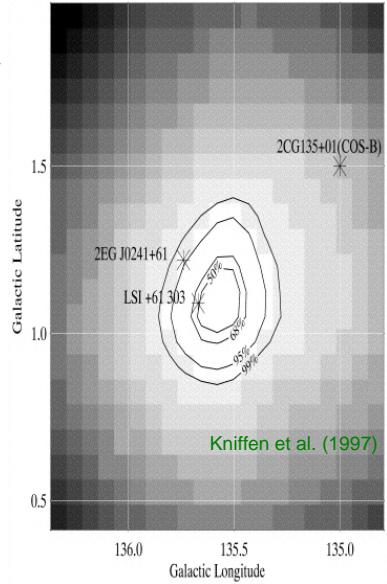}
\caption{{\bf Contour plot in galactic coordinates of 2EG J0421+6119.} (Kniffen
et al. 1997) see text.}
\label{fig:kniffen0}
\end{figure*}
\begin{figure*}[htb]
\centering
\includegraphics[angle=-90, scale=0.4]{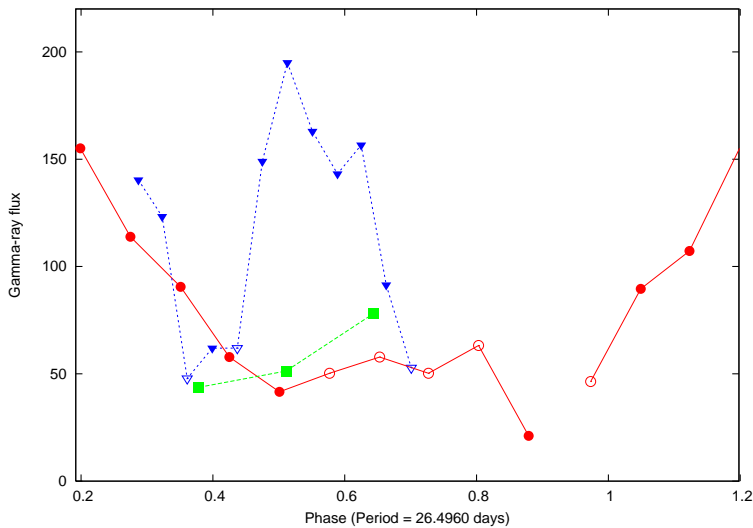}
\caption{{\bf: EGRET data vs. orbital Phase} Massi et al 2004b (see text).}
\label{fig:massiegret}
\end{figure*}

Gregory and Taylor (1978) reported  the discovery of
a radio source (later on associated with \lsi) within the 1$\sigma$
error circle of the COS~B $\gamma$-ray source 2CG~135+01. This association
however remained  controversial because of the presence of the quasar
QSO~0241+622 within the relatively large COS B error-box. 
In the CGRO mission (1991 May- 1995 October)
 the source, given  there as 2EG~J0241+6119,
 was detected by EGRET with a significance of 17 $\sigma$,
 with a time averaged photon flux of $9.2 \pm 0.6 × 10^{-7}$ 
cm$^{-2}$ s$^{-1}$ for energies  $\ge$100 MeV. 
This flux reported by Kniffen et al. (1997) 
is slightly different from those quoted in 
the EGRET Catalogs because of the addition of data 
after the 1993 September cutoff date for the catalog. 
The position of this gamma-ray source  is $l$ =
135$^{\circ}$.58, $b$ = 1$^{\circ}$.13. 
As shown in  Fig.~\ref{fig:kniffen0}
the contour position obtained with additional data is about
11´ from the 2EG catalog position and about 40´
 from the old 2CG catalog position. 
The radius of the 95$\%$ confidence error contour is about 13´,
 ruling out the possible identification with
 QSO 0241+622 at l = 135$\degr$.7, b = 2$\degr$.2, which is 64´ away. 
The position is only  8´ distant from \lsi 
(Kniffen et al. 1997). 
In 1998 Tavani and collaborators established the
possibility of variability of 2CG~135+01 on timescales of days 
(Tavani et al. 1998). Massi (2004) examined the EGRET data as a function
of the orbital phase and noticed the clustering of high flux values around 
periastron passage.
Figure \ref{fig:massiegret} shows 
(Massi et al. 2004)
 the follow-up of the EGRET gamma-ray emission along one
full orbit. At epoch JD~2\,450\,334 (i.e. circles in the plot, with empty
circles indicating upper limits) the orbit has been well sampled at all
phases: A clear peak is centered at periastron passage 0.2 and 1.2. At a
previous epoch (JD~2\,449\,045; triangles in the plot) the sampling is
incomplete, but  the data show an increase near periastron passage at $\phi
\simeq $0.3, and a peak at $\phi\simeq$0.5. The 3 squares refer to a third
epoch (JD~2\,449\,471).

In conclusion, the gamma-ray data strongly support the ejection
at periastron passage predicted by the two-peak model (Massi 2004b).  
During the first ejection
(because of the proximity of the Be star)   
stellar photons  are upscattered by the inverse Compton process 
 by the relativistic electrons of the jet 
(Bosch-Ramon \& Paredes 2004).
The inverse Compton   losses are  so severe that no
electrons survive: radio outbursts indeed never have been observed at periastron
passage in more than 20 years of radio flux measurements (Gregory 2002).
At the second accretion peak the compact object is far enough  away from the Be-
star, so that energetic losses are smaller and electrons can propagate out of the
orbital plane. 
At this point the gamma-ray peak at $\phi\simeq0.5$ is very interesting.
It could originate from a second ejection which occurred still enough close to
the Be-star. In fact, while the first ejection is always at periastron passage,
the second ejection occurs at a varying point in the  orbital phase interval
0.4--0.8.

\section{RADIO INTERFEROMETRY: IMAGING AT HIGH RESOLUTION} \label{vlbi}

Nowadays it is possible to obtain images of jets at  infrared wavelengths 
and in X-rays (Sams et al. 1996; Corbel et al. 2002).
However,
these jets at 
tenths of arcseconds 
are not related to the  
 emitting regions  close to the engine (with  
quite short lifetimes because of their large adiabatic/synchrotron losses)
but require  a re-acceleration mechanism.
The  study of the jet closest as possible 
 to the ``engine'' at a spatial resolution up to
milliarcseconds  (mas)  is  possible at radio wavelengths
thanks  to  Very Long Baseline
Radio Interferometry (VLBI) (Appendix).

\subsection{The Jet Velocity} \label{jet}

\begin{figure*}[htb]
\begin{center}
\includegraphics[viewport=150 50 600 800 width=0.8\textwidth, height=0.8\textwidth]{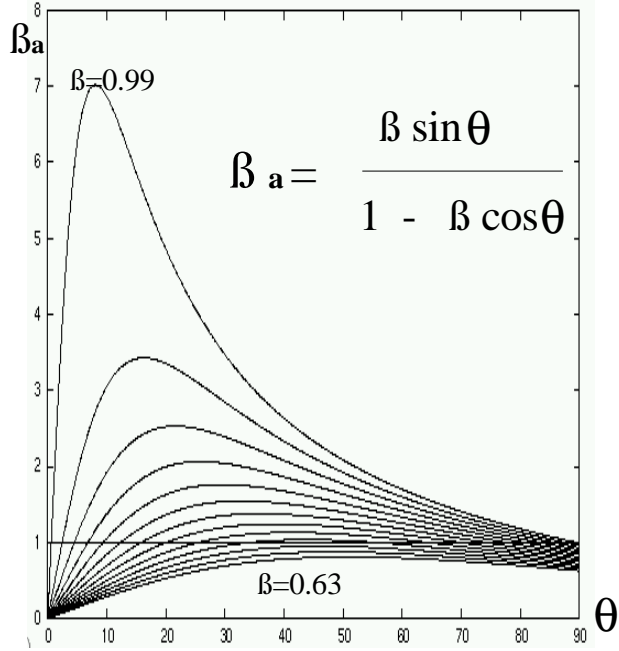}
\caption{{\bf Superluminal motion:} Apparent transverse velocity for the approaching jet as a function of angle $\theta$ for different values of the true velocity, from $\beta$=0.63 to $\beta$=0.99 by steps of 0.03.}
\label{fig:beta0}
\end{center}
\end{figure*}

For  symmetric ejection of two jets at a velocity $\beta$ (i.e. expressed as 
a fraction of $c$), the two  (approaching and receding) jets move with an 
apparent velocity of $\beta_{\rm a,r}$ 
(Mirabel \& Rodr\'{\i}guez 1994,  Fender 2004):
\begin{equation}
\beta_{\rm a,r}={\beta \sin\theta\over 1\mp \beta\cos\theta},
\label{eq:e1}
\end{equation}
$\theta$ is the angle between the direction of motion of the ejecta
and the line of sight to the observer. 
Depending on the angle,
for a jet with $\beta\ge 0.7$
the apparent velocity $\beta_{\rm a}$ of the approaching jet 
can become  greater than 1
(superluminal effect, see  
Fig.~\ref{fig:beta0}).

In order to show how the apparent velocity of the jet is derived, let us 
assume   $\theta$=90. In this case the proper motion, $\mu$, of the jet
on the sky plane is:
\begin{equation}
\mu={170~ \beta \over D},
\label{eq:e2}
\end{equation}
where the distance $D$ is in kpc and $\mu$ is expressed 
 in milliarcseconds per day
(mas/day).
The range of  $\beta$ is about  0.15--0.99 and the range
of the distance is about 1-12.5 kpc.
From the two extreme conditions, i.e
$\beta_{max}\over D_{min}$ and  $\beta_{min}\over D_{max}$, 
the proper motion ranges from  2 mas/day to 170 mas/day.
In order to estimate $\beta$ from multi-epoch observations one has to select
 the proper radio network. One must take into account
that  at $\lambda$=6cm  the VLBI provides a resolution of $\sim 1$ mas,
 MERLIN one of  $\sim $ 50 mas and the VLA in the largest  configuration
one of $\sim$ 100  mas. Therefore high proper motions are best studied 
with MERLIN and the VLA.

Beside multi-epoch observations
even with only one observation it is possible
to recover the quantity $\beta  \cos \theta$, if 
$\theta$ is significantly less than   90$^{\circ}$.
In fact, 
the observed flux densities 
$S_{\rm a,r}$ from the approaching and receding jets,
\begin{equation}
S_{\rm a,r}=S \delta_{\rm a,r}^{k-\alpha},
\label{eq:e3}
\end{equation}
(where $\alpha$ is the spectral index of the emission $S\propto
\nu^{\alpha}$ and $k$ is 2 for a continuous jet and 3 for discrete
condensations) 
are governed by the Doppler factor,
\begin{equation}
\delta_{\rm a,r}={1\over \gamma (1 \mp \beta\cos\theta)},
\label{eq:e4}
\end{equation}
(where $\gamma=(1-\beta^2)^{-1/2}$ is the Lorentz factor) 
and therefore (Mirabel \& Rodr\'iguez 1994)
one can determine the quantity $\beta \cos \theta$ by means of the ratio 
between the flux densities from the approaching and receding jet:
\begin{equation}
{S_{\rm a}\over{S_{\rm 
r}}}=\left({1+\beta\cos\theta\over1-\beta\cos\theta} \right)^{k-\alpha}.
\label{eq:e5}
\end{equation}

Let us  assume an ejection nearly aligned to the line
of sight with  $\theta$=0.5 and 
 with the other parameters:
 $\alpha=-0.5$, k=2 and  $\beta \sim0.6$. 
Using equation \ref{eq:e3} 
 one determines  
$\delta_{\rm a}^{k-\alpha}\simeq 6$ and  $\delta_{\rm r}^{k-\alpha}=0.1$.
As a result, the counter-jet can be rather  faint, 
and if $S_{\rm r}$ results to be  lower than 
the noise limit of the radio image
  the counter-jet will completely disappear.  
In this case the image will show a one-sided jet (the approaching one) 
and only a lower limit for $\beta \cos \theta$ can be estimated
using the noise limit of the image (Massi et al. 2001).

A constant ejection angle  $\theta$ implies a  constant  ratio between the
flux densities from the approaching and receding jet
during the epochs.
An obvious variation of this ratio is interpreted as a
variation of the ejection angle $\theta$ , explained as  jet precession.

\subsection{The Precessing Jet of \lsi} \label{precession} 

\begin{figure*}
\centering
\includegraphics[height=\textwidth, angle=-90.0 ]{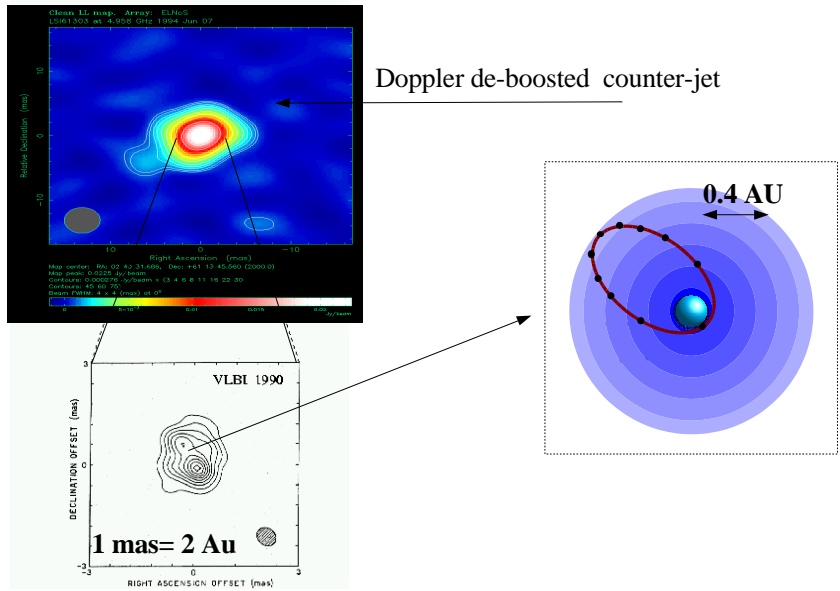}
\caption{{\bf: The relativistic jet of \lsi}
Left-Top:
EVN uniform weighted map of \lsi at 6 cm.
Only the approaching jet is visible, whereas the receding jet is attenuated
below the sensitivity limit of the image. One-sided jets are the observational evidence
that  
the angle between the ejecta and the line of sight is smaller
than 90$\degr$ {\bf and} 
that the jet is relativistic.
Left-Bottom:
VLBI observation of \lsi at 6cm.
A structure at PA$\sim~30^{\circ}$ (and therefore rotated in respect to
the one-sided jet of the EVN image) formed by two components
separated about 2 AU 
is surrounded by an envelope clearly rotated with respect to it.
This envelope could be an older expanding  
jet, previously ejected at another angle (because of precession).
Right: The stellar system. The accretor, whose disk is ''feeding'' the jet (mapped
with  EVN and VLBI) and the companion
Be-star with its equatorial disk.}
\label{fig:evn0}
\end{figure*}
The first VLBI observation resolving the source  
(Massi et al. 1993) in  Fig. \ref{fig:evn0} 
reveals a complex morphology (Fig.\ref{fig:evn0}~Bottom):
A structure at PA$\sim~30^{\circ}$  formed by two components
separated 0.9 mas 
(about 2 AU at the distance of 2 kpc)
is surrounded by an envelope clearly rotated with respect to it.
This envelope could be an older expanding  
jet, previously ejected at another angle (because of precession).

\begin{figure*}
\centering
\includegraphics[width=8.0cm, angle=-90]{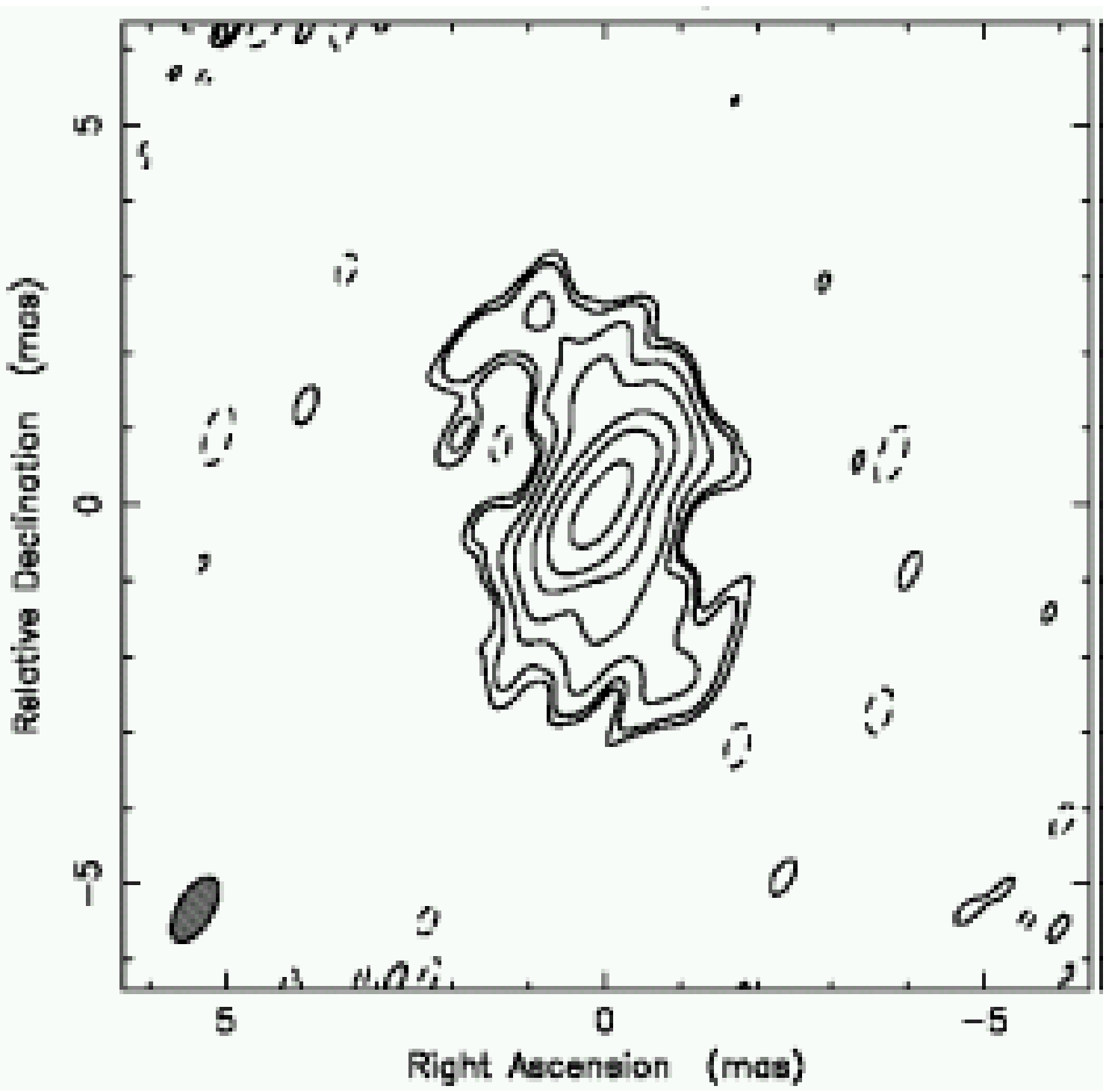}
\caption{{\bf: VLBI observations
in combination with the HALCA orbiting antenna.}Taylor et al. 2000.}
\label{fig:vsop}
\end{figure*}
Taylor and collaborators (2000)
 performing VLBI observations 
in combination with the HALCA orbiting antenna of the VSOP mission,
 imaged a structure {\bf reminiscent of the precessing radio 
jet seen in SS433} (Fig. \ref{fig:vsop}).
On the other hand at a lower resolution with a scale up to tens of AU  (with the EVN), 
Massi and collaborators (2001)
obtained an image that 
for the first time showed an elongation in a clear
direction  without any ambiguity (see Fig. \ref{fig:evn0}).
The most interesting aspect of the EVN map is that, for the first time, we
detected asymmetric emission in the southeast direction. 
Using the noise level ($\sigma$) of the map and the peak value of the
approaching component ($S_{\rm a}^{\rm peak}$)
for k=2 and $\alpha$=-0.5
we determine
 $\beta\cos\theta>0.6$.    
This would correspond to the two limits of $\theta < 53 \degr$ for 
$\beta \sim 1$ and $\beta\ge 0.6$ for $\theta=0 \degr$. 
A  value for the velocity  well within the range
$0.1~c$ to $0.9~c$ found for other
Microquasars
 (Mirabel \& Rodr\'{\i}guez 1999i).

Two observations at still lower resolution have been performed with MERLIN
(see Table 1).
The first MERLIN image shows a double  S-shaped jet 
extending  to about 200~AU on both sides of a central source.
\begin{table}
\begin{center}
\begin{tabular}{lcccc}
\hline \hline \noalign{\smallskip}
Date     & Start MJD & Stop MJD & $\phi_{\rm start}$ & $\phi_{\rm stop}$ \\
\noalign{\smallskip} \hline \noalign{\smallskip}
April 22 & 52021.73  & 52022.10 & 0.670              & 0.684 \\
April 23 & 52022.68  & 52023.17 & 0.706              & 0.724 \\
\noalign{\smallskip} \hline
\end{tabular}
\caption[]{Log of the MERLIN observations. Start and Stop are given in Modified Julian Date (MJD=JD$-$2400000.5). The corresponding orbital phases have been calculated using the new ephemerides, $t_0$=JD\,2443366.775 and $P$=26.4960~d, from Gregory 2002).}
\end{center}
\label{table:log}
\end{table}
\begin{figure*}[]
\centering
\resizebox{1.0\hsize}{!}{\includegraphics{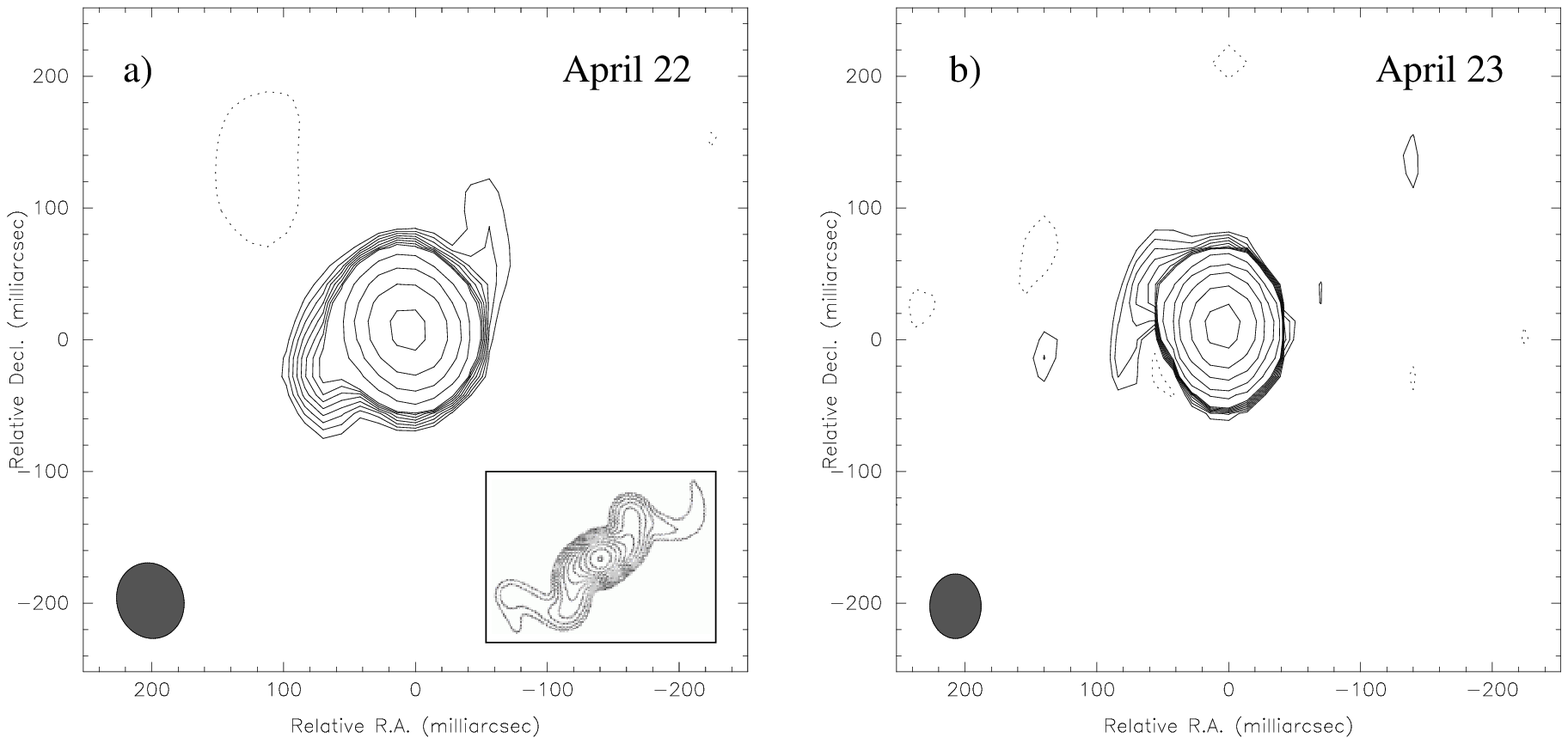}}
\caption{{\bf: The precessing jet of \lsi.} Massi et al 2004. 
a) MERLIN self-calibrated image of \lsi  at 5~GHz
using natural weights, obtained on 2001 April 22. North is up and East is to
the left. The synthesized beam has a size of 51 x 58~mas, with a PA of
17$\degr$. The contour levels are at -3, 3, 4, 5, 6, 7, 8, 9, 10, 20,
 40, 80,
and 160$\sigma$, being $\sigma$=0.14~mJy~beam$^{-1}$. The S-shaped morphology
strongly recalls the precessing jet of SS~433, whose simulated radio
emission (Fig.~6b in Hjellming $\&$ Johnston 1988), (rotated here
for comparison purposes) is given in the small box.
 b) Same as before but for the April 23 run and using uniform weights (see
text). The synthesized beam has a size of 39 x 49~mas, with a PA of
5 $-10^{\circ}$. 
 The contour levels are the same as those used in the April
 22 image but up to 320$\sigma$, with $\sigma$=0.12~mJy~beam$^{-1}$.}
\label{fig:merlin0}
\end{figure*}
The morphology of the MERLIN image (Fig.~\ref{fig:merlin0}a)
has a bent, S-like
structure. In the small box in Fig.~\ref{fig:merlin0}a we show  the simulated
radio emission from the Hjellming \& Johnston 1988 model of the
precessing jet of SS~433 (rotated here for comparison purposes). The
similarity between the MERLIN image of LS~I~+61$^{\circ}$303 and the
precessing model for SS~433 strongly suggests a precession of the jet 
of LS~I~+61$^{\circ}$303.
The precession  becomes evident in the
second MERLIN image, shown in Fig.~\ref{fig:merlin0}b, where a new feature 
oriented to Northeast at a position angle (PA) of 67$\degr$ is present. The
Northwest-Southeast jet of Fig.~\ref{fig:merlin0}a has  PA=124$\degr$.
Therefore, a quite large rotation has occurred in only 24 hours. This fast
precession causes a deformation of the morphology during the second
observation, and the one-sided jet appears bent in Fig~\ref{fig:merlin0}b. Only
3$\sigma$ features can be associated with the double jet of the day before.
The feature at 3$\sigma$ to the East is well compatible with a displacement of
$0.6c\times24$ hours.

\begin{figure*}[]
\includegraphics[width=15.0cm, height=14cm, angle=-90]{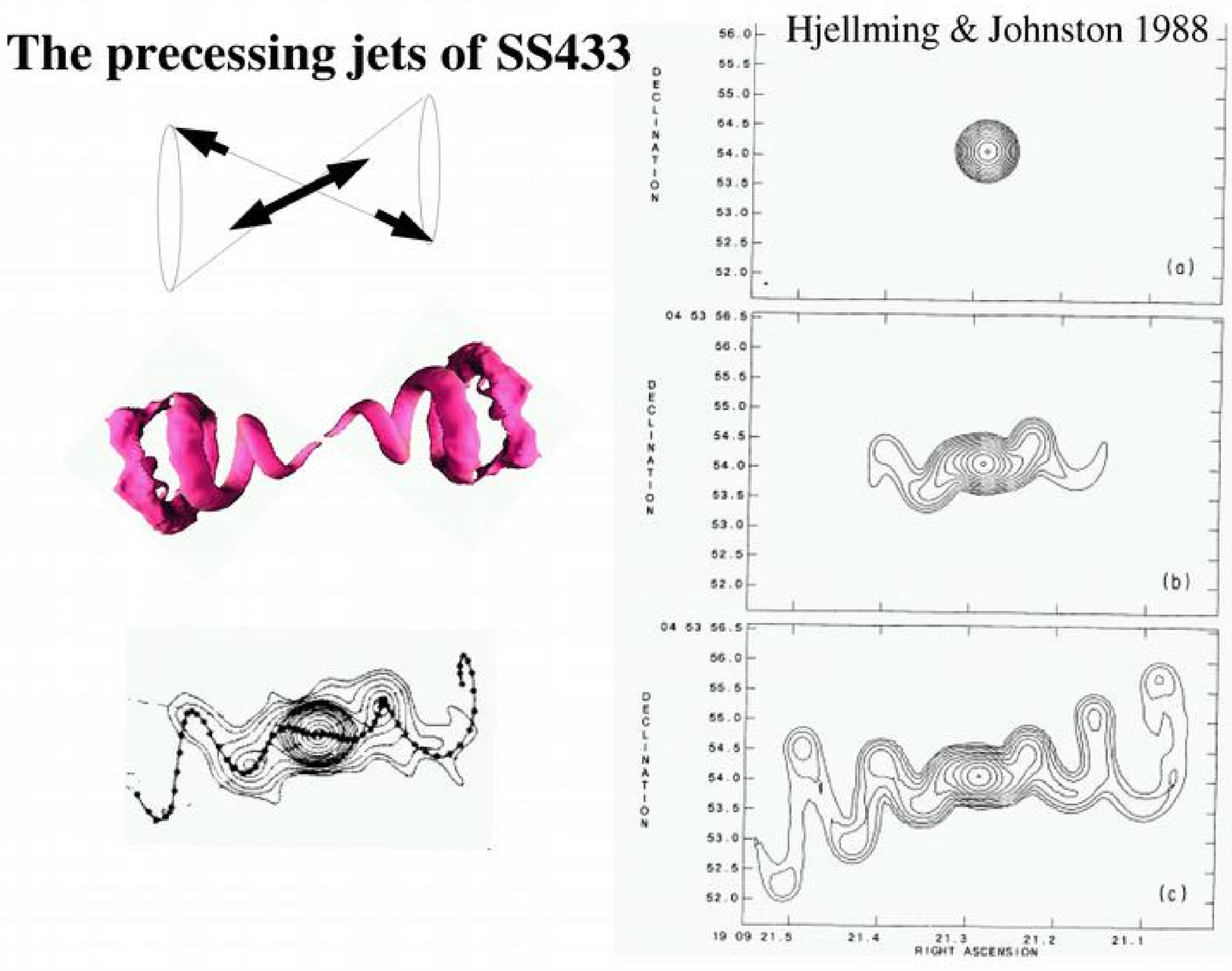}
\caption{{\bf: Morphology of a precessing and expanding jet.} A precessing jet describes a twin-corkscrew pattern that projected in the
sky plane resembles a quasi-sinusoidal path. Hjellming and Johnston (1988) have shown that
only for slow expansion (Right-Bottom) it is possible to observe the jet in its 
twin-corkscrew pattern, whereas for free expansion (i.e. strong adiabatic losses)
the morphology reduces to that on the Right-Top. The morphology in the center is that reproducing
the observed structure (Bottom-left) and corresponds to a slow expansion switched to free expansion
at a proper distance from the core.
}
\label{fig:precessione0}
\end{figure*}

The appearance of 
successive ejections of a precessing jet with  ballistic
motion of each ejection
is, as shown in Fig. \ref{fig:precessione0}, a curved path that, depending on the modality of the  expansion
and therefore on the adiabatic losses, 
seems  to be a  ``twin-corkscrew''  or a simply S-shaped pattern (Hjellming \& Johnston 1988; Crocker et al 2002).
The last one seems to be the case of {LS~I~+61$^{\circ}$303}.
Can we trace any ballistic motion of any jet component ?

\begin{figure*}[]
\includegraphics[width=14.0cm, height=3.5cm]{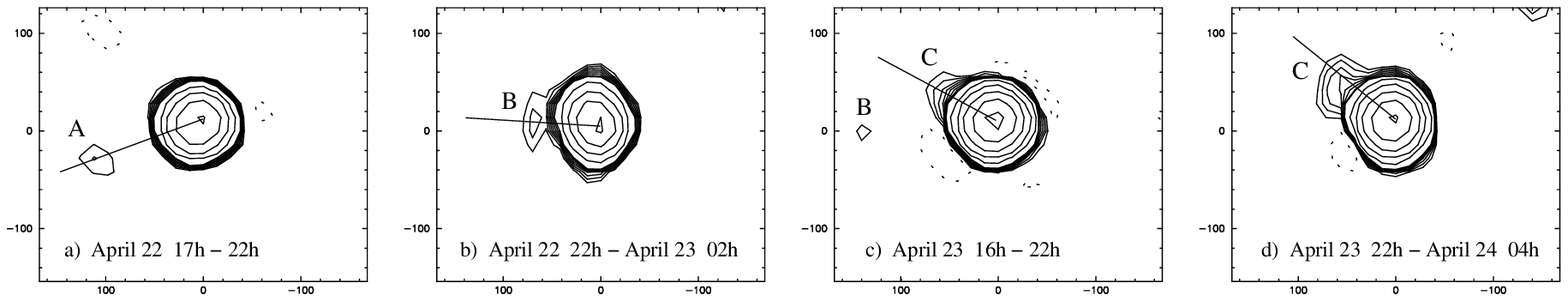}
\caption{
MERLIN self-calibrated images of {LS~I~+61$^{\circ}$303} at 5~GHz using uniform weights, obtained on 2001 April 22 and April 23. The data set of each epoch has been split into two blocks. A convolving beam of 40~mas has been used in
all images for better display. The first contour represents the $3\sigma$ level
in all images except for c), where we start from the $2\sigma$ level to display
the faint B component. The rms noises are $\sigma$=0.13~mJy~beam$^{-1}$, $\sigma$=0.20~mJy~beam$^{-1}$, $\sigma$=0.13~mJy~beam$^{-1}$, and $\sigma$=0.15~mJy~beam$^{-1}$, respectively. The PA of the ejections is indicated by a bar (see text).
}
\label{fig:four}
\end{figure*}
We have splitted the data base  of each MERLIN run into two subsets.
The first map (Fig. \ref{fig:four}-a) represents  the first
four hours of the first run.
 It shows an ejection "A" already quite displaced from
the core. In Fig.\ref{fig:four}-b, there is present a new  ejection "B"  at another PA. 
The combination of these two maps together produce as a consequence a
 "bent" jet, that is the southern jet of Fig.~\ref{fig:merlin0}a. The two counter-jets
for "A" and "B" are (Figs. \ref{fig:four}-a and \ref{fig:four}-b)
 too weak to be detected, and
they become visible only in the more sensitive image of Fig. ~\ref{fig:merlin0}a, where
all the jet and counter jets for A and B form together the "S-shaped" jet.   
The "B" component is still detectable after 9 hours in the
third image (Fig.\ref{fig:four}-c). Its motion is  ballistic: 
the PA is still the same (almost 90$^{\circ}$) as in Fig.\ref{fig:four}-b. 
A new component "C" is present at another PA 
(Fig.~\ref{fig:four}-d) 6 hours later, little rotation of the PA is compatible
with $\Delta$ PA$_{(\rm B-\rm C)}/ 3$ of the previous image.
The Northern elongation in  the higher sensitivity map of
Fig.~\ref{fig:merlin0}b  therefore is  the
result of 
a set of   little ejections  of
 a rotating stream. 
The noise level in this image is still lower than 
that in  Fig.~\ref{fig:merlin0}a, nevertheless the counter-jet is 
not visible. This implies a  decreased $\theta$ due to precession.
 In the case of
the MERLIN image of April 22 we derive $\beta\cos\theta=0.12$, which for
$\beta=0.6$ leads to an ejection angle of $\theta=78\degr$. This is an average
of the ejection angles  $\theta_{\rm A }$ and $\theta_{\rm B}$ of
 features A and B in  Figs. \ref{fig:four}a and \ref{fig:four}b.
A direct estimate of these angles
is prevented because of  the lack of the receding jets.
 Using the r.m.s. noise we derive $\theta_{\rm A} <
90\degr$, $\theta_{\rm B}< 80\degr$ and for the C ejection in
Fig.\ref{fig:four}c, $\theta_{\rm C} < 68 \degr$.

 Therefore, the angle between the jet and the line of
sight, 
$\theta$, has decreased  by more than 10$^{\circ}$  in 24 hours. 
It is this  much narrower alignment of the jet  with the line of sight, 
that causes the counter-jet to get further  Doppler
de-boosted with respect to the first
image and lets it disappear below the sensitivity limit of the image.

\section{CONCLUSIONS } \label{conclusions}

The conclusions of this review of the astronomical methods used for the
investigation of
Microquasars, with an examplary view on  the  source LS~I~+61$^{\circ}$303,
are:

\begin{enumerate}
\item
It is still an open issue, whether the 
compact object in this system is a neutron star or a black hole.
In fact, taking into account the uncertainty in inclination,  
mass of the companion and the mass function, the existence of a black 
hole cannot be ruled out.

\item
The observational results from  X-rays for  \lsi   are 
consistent with  transitions between 
X-ray spectral states   typical for a variable
accretion disk. These transitions are properly related 
to the onset of strong radio emission,
 as expected if the jet is ''fed'' by the disk.
Quasi-periodic oscillations at soft X-rays  
and  radio wavelengths are present in a  strong analogy with those 
observed in  GRS 1915+105.
They occur at  the onset and decay of large radio outbursts.
If confirmed,  this fact might
 indicate that 
at the beginning the matter is ejected  from the disk 
 in the form of discrete condensations (i.e. blob-like),
 then  follows  a steady state  where the matter-supply occurs at a
higher/continuous  rate (i.e. a  continuous jet)
and  finally the ejection  again ends up
in a blob-like  form.
(Hujerat \& Blandford 2003).
\item
At a scale of hundreds of AU the  radio jet   quite strongly changes
its morphology in  short intervals ($<$~24 hours),  evolving 
from an initial double-sided jet into  an one-sided
jet. This  variation
corresponds to a reduction of more than 10$^{\circ}$ in the angle between the jet and
the line of sight. This 
new alignment severely Doppler de-boosts the counter-jet.
Further observational evidence for a precessing jet
is recognizable even  at AU scales.

\item
The same population
of relativistic electrons emitting
 radio-synchrotron radiation
 upscatters - by inverse Compton processes -
 ultraviolet stellar photons  and produces Gamma-ray emission.
Ejections near the periastron passage 
produce  Gamma-ray flares but no radio flares, implying 
severe Compton losses.

\end{enumerate} 
We conclude that, because precession
and variable Doppler boosting
are   the causes of  the  rapid changes in the radio-morphology,
 precession and variable  Doppler boosting  are  likely
to produce    Gamma-ray variations at short time scales.
The amplification due to the Doppler factor
for Compton scattering of stellar photons by the relativistic electrons of the
jet is $\delta^{3-2\alpha}$ (where $\alpha<0$), and therefore is  higher than
that for synchrotron emission, i.e. $\delta^{2-\alpha}$ (Georganopoulos et al. 2001; Kaufman Bernad\'o
et~al. 2002).
 LS~I~+61$^{\circ}$303 becomes therefore an ideal laboratory to test
the recently proposed model for Microblazars with INTEGRAL and MERLIN
observations now and by AGILE and GLAST in the future.

\newpage
\section{SUMMARY}

Because of their accretion disk 
super-massive black holes, with 10$^6$-10$^9$ solar masses,
in the heart of galaxies 
are  the cause
for the most energetic sources of emission in our Universe. 
The centers of such galaxies are called Active Galactic Nuclei (AGN).
Some  AGN, like the  quasars,  
produce "jets" of subatomic particles with  speeds approaching that of light.
A microquasar - as its name suggests - is a miniature version of a quasar:
A disc of a few thousand kilometers radius surrounds a black hole of a few solar masses 
and two relativistic jets are propelled out of the disk  
 by the same process occurring in a Quasar.
 The  Microquasars therefore can serve  as a convenient "laboratory" 
for studying the physics of jets. 
The Microquasars are objects very much closer to us than Quasars  and  
the  study of the  evolution of relativistic jets can be done   
in  a few days only, whereas on the contrary  for far distant Quasars 
observations of many years apart are necessary to obtain appreciable proper motions of the radio jets.
Moreover, concerning the intrinsic variability
of Microquasars, these ``small'' objects  change   more quickly than Quasars: 
Considering as a characteristic time scale for variations 
$\tau\sim R_{Schwarzschild}/c \propto Mass$, 
phenomena of timescales of minutes connected with a Microquasar 
of 10 solar masses would take years  in a AGN of $10^7$ solar masses.
Such an enormous difference is the main reason why Microquasars
got such a great interest and growth in Astrophysics  
 in the last decade.

The Microquasars belong to the class of the X-ray binaries,
where   a compact object (black hole or neutron star)
accretes from a normal companion star. Such systems are
well known since the 1960s.  The X-ray emission originates from the
very hot accretion disk surrounding the compact object.
However, it took a long time to discover that some of  these systems also
have relativistic radio jets like  Quasars. 
For several years, after its discovery in 1979,
SS 433 with  its spectacular jets was thought to be a
unique exotic case, a mere curiosity in our galaxy.
Since the beginning of the
 1990, after  the discovery of other possible candidates 
of the same nature, several groups (including the author of this review)
have begun a systematic research on  
X-ray binaries with radio jets.

Here I  review the astronomical  methods used 
from Gamma-rays over X-rays and optical to radio wavelengths
for the investigation of these objects.
The  description of the methods is accompanied by
directly applying  them to
the system \lsi, one of the  most
enigmatic objects in our galaxy, because it is
associated with a
  variable  high-energetic Gamma-ray emission of unknown origin.

The nature of the accretor - a neutron star or a black hole -
is determined by optical measurements of the Doppler shift
of spectral lines of the normal star orbiting around the invisible
companion.
Observations at X-rays  probably are the most spectacular
ones, in respect to the progress in the knowledge of the accretion disk
and the disk-jet connection. Fitting the X-ray spectra information of
the size of the last stable orbit around the compact object can be derived
and  ejections of matter
into relativistic jets  can be related to variations of the disk.
The results from  X-ray observations for  \lsi   are
consistent with  transitions between 
spectral states  typical for a variable
accretion disk. These transitions are properly related
to the onset of strong radio emission
 as expected for a jet  ''fed'' by the disk.

Onset and decay of some large radio outbursts 
are   modulated with quasi-periodic oscillations  
that correspond to  repetitive ejections of 
  discrete condensations (i.e. blob-like).
Continuous ejections have 
a flat radio spectrum  and Low/Hard  X-ray state.

Speed and  morphology of the
ejections  at high resolution are  studied 
with radio interferometric techniques. 
The results of more than 10 years of VLBI/EVN and MERLIN observations 
of \lsi are presented together with our 
discovery of the relativistic jet and its precession.
Successive ejections are in   ballistic
motion; because of precession  their projected path on the sky plane
draw a bending  jet.

The radio bursts occur around apastron passage, where the low velocity
of the accretor enables it to capture more material of the wind
from the companion star. However, no bursts
are observed at periastron passage, where  accretion theory
predicts another super-accretion event.
There the accretor is  completely embedded in the densest part
of the wind.  I found that this great open question
about \lsi and the other enigma about the association of \lsi
with a variable Gamma-ray source are indeed not two separate questions,
but on the contrary one is the answer to the other: We do not
see a radio outburst at periastron passage, 
because we see a Gamma-ray outburst. 
With LS 5039 for the first time we identified
a Microquasar with an high-energy (E$>$ 100 Mev) source. This fact opens
the perspective that others of the more than one hundred still unidentified
EGRET sources could belong 
to a new class of objects: Gamma-ray Microquasars.

In this review 
I show that the 
variable Gamma-ray emission  
of \lsi is related to 
 the orbit of the system, with   peaks clustering where the companion
star - a strong emitter of ultraviolet photons - is closest (at periastron).
The suggested most probable explanation is that
the ejected relativistic electrons  are not able to emit
synchrotron radiation at radio wavelengths, because at periastron passage they 
are embedded in such a strong UV-field of radiation that  
they loose completely their energy by
inverse  Compton process. 
During the second accretion peak, the
compact object  is much farther away from the companion star and  
inverse Compton losses
are lower: The electrons can propagate out of the orbital
plane and   radio outbursts are observed.

\newpage
\section{ZUSAMMENFASSUNG}

Supermassive Schwarze Löcher mit 10$^6$-10$^9$ Sonnenmassen im Zentrum von 
Galaxien sind wegen ihrer Akkretionscheibe der Grund f\"ur die st\"arksten 
Strahlungsquellen in unserem Universum. Die Zentren solcher Galaxien werden 
Aktive Galaktische Nuklei (AGN) genannt. Einige AGN wie z.B. Quasare 
produzieren  "Jets" von subatomaren Teilchen mit Geschwindigkeiten 
bis nahe an die Lichtgeschwindigkeit. Ein Mikroquasar ist, wie der Name schon 
sagt, die Miniatur eines Quasars: Eine Scheibe von einigen tausend 
Kilometern umgibt ein Schwarzes Loch von einigen Sonnenmassen und zwei 
relativistische Jets werden durch denselben
Prozess wie bei Quasaren aus der Scheibe herausgeschleudert. 
Ein Mikroquasar kann deshalb 
als ein brauchbares ``Labor'' zum Studium der Physik solcher Jets
dienen. Die Mikroquasare sind Objekte, die wesentlich n\"aher zu uns liegen 
als Quasare und die Untersuchung der Evolution von relativistischen Jets
kann in nur ein paar Tagen geschehen,wohingegen man f\"ur weit entfernte
Quasare langj\"ahrige Beobachtungen ben\"otigt, um ausreichende 
Eigenbewegungen zu erhalten. Weiterhin, wenn man die intrinsische Variabilität
von Mikroquasaren betrachtet, ändern sich diese ``kleinen'' Objekte schneller
als Quasare: Nimmt man als charakteristische Zeitskala für Variationen
$\tau\sim R_{Schwarzschild}/c \propto Mass$, nehmen Phänomene mit einer 
Zeitskala von Minuten bei Mikroquasaren von 10 Sonnenmassen 
eine Zeit von  Jahren bei
einem Quasar von $10^7$ Sonnenmassen in Anspruch, wenn man es mit der Masse 
des Akkretors skaliert.Diese enorme Differenz ist der Hauptgrund, weshalb 
Mikroquasare in der letzten Dekade solch ein grosses Interesse und Wachstum 
in der Astrophysik auf sich gezogen haben.

Die Mikroquasare gehören zur Klasse der Röntgen-Doppelsterne, wo ein kompaktes
Objekt ( Schwarzes Loch oder Neutronenstern ) von einem normalen Begleitstern
einen Massenzuwachs erf\"ahrt. Solche Systeme sind seit den 1960er-Jahren gut
bekannt.Die Röntgenstrahlung stammt von der sehr heissen Akkretionsscheibe, 
die das kompakte Objekt umgibt. Allerdings dauerte es eine lange Zeit, bis man
entdeckte, dass einige dieser Systeme ebenso wie Quasare relativistische 
Radio-Jets aussenden. Für viele Jahre nach seiner Entdeckung
1979 galt SS 433 mit seinen spektakulären Jets als ein einzelner exotischer
Fall, eine einzigartige Kuriosität in unserer Galaxie. Anfag der  1990 nach der 
Entdeckung von anderen m\"oglischen Kandidaten derselben Art begannen einige Gruppen
( einschliesslich des Autors dieses Reviews) mit einer systematischen 
Forschungsarbeit an Röntgen-Doppelsternen mit Radio-Jets.

Ich gebe hier einen Überblick über die astronomischen Methoden, die im Bereich
von Gamma-Strahlung über Röntgen-Strahlung und optischen bis hin zu Radio 
Wellenlängen zur Untersuchung dieser Objekte angewandt werden. Die 
Beschreibung dieser Methoden wird unmittelbar begleitet durch die Anwendung
der Methoden auf das System \lsi, einem der rätselhaftesten Objekte in
unserer Galaxie,weil es mit einer variablen hochenergetischen Gamma-
Strahlungsquelle unbekannten Ursprungs verbunden ist.

Die Natur des Akkretors - ein Neutronenstern oder eine Schwarzes Loch - wird
durch optische Messungen anhand der Dopplerverschiebung von Spektrallinien
des normalen Sterns, der sich um seinen unsichtbaren Begleiter bewegt,
ermittelt. Beobachtungen im R\"ontgenbereich sind wahrscheinlich die 
spektakulärsten im Hinblick auf den Fortschritt bez\"uglich der Kenntnisse
über die Akkretionsscheibe und die Scheiben-Jet Verknüpfung. Indem man
die R\"ontgenspektren untersucht, erh\"alt man Informationen \"uber die letzte
stabile Bahn um das kompakte Objekt und ebenso kann man die Variation
dieser Grösse mit  dem Auswurf von Materie in die relativistischen Jets
correlieren. Die Ergebnisse von Röntgenbeobachtugen von \lsi sind konsistent
mit Übergängen zwischen spektralen Zuständen im Röntgenbereich typisch
für eine ver\"anderliche Akkretionsscheibe.Diese \"Uberg\"ange sind passend 
verbunden mit einem Anstieg von starker Radiostrahlung, wie man es für einen
Jet, der von der Scheibe ``gespeist'' wird, erwartet.

Anstieg und Abfall der starken Ausbr\"uche 
sind  moduliert mit quasi-periodischen Oszillationen, die wiederholten 
Auswürfen von diskreten Kondensationen  entsprechen.
Ununterbrochene Ausw\"urfen
haben  Radiospektrum flach und  Röntgenzustand  ``Low/Hard''.

Die Geschwindigkeit und die Morphologie der Ausw\"urfe wereden 
mithilfe der Radiointerferometrie-Technik untersucht.
Die Ergebnisse über mehr als zehn Jahre von Beobachtungen mit VLBI/EVN 
und MERLIN von \lsi werden hier zusammen mit unserer Entdeckung des 
relativistischen Jets und seiner Pr\"azession dargestellt. Aufeinander
folgende Ausw\"urfe folgen ballistischer Bewegung; wegen der Pr\"azession
bildet ihr auf die Himmelsebene projezierter Weg einen gebogenen 
 Jet.

Der Radioausbruch geschieht um den Apoastron-Durchgang herum, wobei die
geringe Geschwindigkeit es dem Akkretor erlaubt, mehr Material vom Wind
des Begleitstern einzufangen. Allerdings werden keine Ausbr\"uche beim
Periastron-Durchgang beobachtet, wo die Akkretions-Theorie ein
weiteres super-akkretives Ereignis vorhersagt. 
Dort ist der Akkretor vollständig vom dichtesten Teil des
Windes umgeben. Ich fand heraus, dass diese grosse offene Frage
über \lsi und das andere Rätsel bez\"uglich der Verbindung von \lsi mit der
veränderlichen Gamma-Strahlungsquelle in Wirklichkeit keine zwei
getrennten Fragen sind, sondern im Gegenteil ist die eine die Antwort
auf die andere: Wir beobachten keinen Radioausbruch beim Periastron-
Durchgang, weil wir einen Gamma-Strahlungsausbruch sehen.
Mit LS 5039 haben wir zum ersten Mal einen Mikroquasar mit einer
hochenergetischen (E$>$ 100 Mev) Quelle identifiziert. Diese 
Tatsache eröffnet die Perspektive, dass andere der mehr als hundert 
noch nicht identifizierten EGRET Quellen zu einer neuen Objektklasse 
geh\"oren k\"onnen: Gamma-Strahlungs-Mikroquasare.

In diesem Review zeige ich, dass die veränderliche Gamma-Strahlung von
\lsi mit der Umlaufbahn des stellaren Systems verbunden ist, mit 
sich l\"aufenden Spitzen dann, wenn der Begleitstern - ein starker Strahler von 
Ultraviolett-Photonen - am nächsten ist (beim Periastron-Durchgang). 
Die vorgeschlagene wahrscheinlichste Erkl\"arung ist, dass die 
ausgeworfenen relativistischen Elektronen nicht in der Lage sind,
Synchrotonstrahlung im Radiobereich auszusenden, weil sie beim 
Periastron-Durchgang in solch einem starken UV-Strahlungsfeld 
eingebettet sind, dass sie  vollstandig  ihre Energie
wegen des  inversen 
Compton Prozesses verlieren.
 W\"ahrend der zweiten
Akkretions-Spitze ist das kompakte Objekt wesentlich weiter weg 
von seinem Begleitstern und so sind Compton-Verluste geringer:
Die Elektronen können aus der Bahnebene heraustreten und man 
beobachtet Radioausbrüche.

\newpage

\section{APPENDIX: 
 Theory of  Very-Long-Baseline-Interferometry Data Analysis}

The lack of phase information had prevented VLBI from being a true imaging technique until Rogers and his co-workers (1974)
 applied a phase closure relationship. 
The introduction of the closure phase concept marks 
the beginning of a new era in VLBI. Many authors developed methods,  reviewed by Pearson and  Readhead (1984), which  explicitly or 
implicitly use this quantity. 
Massi (1989) showed how the methods  explicitly
 using the closure phase can be unified in one equation.
 In an attempt to unify all methods together Massi \& Comoretto
 (1990)
 found that all  methods turn out to be  particular cases of
 the method proposed by Schwab (1980) depending on a proper
 scheme of  baseline  weighing.

\begin{figure*}[]
\centering
\includegraphics[width=8cm]{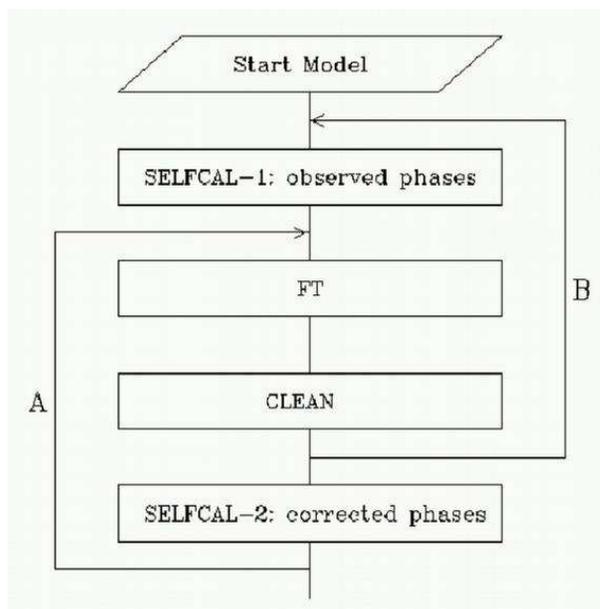}
\caption{{\bf: Hybrid mapping.} The loop "A" is 
the standard hybrid mapping scheme. Massi and Aaron (1999) have shown as 
the simple use of loop "B" 
eliminates the spurious structures created by standard hybrid mapping.}
\label{fig:aaron}
\end{figure*}

 Using   Schwab's method,   called self-calibration,a map 
of the radio source can be obtained by using an algorithm 
 which includes fourier transform and CLEAN,
 following an iterative  procedure first indicated
 by Readhead and Wilkinson (1978) and called Hybrid mapping (Fig. ~\ref{fig:aaron}).
 That this procedure can converge on wrong solutions has been pointed out by many authors in the past: Walker (1986) indicated the bias in the resultant data due to the use of a point source as starting model. B\aa \aa th (1989) suggested the use of the original data set  in each iteration of self calibration. Linfield (1986) analysed the role of the (u-v) coverage and lack of intermediate spacing. Generally,  the full procedure to avoid false features is  not clear and only experience with imaging helps  the user to avoid them.

Massi \& Aaron (1999)
demonstrated that the problem is connected with the
 non-linear nature of self calibration which leaves
 initial wrong assumptions frozen in the final solution. 
We demonstrated that the general precondition to avoid false
 structures in the map is that the errors
 (or more precisely their cube) of the model should  be smaller
 than the observed  closure phases. This condition, 
 generally satisfied  for a standard earth based array,
 is violated if  one telescope of the array is very displaced
 from the others, as it is for an array including a telescope
 mounted on a satellite. In this case one   should avoid the use
 of a point like model as starting model.
 Moreover, one should at each  iteration of self calibration,  
adopt the model derived by CLEAN directly on  the original data
 and not on the corrected data biased by  previous wrong  solutions
 (Fig. ~\ref{fig:aaron}).

Using self calibration  it is  assumed that the baseline based errors are negligible. In spite of the fact that these baseline errors are quite small their effect on the map's quality is rather serious. Tests were performed to determine at which level errors limit the obtainable dynamic range with the VLA (Perley 1986), with the VLBA (Briggs et al. 1994), with  the Nobeyama Radioheliograph  (Koshiishi et  al., 1994).
Massi and collaborators 
 have performed such an  analysis for the European VLBI
 Network (EVN).
The result was that the instrumental polarization (D terms)
of the telescopes of the network
 had an average value of  9 percent  arriving at some telescopes 
 at values of  20 percent.
 For comparison the values of  VLBA  telescopes were below 2 percent. 
The instrumental polarization was therefore the main reson for the lower performance of the EVN damaging the dynamic range of the images up a factor of 7
(Massi et al. 1991; 1996; 1997; 1997b; 1997c;
 1997d; 1997e; 1998; Massi 1999b).
\newpage
\section{REFERENCES}

Apparao, K.M.V.
2000, A\&A, 356, 972 
\\
\\
B\aa \aa th, 1989,  Very Long Baseline
Interferometry. Techniques and Applications. ed. M. Felli and  R. Spencer.
Klewer Academic Publishers. NATO ASI Series C. Vol.283, pag.206
\\
\\
Belloni, T., Klein-Wolt, M., Méndez, M., van der Klis, M. \& van Paradijs, J.
2000, A\&A, 355,27
\\
\\
Belloni, T.,
M\'endez,M., King, A.R., van der Klis, M. \&  van Paradijs,
J. 1997a, ApJ., 479, L145 
\\
\\
Belloni, T.,
M\'endez, M., King, A.R., van der Klis, M. \& van Paradijs,
J. 1997b, ApJ., 488, L109
\\
\\
Bondi, H, 1952, MNRAS, 112, 195
\\
\\
Bosch-Ramon, V. \& Paredes, J. M.
2004, A\&A, 425,1069
\\
\\
Briggs, D.S., Davis, R.J., Conway, J. E. \& Walker, R.C.
1994, July 25, VLBA memo 697
\\
\\
Brocksopp, C., Fender, R. P., McCollough, M., Pooley, G. G., Rupen, M. P., Hjellming, R. M., de la Force, C. J., Spencer, R. E., Muxlow, T. W. B., Garrington, S. T. \& Trushkin, S.
 2002, MNRAS, 331, 765
\\
\\
 Cadolle Bel, M., Goldwurm, A.,Rodriguez, J. et al. 2004. 
A\&A, 426, 659
\\
\\
Casares, J., Ribas, I., Paredes, J.~M., Mart\'{\i}, J. \&
C. Allende Prieto
2004, MNRAS,  submitted
\\
\\
Charles, P.A., \& Wagner, R.M.
1996, Sky \& telescope, May, 38.
\\
\\
Coppi, P. S. 2002
Bulletin of the American Astronomical Society, Vol. 32, p.1217
\\
\\
Corbel, S., Fender, R. P., Tzioumis, A. K.,
Tomsick, J. A., Orosz, J. A., Miller, J. M.,
Wijnands, R., \&  Kaaret, P.
2002, Science, 298, 196
\\
\\
Crocker, M.,M., Davis, R., J., Spencer, R., E., Eyres, S., P., S., Bode, M., F.,\& Skopal, A.
2002, MNRAS, 335 , 1100
\\
\\
Dulk, G. A., 1985, ARA\&A, 23, 169
\\
\\
Eikenberry, S.S., Matthews, K., Morgan, E., Remillard, R.A.,
\& Nelson, W. R.  1998, ApJ., 494, L61
\\
\\
Esin, A.A, Narayan, R.,Cui, W., Grove, J.E., \& Zhang, S.N.
1998, ApJ, 505, 854
\\
\\
Falcke, H., Körding, E., \& Markoff, S.
 2004, A\&A, 414, 895
\\
\\
Fender, R.P
2004, Compact Stellar X-Ray Sources,  W.H.G. Lewin \& M. van der Klis (Ed.),
 Cambridge University Press, Cambridge, astro-ph/0303339
\\
\\
Fender, R.P. \& Belloni, T.
2004, ARA\&A, 42, 317
\\
\\
Fender, R.P., Garrington, S.T., McKay, D.J.,
Muxlow, T.W.B., Pooley, G.G., Spencer, R.E., Stirling, A.M., \& Waltman, E.B.  
1999, MNRAS, 304, 865
\\
\\
Fender, R. P., Hjellming, R. M., Tilanus, R. P. J., Pooley, G. G., Deane, J. R., Ogley, R. N. \& Spencer, R. E.
2001, MNRAS, 322 , L23
\\
\\
Fender, R.P., Rayner, D., Trushkin, S.A., O'Brien, K., Sault, R.J., Pooley,
2002, Lect. Notes Phys. 589,  101 astro-ph/0109502
\\
\\
Fender, R.P., Rayner, D., Trushkin, S.A., O'Brien, K., Sault, R.J., Pooley,
G.G., \& Norris, R.P. 
2002, MNRAS, 330, 212
\\
\\
Fender, R. P., Roche, P., Pooley, G. G., 
 Chakrabarty, D.,  Tzioumis, A. K.,  Hendry, M. A.,   Spencer, R.E.
1996
Proceedings of 2nd INTEGRAL workshop : The Transparent Universe', ESA SP-382
astro-ph/9612080
\\
\\
 Fender, R. P., Spencer, R. E., Newell, S. J. \& Tzioumis, A. K.
 1997, MNRAS, 286, L29
\\
\\
Filippenko, A. V., Leonard, D. C., Matheson, T., Li, W., Moran, E. C., Riess, A. G. 1999, PASP, 111, 969
\\
\\
Fomalont, E. B., Geldzahler, B. J. \&  Bradshaw, C. F.
2001, ApJ, 558, 283
\\
\\
Frail, D.A., \& Hjellming, R.M.
1991, AJ, 101, 2126
\\
\\
Frank, J., King, A. \& Raine, D.J. 2002,
Accretion Power in Astrophysics, ARI, CUP
\\
\\
Gallo, E., Corbel, S., Fender, R. P., Maccarone, T. J. \& Tzioumis, A. K
2004, MNRAS, 347, 52L
\\
\\
Gallo, E., Fender, R. P. \&  Pooley, G. G.
2003, MNRAS, 344, 60
\\
\\
Geldzahler, B. J., Johnston, K. J.,  Spencer, J. H.,
 Klepczynski, W. J., Josties, F. J., Angerhofer, P. E., Florkowski, D. R., McCarthy, D. D., Matsakis, D. N.\& Hjellming, R. M.
1983, ApJ, .273, 65L
\\
\\
Georganopoulos, M., Kirk, J.G.,\&  Mastichiadis, A.
2001, ApJ, 561, 111
\\
\\
Gregory, P.C.  2002, ApJ, 575, 427.   
\\
\\
Gregory, P.C. \&  Neish, C. 2002, Ap. J. 580,  1133 
\\
\\
Gregory, P.C., \& Taylor, A.R.  1978, Nature, 272, 70
\\
\\
Greiner, J., \& Rau, A.  2001, A\&A 375,145
\\
\\
Hannikainen, D.,  Wu, K., Campbell-Wilson, D., Hunstead, R., Lovell, J., McIntyre, Vi., Reynolds, J., Soria, R. \& Tzioumis, T.
 2001, Exploring the gamma-ray universe:, Proc. A. Gimenez, V. Reglero \& C. Winkler. (Ed.),  ESA SP-459, Noordwijk: ESA Pub. Division,
 ISBN 92-9092-677-5,  291
\\
\\
Harrison, F.A., Ray, P.S., Leahy, D.A., Waltman,
E.B.,\&  Pooley, G.G. 2000,  ApJ, 528, 454
\\
\\
Hartman, R.C., Bertsch, D.L., Bloom, S.D., etal.
1999, ApJS, 123, 79
\\
\\
Heinz, S. \& Sunyaev, R. A.
2003, MNRAS, 343, 59
\\
\\
Hjellming, R.~M., \&  Johnston, K.~J.  1988, ApJ, 328, 600.
\\
\\
 Hjellming, R. M. \& Rupen, M. P.
 1995, Nature, 375, 464
\\
\\
Hjellming, R. M., Rupen, M. P., Hunstead, R. W., Campbell-Wilson, D., Mioduszewski, A. J., Gaensler, B. M., Smith, D. A., Sault, R. J., Fender, R. P., Spencer, R. E., de la Force, C. J., Richards, A. M. S., Garrington, S. T., Trushkin, S. A., Ghigo, F. D., Waltman, E. B. \&  McCollough, M.
2000, ApJ, 544, 977
\\
\\
Hjellming, R. M., Rupen, M. P., Mioduszewski, A. J., Smith, D. A., Harmon, B. A., Waltman, E. B., Ghigo, F. D. \& Pooley, G. G.
 1998, AAS, 193, 103.08 (Bull. AAS 30, 1405)
\\
\\
Hujeirat, A.,\&  Blandford, R. astro-ph/0307317
\\
\\
Hutchings, J.B., \& Crampton, D.  1981, PASP, 93, 486
\\
\\
Hutchings, J.B., Nemec, J.M., \& Cassidy, J.
1979, PASP, 91, 313
\\
\\
Kaufman Bernad\'o, M.M., Romero, G.E., \& Mirabel, I.F.
2002, A\&A, 385, L10
\\
\\
King, A. 
1996,  X-Ray Binaries,  W.H.G. Lewin, J.
  van Paradijs \& M. van der Klis (Ed.),
 Cambridge University Press, Cambridge,  419.
\\
\\
 Kniffen, D. A., Alberts, W. C. K., Bertsch, D. L., Dingus, B. L., Esposito, J. A., Fichtel, C. E., Foster, R. S., Hartman, R. C., Hunter, S. D., Kanbach, G., Lin, Y. C., Mattox, J. R., Mayer-Hasselwander, H. A., Michelson, P. F., von Montigny, C., Mukherjee, R., Nolan, P. L., Paredes, J. M., Ray, P. S., Schneid, E. J., Sreekumar, P., Tavani, M. \&  Thompson, D. J.
 1997,ApJ, 486, 126
\\
\\
Kogure, T.  1969, PASJ, 21, 71      
\\
\\
Koshiishi, H., Enome, S., Nakajima, H., Shibasaki, K., Nishio, M.,
Takano, T., Hanaoka, Y., Torii, C., Sekiguchi, H., Kawashima, S.,
Bushimata, T., Shinohara, N., Irimajiri, Y. \& Shiomi, Y.
1994, PASJ, 46, L33
\\
\\
Leahy, D. A. 2001 A\&A, 380,516    
\\
\\
Leahy, D.A.,  Harrison, F.A.,\&   Yoshida, A. 1997, ApJ, 475, 823  
\\
\\
Linfield R. P., 1986, A. J. 92, 21
\\
\\
Liu, Q.Z., Hang, H.R., Wu, G.J., Chang, J.,\&   Zhu, Z.X.
2000, A\&A, 359, 646
\\
\\
Liu, Q.,Z., van Paradijs, J., \& van den Heuvel, E.P.J. 
2000, A\&AS, 147, 25
\\
\\
Liu, Q.,Z., van Paradijs, J.,\&  van den Heuvel, E.P.J.
 2001, A\&A, 368, 1021
\\
\\
Longair, M.S. 1994,
High Energy Astrophysics, Vol. 2,
Stars, the Galaxy and the interstellar medium, 
 Cambridge University Press, Cambridge, 135.
\\
\\
Maccarone, T. J.  2004,
MNRAS 351,  1049
\\
\\
Maraschi, L.,\&  Treves, A.  1981, MNRAS, 194, 18     
\\
\\
Margon, B.A. 
1979 IAUC, 3345, 1
\\
\\
Margon, B.A. 
1980, Sci. Am. 243, 54
\\
\\
Margon, B.A. 1984, ARA\&A, 22, 507
\\
\\
Mart\'{\i}, J., \& Paredes, J.M.  1995, A\&A, 298, 151 
\\
\\
Massi, M.
1989, A\&A, 208, 392
\\
\\
Massi, M.
1999, Dissertation, The Dynamo and the Emission Processes in the Stellar System UX Arietis. University Bonn
\\
\\
Massi, M.
1999b, EVN Doc. n.91
\\
\\
Massi, M.
2003, Recent  Research Developments in Astronomy \& Astrophysics,
  eds. A.  Gayathri (Ed.), Kerala, India  700-712 (2003)
\\
\\
Massi, M.
2004, A\&A, 422, 26 
\\
\\
Massi, M.
2004b, proc. 7th EVN Symposium. Bachiller,Colomer,Desmurs,de Vicente (eds) October 12th-15 2004, Toledo, Spain astro-ph/0410502  
\\
\\
Massi, M., \&  Aaron, S.
1997c, EVN Doc. n. 75
\\
\\
 Massi, M., \&  Aaron, S.
1997d, EVN Doc. n. 77
\\
\\
Massi, M.\& Aaron, S.
1999, A\&AS, 136, 211
\\
\\
Massi, M., \&  Comoretto, G.
 1990, A\&A, 228,569
\\
\\
Massi, M., Comoretto, G.,   Rioja, M., \&  Tofani G.
1996, A\&A Suppl., 116, 167
\\
\\
Massi, M.,  Menten, K.,\&  Neidhöfer, J. 2002, A\&A, 382, 152.
\\
\\
Massi, M., Paredes, J.M., Estalella, R.,\&  Felli, M.
1993, A\&A, 269, 249
\\
\\
Massi, M., Rib\'o, M., Paredes, J.M., Peracaula, M., \& Estalella, R.
2001, A\&A, 376, 217
\\
\\
Massi, M., Rib\'o, M., Paredes, J.M., Garrington, S.T.,
Peracaula, M., \& Mart\'i, J.
2004, A\&A, 414, L1
\\
\\
Massi, M., Rib\'o, M., Paredes, J.M., Garrington, S.T.,
Peracaula, M., \& Mart\'i, J.
2004b, proc. of the Symposium on High-Energy Gamma-Ray Astronomy, Heidelberg, July 26-30, 2004 (AIP Proceedings Series)
astro-ph/0410504 
\\
\\
Massi, M., Rib\'o, M., Paredes, J.M.,
Peracaula, M., Mart\'i, J., \& Garrington, S.T.
2002b,  The 4th Microquasar Workshop, Ph. Durouchoux, Y. Fuchs \& J. Rodriguez,
(Ed.),  Center for Space Physics, Kolkata, 238
\\
\\
Massi,  M. Rioja, M.,   Gabuzda, D.,  Leppanen,K.
  Sanghera,H.,   Ruf, K., \&  Moscadelli L.
1997, A\&A, 318, L32
\\
\\
 Massi, M. Rioja, M. Gabuzda, D. Lepp"anen, K.  Sanghera, H.  Ruf, \& K. Moscad
elli, L.
1997b, Vistas in astronomy, vol. 41, Part 2
\\
\\
Massi, M.,  Ruf, K., \&  Orfei S.
1998, EVN Doc. n.85
\\
\\
Massi, M. Tofani, G. \&  Comoretto, G.
1991,A\&A, 251, 732
\\
\\
  Massi, M., Tuccari, G., \&  Orfei, S.
1997e, EVN Doc 81
\\
\\
Matsumoto, R., Uchida, Y.,Hirose, S., Shibata, K., Hayashi, M.R., Ferrari, A.,
\& Bodo,G. 
1996, ApJ, 461, 115.
\\
\\
McClintock, J.E., \& Remillard,R.A.
2004, Compact Stellar X-Ray Sources,  W.H.G. Lewin \& M. van der Klis (Ed.),
 Cambridge University Press, Cambridge,  astro-ph/0306213
\\
\\
Meier, D. L. 2001
ApJ, 548, 9
\\
\\
Meier, D. L., Koide, S.,\&  Uchida, Y.  2001, Science, 291, 84.
\\
\\
Mendelson, H., \& Mazeh, T.  1989 MNRAS 239, 733
\\
\\
Merloni, A., Fabian, A. C. \&  Ross, R. R.
2000, MNRAS, 313, 193
\\
\\
Merloni, A., Heinz, S., \& di Matteo, T.
2003, MNRAS 345, 1057
\\
\\
Meyer, F.,  Liu, B.F \&  Meyer-Hofmeister, E.
2000, A\&A 354, L67
\\
\\
 Mioduszewski, A. J., Hjellming, R. M. \& Rupen, M. P.
 1998, AAS, 192, 7402
\\
\\
Mirabel, I.F., Dhawan, V., Chaty, S.,  Rodr\'iguez, L.F.,
Mart\'i, J., Robinson, C.R., Swank, J., \& Geballe, T.R.
1998, A\&A, 330, L9
\\
\\
Mirabel, I.F.,\&    Rodr\'iguez, L.F.
1994, Nature, 371, 46
\\
\\
Mirabel, I.F., \& Rodr\'iguez, L.F.  1999, ARA\&A, 37, 409
\\
\\
Mirabel, I. F., Rodriguez, L. F., Cordier, B.; Paul, J.;\& Lebrun, F.
1992, Nature, 358, 215
\\
\\
Mitsuda, K., Inoue, H., Koyama, K., Makishima, K., Matsuoka, M.,
Ogawara, Y.,Shibazaki, N., Suzuki, K.,\&  Tanaka, Y.
1984, PASJ 36, 741
\\
\\
Narayan, R., \& Heyl, J.S.
2002, ApJ, 574, 139
\\
\\
Paredes, J.M., Estalella, R. \& Rius, A. 1990,
A\&A, 232, 377
\\
\\
Paredes, J.M. \& Figueras, F. 1986
  A\&A, 154, L30
\\
\\
Paredes, J.M., Marti, J., Estalella, R. \& Sarrate, J.
 1991,  A\&A, 248, 124
\\
\\
Paredes, J.M., Marti, J., Peracaula \& M., Ribo, M. 1997,
 A\&A, 320  ,L25
\\
\\
Paredes, J.~M., Mart\'{\i}, J., Rib\'o, M., \& Massi, M.
2000, Science, 288, 2340 
\\
\\
Paredes, J.M., Massi, M., Estalella, R., \& Peracaula, M.
1998, A\&A, 335, 539
\\
\\
Pearson, T. J. \& Readhead, A. C. S.
 1984, ARA\&A, 22, 97
\\
\\
Peracaula, M., Gabuzda, D.~C., \& Taylor, A.~R.
1998, A\&A, 330, 612
\\
\\
Peracaula, M.,  Mart\'{\i}, J.,\& Paredes, J.M.
1997, A\&A, 328, 283
\\
\\
Perley, R. A. 
1986, "Syntesis Imaging" proc. NRAO eds R.A. Perley,
F.R. Schwab \& A. H. Bridle, p.290
\\
\\
Punsly, B.  1999, ApJ, 519, 336
\\
\\
Readhead, A. C. S.  \&Wilkinson, P. N.
 1978, ApJ, 223, 25
\\
\\
Rhoades, C.E. \&   Ruffini, R. 1974
Physical Review Lett., 32, 324
\\
\\
Rodriguez, L. F., Mirabel, I. F., \&  Marti, J.
1992, ApJ, 401, L15
\\
\\
Rogers, A. E. E., Hinteregger, H. F., Whitney, A. R., Counselman, C. C., Shapiro, I. I., Wittels, J. J., Klemperer, W. K., Warnock, W. W., Clark, T. A. \& Hutton, L. K.
1974, ApJ, 193, 293
\\
\\
Sams, B.,Eckart, A., \& Sunyaev, R. 1996, Nature, 382, 47
\\
\\
 Schalinski, C. J., Johnston, K. J., Witzel, A. 
 Parsec-scale radio jets,  Proc.
\\
\\
 Schalinski, C. J., Johnston, K. J., Witzel, A., Spencer, R. E., Fiedler, R., Waltman, E., Pooley, G. G., Hjellming, R. \&  Molnar, L. A.
 1995, ApJ, 447, 752S
\\
\\
Schwab F. R., 1980, Proc. Soc. Photo-Opt. Inst. Eng. 231,18
\\
\\
Shakura, N.I.,\&  Sunyaev, R.A.
1973,A\&A,24, 337
\\
\\
Spencer, R. E.  1979 Nature, 282, 483
\\
\\
Spencer, R. E., Swinney, R. W., Johnston, K. J., \&  Hjellming, R. M. 
1986, ApJ, 309, 694 
\\
\\
Stewart, R. T., Caswell, J. L., Haynes, R. F. \&   Nelson, G. J.
1993, MNRAS. 261, 593 
\\
\\
Stirling, A.M., Spencer, R.E., De la Force, C.J., et~al.
2001, MNRAS, 327, 1273
\\
\\
Tanaka,Y.
1997, Accretion Disks-New Aspects, E. Meyer-Hofmeister \& H. Spruits (Ed.),
Lecture Notes in Physics 487.
Springer-Verlag Berlin Heidelberg New York, 1
\\
\\
Taylor, A.R., Dougherty, S.M., Scott, W.K., Peracaula, M., \& Paredes, J.M.
2000,  Astrophysical Phenomena Revealed by
Space VLBI,
H. Hirabayashi, P.G. Edwards, \& D.W. Murphy (Ed.), ISAS, 223
\\
\\
Taylor, A.R., \& Gregory, P.C.  1982, ApJ, 255, 210. 
\\
\\
Taylor, A.R.,  Kenny, H.T., Spencer, R. E.,\&  Tzioumis, A. 1992,
 ApJ, 395,268
\\
\\
Taylor, A.R., Young, G., Peracaula, M., Kenny, H.T.,\&  Gregory, P.C.
1996, A\&A, 305, 817
\\
\\
Tavani, M., Kniffen, D., Mattox, J.R., Paredes, J.M., \& Foster, R.S.
1998, ApJ, 497, L81
\\
\\
Tennant, A. F., Fabian, A. C., \& Shafer, R. A.
1986, MNRAS, 221, 27 
\\
\\
 Tingay, S. J., Jauncey, D. L., Preston, R. A., Reynolds, J. E., Meier, D. L., Murphy, D. W., Tzioumis, A. K., McKay, D. J., Kesteven, M. J., Lovell, J. E. J., Campbell-Wilson, D., Ellingsen, S. P., Gough, R., Hunstead, R. W., Jones, D. L., McCulloch, P. M., Migenes, V., Quick, J., Sinclair, M. W. \& Smits, D
 1995, Nature, 374, 141 
\\
\\
Torricelli, G.  Franciosini, E.,   Massi, M.,  Neidh\"ofer, J.
1998,  A\&A, 333, 970
\\
\\
Ulrich, M., Maraschi, L., \& Urry, C.M.
1997, ARAA, 35, 445 
\\
\\
Van der Klis, M.
2004, Compact Stellar X-Ray Sources,  W.H.G. Lewin \& M. van der Klis (Ed.),
Cambridge University Press, Cambridge, astro-ph/0410551
\\
\\
Van der Laan, H. 1966, Nature, 211, 1131
\\
\\
Van Paradijs, J. \& McClintock, J.E. 1996
1996,  X-Ray Binaries,  W.H.G. Lewin, J.
  van Paradijs \& M. van der Klis (Ed.),
 Cambridge University Press
\\
\\
Verbunt, F., \& van den Heuvel, E.P.J. 
1996,  X-Ray Binaries,  W.H.G. Lewin, J.
  van Paradijs \& M. van der Klis (Ed.),
 Cambridge University Press
\\
\\
Wallace, P.M., Griffis, N.J., Bertsch, D.L., Hartman, R.C., Thompson, D.J.,
Kniffen, \& D.A., Bloom, S.D.
2000, ApJ, 540, 184
\\
\\
Waters, L.B.F.M., van den Heuvel, E.P., Taylor, A.R., Habets, G.M.H.J.,
\&  Persi, P.  1988, A\&A, 198, 200.
\\
\\
White, N.E., Nagase, F., \& Parmar, A.N.
1996,  X-Ray Binaries,  W.H.G. Lewin, J.
  van Paradijs \& M. van der Klis (Ed.),
 Cambridge University Press, Cambridge,  1, 33, 6
\\
\\
Zamanov, R.~K.
1995, MNRAS, 272, 308
\\
\\
Zamanov, R. K., Reig, P., Martí, J., Coe, M. J., Fabregat, J., Tomov, N. A., Valchev, T.
2001, A\&A, 367, 884
\\
\\
Zamanov, R.K., Mart\'{\i}, J.  2000, A\&A, 358, L55                            
\\
\\
Zdziarski, A. A., Grove, J. E., Poutanen, J., Rao, A. R. \& Vadawale, S. V.
2001, ApJ, 554, 45L
\\
\\
Zhang, S. N., Cui, W., Harmon, B. A., Paciesas, W. S., Remillard,
R. E., \& van Paradijs, J. 1997, ApJ, 477, L95
\\
\\
 
\newpage
\section{DANKSAGUNG}

Mein besonderen Dank gilt Prof. Ulrich Mebold f\"ur die Gelegenheit
diese Arbeit durchzuf\"uhren 
und Prof. Karl Menten f\"ur seine stetige 
Unterst\"utzung und sein kontinuierliches Interesse an diesem Projekt.

Die Resultate dieser Arbeit  wurden \"uber die letzten Jahre in Zusammenarbeit
mit vielen Kollegen gewonnen; f\"ur die Arbeit mit dem VLBI-, EVN- und
MERLIN- Netzwerk sei 
besonders Marc Rib\'o, Prof. Josep Paredes,  Prof. Josep Mart\'{\i},
 Simon Garrigton and Marta Peracaula gedankt. 

Ich m\"ochte Prof. Ralph Spencer  f\"ur seine
hilfreichen Kommentare, Prof. Rolf Chini  f\"ur seine Unterst\"utzung und
J\"urgen Kerp  f\"ur alle seine  wichtigen 
praktischen Ratschl\"age danken.

Meinem Mann, J\"urgen Neidh\"ofer gilt mein Dank f\"ur
die zahlreichen  Diskussionen, die unseren gemeinsamen Interessen 
an den physikalischen Prozessen von  Doppelstern-System and ihren
Periodizit\"aten galten,
und f\"ur die kritische Durchsicht des Manuskripts. 

Danken m\"ochte ich auch meinen beiden S\"ohnen Guido und Claudio
f\"ur ihre Geduld, wenn ich einige "sekund\"are" Sachen wie 
"die W\"ascheb\"ugeln"
wochen lang verschoben habe.

\newpage
\section{ABSTRACT}

The Astrophysics of microquasars - galactic miniatures
of the far distant quasars - has become
one of the most active fields of modern Astronomy  in  recent years.
Here I  review the astronomical  methods used
for the investigation of these objects,
from Gamma-rays over X-rays and optical to radio wavelengths.
The description of each astronomical method
is always followed by an examplary application
on the  source LS I+61 303, one of the most observed  
Be/X-ray binary systems
because of its   periodical  radio emission and 
strong,  variable Gamma-ray emission.

\end{document}